%% file: main_LoP.tex
\documentclass[a4paper,11pt]{article}
\pdfoutput=1 
\usepackage{jcappub} 
\usepackage[T1]{fontenc} 

\title{The anomaly of the CMB power with the latest \textit{Planck} data}

\author[a,b]{M.~Billi}
\author[a]{R.B.~Barreiro}
\author[a]{E.~Martínez-González}

\affiliation[a]{
Instituto de Física de Cantabria, CSIC-Universidad de Cantabria, Av. de Los Castros s/n, 39005 Santander, Spain
}
\affiliation[b]{
Istituto Nazionale di Astrofisica - Osservatorio di Astrofisica e Scienza dello Spazio di Bologna, via Gobetti 101, I-40129 Bologna, Italy
}

\emailAdd{matteo.billi2@studio.unibo.it, barreiro@ifca.unican.es, martinez@ifca.unican.es}

\abstract{
The lack of power anomaly is an unexpected feature observed at large angular scales in the maps of Cosmic Microwave Background (CMB) produced by the COBE, WMAP and \textit{Planck} satellites. This signature, which consists in a missing of power with respect to that predicted by the $\Lambda$CDM model, might hint at a new cosmological phase before the standard inflationary era. 
The main point of this paper is taking into account the latest \textit{Planck} polarisation data to investigate how the CMB polarisation improves the understanding of this feature. With this aim, we apply to the latest \textit{Planck} data, both PR3 (2018) and PR4 (2020) releases, a new class of estimators capable of evaluating this anomaly  by considering temperature and polarisation data both separately and in a jointly way. This is the first time that the PR4 dataset  has been used to study this anomaly. To critically evaluate this feature, taking into account the residuals of known systematic effects present in the \textit{Planck} datasets, we analyse the cleaned CMB maps using different combinations of sky masks, harmonic range and binning on the CMB multipoles. 
Our analysis shows that the estimator based only on temperature data confirms the presence of a lack of power with a lower-tail-probability (LTP), depending on the component separation method, $\leq 0.33\%$ and $\leq 1.76\%$ for PR3 and PR4, respectively. To our knowledge, the LTP$\leq 0.33\%$ for the PR3 dataset is the lowest one present in the literature obtained from \textit{Planck} 2018 data, considering the \textit{Planck} confidence mask. We find significant differences between these two datasets when polarisation is taken into account most likely due to a different level of systematics. Especially, the analysis with PR3 data, unlike that with PR4, seems to point towards a lack of power at large scales also for polarisation. 
Moreover, we also show that for the PR3 dataset the inclusion of the subdominant polarisation information provides estimates  that are less likely accepted in a $\Lambda$CDM cosmological model than the only-temperature analysis  over the entire harmonic-range considered.  In particular, at $\ell_{max} = 26$, we found that no simulation has a value as low as the data for all the pipelines.}

\begin{document}
\maketitle
\flushbottom


\input{1intro}

\input{2MathFramework}
\input{3dataset}

\input{4results}
\input{5conclusions}

\section*{Acknowledgements}
We thank J.D. Bilbao-Ahedo, C. Gimeno-Amo, A. Gruppuso, R. Keskitalo and P.Vielva, for valuable advice and comments.
MB would like to thank the Angela Della Riccia Foundation for the financial support provided under the ADR fellowships 2022 and 2023. 
We acknowledge partial financial support from the grant PID2019-110610RB-C21 funded by MCIN/AEI/$10.13039/501100011033$ and from the Red de Investigaci\'on RED2022-134715-T funded by MCIN/AEI/10.13039/5011000011033.
This research used resources of the National Energy Research Scientific Computing Center (NERSC), a U.S. Department of Energy Office of Science User Facility located at Lawrence Berkeley National Laboratory, operated under Contract No. DEAC02-05CH11231.
This work based on observations obtained with \textit{Planck} (\url{https://www.esa.int/Planck}), an ESA science mission with instruments and contributions directly funded by ESA Member States, NASA, and Canada. Some of the results in this paper have been derived using the HEALPix \citep{Gorski:2004by} package and {\sc ECLIPSE} \footnote{\url{https://github.com/CosmoTool/ECLIPSE}} code \cite{Bilbao-Ahedo:2021jhn}.



\bibliographystyle{JHEP}
\bibliography{bibliography}



\appendix
\input{appendixA}
\input{appendixB}
\input{appendixC}

\end{document}

%% file: 1intro.tex
\section{Introduction}
\label{sec:intro}

The CMB anomalies are unexpected features at large angular scales in the CMB maps, observed by the Cosmic Background Explorer (COBE), the Wilkinson Microwave Anisotropy Probe (WMAP) and the \textit{Planck} satellites, that deviate from the $\Lambda$CDM cosmological model  with a statistical significance typically around 2-3 $\sigma$ confidence level (C.L.). These anomalies are not independent, showing different levels of correlation between them \cite{Muir:2018} (notice however that these correlations are not present in the tails of the distributions as shown by \cite{Jones:2023ncn}). In particular, the maps of temperature anisotropies show low variance \cite{Monteserin:2007fv}, a lack of correlation at the largest angular scales \cite{Schwarz:2015cma}, a preference for odd parity modes \cite{Land:2005jq}, a hemispherical power asymmetry \cite{Eriksen:2003db, Gimeno-Amo:2023jgv}, an alignment between several low multipole moments \cite{deOliveira-Costa:2003utu}, an alignment between these low multipole moments and the motion and geometry of the Solar System \cite{Copi:2006tu}, and an unexpected large cold spot in the Southern hemisphere \cite{Vielva:2003et}. Historically, the first observed anomalous feature, already within the COBE data, was the smallness of the quadrupole moment. It was confirmed to be low when WMAP released its data \cite{WMAP:2003ivt}. However it was also shown that the cosmic variance allows for such a small value. Another rediscovery in the first release of WMAP \cite{WMAP:2003ivt} was that the angular two-point correlation function is unexpectedly close to zero at angular scales larger than 60 degrees, where a significant correlation signal is expected. This feature had already been observed by COBE \cite{Hinshaw:1996ut}, and was confirmed by WMAP and \textit{Planck}.

Here we focus on the lack of power anomaly\footnote{ It should be noted that the lack of power anomaly and the lack of correlation, although related, are distinct phenomena. In general, the latter cannot be simply explained by reducing the power through different models.}, an intriguing feature that seems to be correlated with a low quadrupole, although it cannot be explained by a lack of quadrupole power alone. This anomaly consists in a missing of power with respect to that predicted by the $\Lambda$CDM model.  
An early fast-roll phase of the inflation could naturally explain such a lack of power, see e.g. \cite{Contaldi:2003zv,Cline:2003ve,Destri:2009hn,Cicoli:2013oba,Dudas:2012vv}: this anomaly might then witness a new cosmological phase before the standard inflationary era (see e.g. \cite{Gruppuso:2015zia,Gruppuso:2015xqa,Gruppuso:2017nap}).
There are three possible explanations for these large-scale CMB features: they could have cosmological origins \cite{Contaldi:2003zv}, they could be artefacts of astrophysical foregrounds and instrumental systematics, or they could be statistical flukes. WMAP and \textit{Planck} are in good agreement on this feature, so it is very hard, albeit not impossible, to attribute this anomaly to instrumental effects. 
Moreover, it is also difficult to believe that a lack of power could be generated by residuals of astrophysical emission, since the latter is not expected to be correlated with the CMB and therefore an astrophysical residual should increase the total power rather than decrease it. In addition, in the case of a positive correlation between CMB photons and extra-galactic foregrounds, as seen in \cite{Luparello:2022kqb}, a correction for this effect would increase the discrepancy with the model, confirming the CMB lack of power \cite{Hansen:2023gra}. Hence, it appears natural to accept this as a real feature present in the CMB pattern. 

This effect has been studied with the variance estimator in WMAP \cite{Monteserin:2007fv,Cruz:2010ud,Gruppuso:2013xba} and in \textit{Planck} 2013 \cite{Planck:2013lks}, \textit{Planck} 2015 \cite{Planck:2015igc} and \textit{Planck} 2018 \cite{Planck:2019evm} data, measuring a lower-tail-probability (LTP) of the order of a few per cent. Such a percentage can be even smaller, below $1 \%$, if only regions at high Galactic latitude are considered \cite{Gruppuso:2013xba}. However, when using only temperature data, this anomaly is not sufficiently significant to claim new physics beyond the standard cosmological model.
In a previous work \cite{Billi:2019vvg}, the former statement was revisited by also considering polarisation data. The authors proposed a new one-dimensional estimator for the lack of power anomaly capable of taking both temperature and polarisation into account. 
By applying this estimator to \textit{Planck} 2015 data (PR2), they found the probability that a random realisation is statistically accepted decreases by a factor of two when polarisation is taken into account. Moreover, they forecasted that future CMB polarisation data at low-$\ell$ might increase the significance of this anomaly.

The main point here is to deeply investigate how CMB polarisation improves the understanding of this feature, expanding on \cite{Billi:2019vvg}.
With this aim, we apply to the latest \textit{Planck} data, both PR3 (2018) and PR4 (2020) releases, a new class of estimators, defined starting by the mathematical framework assessed in \cite{Billi:2019vvg}, able to evaluate this anomaly considering temperature and polarisation data both separately and in a jointly way.
Up to now, lack of power has never been studied with PR4 dataset.
For other studies on the CMB anomalies using the PR3 dataset see \cite{Planck:2019evm}, \cite{Chiocchetta:2020ylv} and \cite{Shi:2022hxc}, and using both PR3 and PR4 see \cite{Gimeno-Amo:2023jgv}.
As described in \cite{Planck:2020olo}, compared to the previous releases, the PR3 but especially
the PR4 data better describe the component in polarisation of CMB: in fact PR3 and PR4 show reduced levels of noise and systematics at all angular scales and an improved consistency across frequencies, particularly in polarisation \cite{Planck:2020olo}. 
In order to critically evaluate this feature, taking into account the residuals of known systematic effects present in the \textit{Planck} datasets, we consider the foreground-cleaned CMB maps (data and end-to-end simulations) studying different combinations of sky-masks, harmonic range and binning on the CMB multipoles. 

The paper is structured as follows: in Section \ref{sec:MathFramework} we introduce our new class of estimators for evaluating the lack of power; in Section \ref{sec:dataset} we describe the datasets and the simulations used; in Section \ref{sec:Results} we present the results on \textit{Planck} PR3 and PR4 data; the conclusions are given in Section \ref{sec:Conclusion}. In addition, in appendix \ref{sec:appendixA} we give the full derivation of our joint estimator, in appendix \ref{sec:appendixB} we discuss in more detail the transfer function affecting the PR4 polarised signal at the largest angular scales, and in appendix \ref{sec:appendixC} we show the application to latest \textit{Planck} {\tt SEVEM} datasets of the lack of power estimator developed in \cite{Billi:2019vvg}.

%% file: 2MathFramework.tex
\section{Mathematical Framework}
\label{sec:MathFramework}

In this section, we first briefly recall the definitions of normalised angular power spectra (NAPS)\cite{Billi:2019vvg} and then show how they can be combined to derive a new class of estimators for the lack of power, based on the usual definition of the variance of a power spectrum in harmonic space.

\subsection{Normalised Angular Power Spectra (NAPS)}
\label{subsec:NAPS}

In section 2 of \cite{Billi:2019vvg}, starting from the usual equations used to simulate temperature and E-mode CMB maps (Eqs. (1) and (2) in \cite{Billi:2019vvg}), the normalised angular power spectra (henceforth NAPS) $x^{(\mathrm{TT})}_{\ell}$ and $x^{(\mathrm{uEE})}_{\ell}$ were defined as follows\footnote{Note that in \cite{Billi:2019vvg} $x^{(\mathrm{TT})}_{\ell}$ and $x^{(\mathrm{uEE})}_{\ell}$ are called $x^{(1)}_{\ell}$ and $x^{(2)}_{\ell}$, respectively.}:

\begin{eqnarray}
 x^{(\mathrm{TT})}_{\ell} & \equiv & \frac{C_{\ell}^{\mathrm{TT}} }{C_{\ell}^{\mathrm{TT,th}}} \, , \label{eq:defNAPS1} \\
x^{(\mathrm{uEE})}_{\ell} & \equiv & \frac{C_{\ell}^{\mathrm{EE}}}{a_{\ell}^{2}} C_{\ell}^{\mathrm{TT,th}} - \frac{C_{\ell}^{\mathrm{TT,th}}}{a_{\ell}^{2}} \left(\frac{C_{\ell}^{\mathrm{TE,th}} }{C_{\ell}^{\mathrm{TT,th}}}\right)^{2} C_{\ell}^{\mathrm{TT}} \nonumber \\                                                                      
                                                                         && - 2 \frac{C_{\ell}^{\mathrm{TE,th}}}{a_{\ell}^{2}}\left[C_{\ell}^{\mathrm{TE}} - \frac{C_{\ell}^{\mathrm{TE,th}} }{C_{\ell}^{\mathrm{TT,th}}} C_{\ell}^{\mathrm{TT}}\right] \, ,
\label{eq:defNAPS2}
\end{eqnarray}
where $a_{\ell}$ is:
\begin{equation}
a_{\ell}  \equiv  \sqrt{C_{\ell}^{\mathrm{EE,th}} C_{\ell}^{\mathrm{TT,th}} -  (C_{\ell}^{\mathrm{TE,th}})^{2}} \, .
\label{a}
\end{equation}
Therefore the NAPS $x^{(\mathrm{TT})}_{\ell}$ and $x^{(\mathrm{uEE})}_{\ell}$ represent a normalised version of the temperature and the uncorrelated E-component power spectra, respectively. One can interpret $C_{\ell}^\mathrm{{TT}}$, $C_{\ell}^{\mathrm{EE}}$ and $C_{\ell}^{\mathrm{TE}}$ as the CMB Angular Power Spectra (APS) estimated from a CMB experiment under realistic conditions, i.e. including noise residuals, incomplete sky fraction, finite angular resolution and also residuals of systematic effects. While the spectra $C_{\ell}^{\mathrm{TT, th}}$, $C_{\ell}^{\mathrm{EE, th}}$ and $C_{\ell}^{\mathrm{TE, th}}$ represent the underlying theoretical model.
The expected values for NAPS are:
\begin{eqnarray}
\langle x^{(\mathrm{TT})}_{\ell} \rangle= & \langle x^{(\mathrm{uEE})}_{\ell} \rangle & = 1 \, , 
\label{eq:expectedNAPS}
\end{eqnarray}
for each $\ell$. Therefore, the advantage of using NAPS instead of the standard CMB APS, i.e $C_{\ell}$, is that they are dimensionless and have similar amplitude numbers. For these reasons, they can be easily combined to define 1-D estimators in harmonic space  that depend on temperature, E-mode polarisation and their cross-correlation. In \cite{Billi:2019vvg}, the authors combined the NAPS $x^{(\mathrm{TT})}_{\ell}$ and $x^{(\mathrm{uEE})}_{\ell}$ to define an optimal, i.e. minimum variance, estimator called $\mathrm{\tilde P}$. This estimator, which can be interpreted as a dimensionless normalised mean power, jointly combines the temperature and polarisation data to evaluate the lack of power. Appendix \ref{sec:appendixC} shows its application to the latest \textit{Planck} dataset. 

Finally, here we define two new NAPS:
\begin{eqnarray}
x^{(\mathrm{EE})}_{\ell} & \equiv & \frac{C_{\ell}^{\mathrm{EE}} }{C_{\ell}^{\mathrm{EE,th}}} \, , \label{eq:defNAPSEE} \\
x^{(\mathrm{TE})}_{\ell} & \equiv & \frac{C_{\ell}^{\mathrm{TE}} }{C_{\ell}^{\mathrm{TE,th}}} \,, \label{eq:defNAPSTE} 
\end{eqnarray}
with the expected values equal to one. The NAPS $x^{(\mathrm{EE})}_{\ell}$ and $x^{(\mathrm{TE})}_{\ell}$, which depend only on the EE and TE CMB spectra respectively, allow us to define specific estimators able to test the lack of power by considering only the E-modes and only their cross-correlation with temperature. It should be noted that $x^{(\mathrm{EE})}_{\ell}$ takes into account both the uncorrelated and the correlated E-mode components.

\subsection{Variance-based estimators}
The idea is to look for estimators sensitive to the variance that are linear in the observed NAPS and also minimum variance.

Starting from the usual definition of the variance of a power spectrum, we define the joint estimator $E_\mathrm{V}^{\mathrm{joint}} $, based on the NAPS $x^{(\mathrm{TT})}_{\ell}$ and $x^{(\mathrm{uEE})}_{\ell}$:
\begin{equation}
E_\mathrm{V}^{\mathrm{joint}}  = \sum_{\ell=2}^{\ell_{\mathrm{max}} } \frac{2\ell+1}{4\pi } \left( \gamma_{\ell} x^{(\mathrm{TT})}_{\ell} + \epsilon_{\ell} x^{(\mathrm{uEE})}_{\ell} \right) \,,
\label{eq:EVjoint}
\end{equation}
where $\gamma_{\ell}$ and $\epsilon_{\ell}$ are appropriate coefficients associated with each NAPS. 
Taking the ensemble average of Eq.~(\ref{eq:EVjoint}) we obtain:
\begin{eqnarray}
\langle E_\mathrm{V}^{\mathrm{joint}}  \rangle &=& \langle \sum_{\ell=2}^{\ell_{\mathrm{max}} } \frac{2\ell+1}{4\pi } \left( \gamma_{\ell} x^{(\mathrm{TT})}_{\ell} + \epsilon_{\ell} x^{(\mathrm{uEE})}_{\ell} \right)  \rangle \nonumber \\
& = & \sum_{\ell=2}^{\ell_{\mathrm{max}} } \frac{2\ell+1}{4\pi } \left(\gamma_{\ell} + \epsilon_{\ell} \right) = \mathrm{const} = 1 \, .
\label{eq:mean_hatE}
\end{eqnarray}
regardless the maximum multipole considered in the analysis. It should be noted that the choice of the value for the constant is completely arbitrary\footnote{It is typical to set the expectation value of an estimator whose variance is to be minimised at 1 for simplicity's sake. In fact, this value does not affect the expression of the coefficients that minimise the variance, which can be found using Lagrange's multiplier method.}. 

It can be shown that the signal-to-noise ratios of the two NAPS ($x^{(\mathrm{TT})}_{\ell}$ and $x^{(\mathrm{uEE})}_{\ell}$) are different even in the cosmic variance limit case: one may therefore wonder which are the best coefficients, $\gamma_{\ell}$ and $\epsilon_{\ell}$,  that make the estimator optimal, i.e. minimum variance. Applying the method of the Lagrange multipliers  to minimise the variance of $E_\mathrm{V}^{\mathrm{joint}} $,  keeping its expected value fixed (see appendix \ref{sec:appendixA} for the full derivation), we find that: 
\begin{eqnarray}
\gamma_{\ell} &=& \frac{1 }{\Gamma_{\ell_{\mathrm{max}} }} \frac{\mathrm{var(} x^{(\mathrm{uEE})}_{\ell})-\mathrm{cov(} x^{(\mathrm{TT})}_{\ell} ,x^{(\mathrm{uEE})}_{\ell})} {\mathrm{var(}  x^{(\mathrm{TT})}_{\ell} )\mathrm{var(}  x^{(\mathrm{uEE})}_{\ell} )-[\mathrm{cov(} x^{(\mathrm{TT})}_{\ell} ,x^{(\mathrm{uEE})}_{\ell})]^2 } \, , \label{eq:gammaell} \\
\epsilon_{\ell} &=& \frac{1}{\Gamma_{\ell_{\mathrm{max}} }} \frac{\mathrm{var(} x^{(\mathrm{TT})}_{\ell})-\mathrm{cov(} x^{(\mathrm{TT})}_{\ell} ,x^{(\mathrm{uEE})}_{\ell})} {\mathrm{var(}  x^{(\mathrm{TT})}_{\ell} )\mathrm{var(}  x^{(\mathrm{uEE})}_{\ell} )-[\mathrm{cov(} x^{(\mathrm{TT})}_{\ell} ,x^{(\mathrm{uEE})}_{\ell})]^2 } \, , \label{eq:epsilonell} 
\end{eqnarray}
where $\mathrm{var(} x^{(\mathrm{TT})}_{\ell})$ and $\mathrm{var(} x^{(\mathrm{uEE})}_{\ell})$ are the variances of $x^{(\mathrm{TT})}_{\ell}$ and $x^{(\mathrm{uEE})}_{\ell}$ respectively, and $\mathrm{cov(} x^{(\mathrm{TT})}_{\ell},x^{(\mathrm{uEE})}_{\ell})$ their covariance:
\begin{equation}
\mathrm{cov(} x^{(\mathrm{TT})}_{\ell},x^{(\mathrm{uEE})}_{\ell}) = \langle(x^{(\mathrm{TT})}_{\ell} - \langle x^{(\mathrm{TT})}_{\ell} \rangle)(x^{(\mathrm{uEE})}_{\ell} -\langle x^{(\mathrm{uEE})}_{\ell}\rangle) \rangle.
\end{equation}
The quantity $\Gamma_{\ell_{\mathrm{max}} }$ is defined as:
\begin{eqnarray}
\Gamma_{\ell_{\mathrm{max}} } &=& \sum_{\ell=2}^{\ell_{\mathrm{max}} } \frac{2\ell+1}{4\pi } \frac{\mathrm{var(} x^{(\mathrm{TT})}_{\ell}) + \mathrm{var(}  x^{(\mathrm{uEE})}_{\ell})-2\mathrm{cov(} x^{(\mathrm{TT})}_{\ell} ,x^{(\mathrm{uEE})}_{\ell})} {\mathrm{var(}  x^{(\mathrm{TT})}_{\ell} )\mathrm{var(}  x^{(\mathrm{uEE})}_{\ell} )-[\mathrm{cov(} x^{(\mathrm{TT})}_{\ell} ,x^{(\mathrm{uEE})}_{\ell})]^2 } \,.
\end{eqnarray}

In addition, in order to assess the relative contribution of each NAPS, a 
procedure similar to the one used to derive the joint estimator is 
used to construct four analogous optimal estimators, which separately consider $x^{(\mathrm{TT})}$, $x^{(\mathrm{uEE})}$, $x^{(\mathrm{EE})}$ and $x^{(\mathrm{TE})}$: 
\begin{eqnarray}
E_\mathrm{V}^{(\mathrm{TT})} &=& \sum_{\ell=2}^{\ell_{\mathrm{max}} } \frac{2\ell+1}{4\pi } \frac{\left[\mathrm{var(} x^{(\mathrm{TT})}_{\ell}) \right]^{-1}}{\mathrm{var}^{\mathrm{TOT}}_{(\mathrm{TT})}}  x^{(\mathrm{TT})}_{\ell} \,, \label{eq:EV1} \\
E_\mathrm{V}^{(\mathrm{uEE})} &=& \sum_{\ell=2}^{\ell_{\mathrm{max}} } \frac{2\ell+1}{4\pi } \frac{\left[\mathrm{var(} x^{(\mathrm{uEE})}_{\ell})\right]^{-1}}{\mathrm{var}^{\mathrm{TOT}}_{(\mathrm{uEE})}}x^{(\mathrm{uEE})}_{\ell}  \,, \label{eq:EV2}\\
E_\mathrm{V}^{(\mathrm{EE})} &=& \sum_{\ell=2}^{\ell_{\mathrm{max}} } \frac{2\ell+1}{4\pi }  \frac{\left[\mathrm{var(} x_{\ell}^{(\mathrm{EE})})\right]^{-1}}{\mathrm{var}^{\mathrm{TOT}}_{(\mathrm{EE})}} x^{(\mathrm{EE})}_{\ell} \,, \label{eq:EVEE} \\
E_\mathrm{V}^{(\mathrm{TE})} &=& \sum_{\ell=2}^{\ell_{\mathrm{max}} } \frac{2\ell+1}{4\pi } \frac{\left[\mathrm{var(} x_{\ell}^{(TE)})\right]^{-1}}{\mathrm{var}^{\mathrm{TOT}}_{(\mathrm{TE})}} x^{(\mathrm{TE})}_{\ell}  \,, \label{eq:EVTE}
\end{eqnarray}
with:
\begin{equation}
\mathrm{var}^{\mathrm{TOT}}_{(\mathrm{k})}= \sum_{\ell=2}^{\ell_{\mathrm{max}} }  \frac{2\ell+1}{4\pi } \frac{1}{\mathrm{var(} x_{\ell}^{(\mathrm{k})})} \,.
\end{equation}
where $\mathrm{k}$ refers to $\mathrm{TT}$, $\mathrm{uEE}$, $\mathrm{EE}$ or $\mathrm{TE}$.
Their expected values are:
\begin{equation}
\langle E_\mathrm{V}^{(\mathrm{TT})} \rangle = \langle E_\mathrm{V}^{(\mathrm{uEE})} \rangle = \langle E_\mathrm{V}^{(\mathrm{EE})} \rangle = \langle E_\mathrm{V}^{(\mathrm{TE})} \rangle = 1  \, ,
\label{eq:meanEV12}
\end{equation}
regardless the maximum multipole considered in the analysis.

\subsection{Implemented formalism}
The final expressions of the optimal estimators, given by Eqs.~(\ref{eq:EVjoint}), (\ref{eq:EV1}),(\ref{eq:EV2}), (\ref{eq:EVEE}) and (\ref{eq:EVTE}),
are obtained starting from the NAPS at single-$\ell$, Eqs. (\ref{eq:defNAPS1}),(\ref{eq:defNAPS2}),(\ref{eq:defNAPSEE}) and(\ref{eq:defNAPSTE}). However, in order to reduce any possible correlation effect between different multipoles arising from using a sky mask, we apply a binning to the CMB multipoles. 

We estimate the binned expressions of the NAPS by applying the standard operator $P$ (see e.g. \cite{Hivon:2001jp}) \footnote{For a set of $n_{\mathrm{bins}}$ bins, indexed by $b$, with respective boundaries $\ell^{(b)}_{\mathrm{low}}<\ell^{(b)}_{\mathrm{high}} = \ell^{(b+1)}_{\mathrm{low}}-1$, where '$\mathrm{low}$' and '$\mathrm{high}$' mean the lower and the higher multipole in the bin $b$, one can define the binning
operator $P$ as follows: 
\begin{eqnarray}
P_{b\ell} &=&  \left \{ 
\begin{array}{l}
\frac{1}{\ell^{(b+1)}_{\mathrm{low}} -\ell^{(b)}_{\mathrm{low}}} \, ,\qquad (if \,\, \ell_{\mathrm{min}} \leq\ell^{(b)}_{\mathrm{low}} \leq \ell< \ell^{(b+1)}_{\mathrm{low}})\\
\\
0 \, ,\qquad \qquad \quad \,\,\,\, (\mbox{otherwise})
\end{array} \right.
\nonumber
\end{eqnarray}
where $\ell_{\mathrm{min}}$ stands for the minimum multipole considered, which for the CMB it is typically $\ell_{\mathrm{min}}=2$.}:
\begin{eqnarray}
x_{b}^{(\mathrm{TT})} &=& \sum_{\ell} P_{b\ell} \, x^{(\mathrm{TT})}_{\ell} \, ,\label{eq:defNAPS1bin} \\
x_{b}^{(\mathrm{uEE})} &=& \sum_{\ell} P_{b\ell} \, x^{(\mathrm{uEE})}_{\ell} \, .\label{eq:defNAPS2bin} \\
x_{b}^{(\mathrm{EE})} &=& \sum_{\ell} P_{b\ell} \, x_{\ell}^{\mathrm{(EE)}} \, ,\label{eq:defNAPSEEbin} \\
x_{b}^{(\mathrm{TE})} &=& \sum_{\ell} P_{b\ell} \, x_{\ell}^{(\mathrm{TE})} \, .\label{eq:defNAPSTEbin}
\end{eqnarray}

Now we use these modified versions of the NAPS to recompute the expressions of the estimators. In particular, the joint estimator becomes: 
\begin{equation}
E_\mathrm{V}^{\mathrm{joint}}  = \sum_{b} \frac{2\ell_{\mathrm{eff}}^{(b)}+1}{4\pi } \left( \gamma_{b} x^{(\mathrm{TT})}_{b} + \epsilon_{b} x^{(\mathrm{uEE})}_{b} \right) \,,
\label{eq:EVjoint_bin}
\end{equation}
with:
\begin{eqnarray}
\gamma_{b} &=& \frac{1 }{\Gamma_{b_{\mathrm{max}}}} \frac{\mathrm{var(} x^{(\mathrm{uEE})}_{b})-\mathrm{cov(} x^{(\mathrm{TT})}_{b} ,x^{(\mathrm{uEE})}_{b})} {\mathrm{var(}  x^{(\mathrm{TT})}_{b} )\mathrm{var(}  x^{(\mathrm{uEE})}_{b} )-[\mathrm{cov(} x^{(\mathrm{TT})}_{b} ,x^{(\mathrm{uEE})}_{b})]^2 } \, , \label{eq:gammab} \\
\epsilon_{b} &=& \frac{1 }{\Gamma_{b_{\mathrm{max}}}} \frac{\mathrm{var(} x^{(\mathrm{TT})}_{b})-\mathrm{cov(} x^{(\mathrm{TT})}_{b} ,x^{(\mathrm{uEE})}_{b})} {\mathrm{var(}  x^{(\mathrm{TT})}_{b} )\mathrm{var(}  x^{(\mathrm{uEE})}_{b} )-[\mathrm{cov(} x^{(\mathrm{TT})}_{b} ,x^{(\mathrm{uEE})}_{b})]^2 } \, , \label{eq:epsilonb} \\
\Gamma_{b_{\mathrm{max}}} &=& \sum_{b} \frac{2\ell_{\mathrm{eff}}^{(b)}+1}{4\pi } \frac{\mathrm{var(} x^{(\mathrm{TT})}_{b}) + \mathrm{var(}  x^{(\mathrm{uEE})}_{b})-2\mathrm{cov(} x^{(\mathrm{TT})}_{b} ,x^{(\mathrm{uEE})}_{b})} {\mathrm{var(}  x^{(\mathrm{TT})}_{b} )\mathrm{var(}  x^{(\mathrm{uEE})}_{b} )-[\mathrm{cov(} x^{(\mathrm{TT})}_{b} ,x^{(\mathrm{uEE})}_{b})]^2 } \label{eq:Gammab} \,.
\end{eqnarray}
In Eqs. (\ref{eq:EVjoint_bin}) and (\ref{eq:Gammab}), for each bin $b$, $\ell_{\mathrm{eff}}^{(b)}$ is defined as:
\begin{equation}
\ell_{\mathrm{eff}}^{(b)} = \frac{1}{N_{\ell}^{(b)}} \sum_{\ell \in b} \ell \, , \label{eq:ell_eff}
\end{equation}
with $N_{\ell}^{(b)}$ the number of multipoles included in the bin $b$. 

Analogously the binned versions of the estimators $E_\mathrm{V}^{(\mathrm{TT})}$, $E_\mathrm{V}^{(\mathrm{uEE})}$,$E_\mathrm{V}^{(\mathrm{EE})}$ and $E_\mathrm{V}^{(\mathrm{TE})}$ are:
\begin{eqnarray}
E_\mathrm{V}^{(\mathrm{TT})} &=& \sum_{b} \frac{2\ell_{\mathrm{eff}}^{(b)}+1}{4\pi } \frac{\left[\mathrm{var(} x_{b}^{(\mathrm{TT})})\right]^{-1}}{\mathrm{var}^{\mathrm{TOT}}_{(\mathrm{TT})}}  x^{(\mathrm{TT})}_{b} \,, \label{eq:EV1_bin} \\
E_\mathrm{V}^{(\mathrm{uEE})} &=& \sum_{b} \frac{2\ell_{\mathrm{eff}}^{(b)}+1}{4\pi } \frac{\left[\mathrm{var(} x_{b}^{(\mathrm{uEE})})\right]^{-1}}{\mathrm{var}^{\mathrm{TOT}}_{(\mathrm{uEE})}}  x^{(\mathrm{uEE})}_{b} \,, \label{eq:EV2_bin} \\
E_\mathrm{V}^{(\mathrm{EE})} &=& \sum_{b} \frac{2\ell_{\mathrm{eff}}^{(b)}+1}{4\pi } \frac{\left[\mathrm{var(} x_{b}^{(\mathrm{EE})})\right]^{-1}}{\mathrm{var}^{\mathrm{TOT}}_{(\mathrm{EE})}}  x^{(\mathrm{EE})}_{b} \,, \label{eq:EVEE_bin} \\
E_\mathrm{V}^{(\mathrm{TE})} &=& \sum_{b} \frac{2\ell_{\mathrm{eff}}^{(b)}+1}{4\pi } \frac{\left[\mathrm{var(} x_{b}^{(\mathrm{TE})})\right]^{-1}}{\mathrm{var}^{\mathrm{TOT}}_{(\mathrm{TE})}} x^{(\mathrm{TE})}_{b} \,, \label{eq:EVTE_bin} 
\end{eqnarray}
with:
\begin{equation}
\mathrm{var}^{\mathrm{TOT}}_{(\mathrm{k})}= \sum_{b} \frac{2\ell_{\mathrm{eff}}^{(b)}+1}{4\pi } \frac{1}{\mathrm{var(} x_{b}^{(\mathrm{k})})} \,.
\end{equation}
where k refers to TT, uEE, EE or TE.

The binned versions of the estimators, defined in Eqs.(\ref{eq:EVjoint_bin}), (\ref{eq:EV1_bin}), (\ref{eq:EVEE_bin}) and (\ref{eq:EVTE_bin}), are actually applied to the dataset described in the following section. 
For the sake of brevity, the results for $x^{(\mathrm{uEE})}_{\ell}$ and for $E_\mathrm{V}^{(\mathrm{uEE})}$ are not shown in the following, since they give qualitatively similar results to $x_{\ell}^{(\mathrm{EE})}$ and $E_\mathrm{V}^{(\mathrm{EE})}$, respectively. 
This reflects the fact that the uncorrelated E-component provides the dominant contribution 
to the EE spectra.

%% file: 3dataset.tex
\section{\textit{Planck} Datasets}
\label{sec:dataset}
We make use of the foreground-cleaned CMB maps, data and end-to-end simulations, provided by the latest \textit{Planck} releases, both PR3 (2018) \cite{Planck:2018nkj} and PR4 (2020), also referred to as NPIPE \cite{Planck:2020olo}. To separate the sky maps into their contributing signals and to clean the CMB maps from foregrounds, the \textit{Planck} team used for the PR3 dataset, data and simulations, all the four official component separation methods {\tt Commander}, {\tt NILC}, {\tt SEVEM} and {\tt SMICA} \cite{Planck:2018yye}, while only the {\tt Commander} and {\tt SEVEM} \cite{Planck:2020olo} pipelines were applied to the NPIPE maps. To test the robustness of our analysis we perform our estimators on each of the \textit{Planck} 2018 and 2020 pipelines.

The PR3 Monte Carlo (MC) simulations, referred to as FFP10, see e.g.~\cite{Planck:2019evm}, are an updated version of the full focal plane simulations described in \cite{Planck:2015txa}. The FFP10 dataset contains the most realistic simulations produced by the \textit{Planck} collaboration to characterise  its 2018 data. They consist of 999 CMB maps extracted from the current \textit{Planck} $\Lambda$CDM best-fit model, which are beam smoothed and  include residuals of beam leakage \cite{Planck:2015txa}. These maps are complemented by 300 instrumental noise simulations for the full mission and for the two data splits: the odd-even (OE) rings and the two half-mission (HM) data. 
These noise simulations, which are provided for each frequency channel, also include residual systematic effects as beam leakage again, analogue-to-digital converter (ADC) \footnote{The ADC non linearity is a systematic effect that affects the modulated signal of the bolometers at high frequency.} non linearities, thermal fluctuations (dubbed 4K fluctuations), band-pass mismatch and others \cite{Planck:2018lkk}. The FFP10 simulations are processed through the component separation algorithms in the same way as data. 

The NPIPE maps represent an evolution of the PR3 dataset: compared to the latter, the PR4 dataset provides a better description of the CMB polarisation component,  with reduced levels of noise and systematics at all angular scales and an improved consistency across frequencies \cite{Planck:2020olo}. 
In addition to the data, 400 and 600 MC simulations are provided for {\tt Commander} 
\footnote{Note that in the case of {\tt Commander}, 400 MC simulations are provided for the Q,U maps. For temperature only 100 simulated maps are produced and only for the full mission.} 
and {\tt SEVEM}, respectively. While the NPIPE {\tt Commander}\ team provides the signal-plus-noise simulations, the NPIPE {\tt SEVEM}\ team  has separately released the only-signal and the only-noise simulated maps for both detector-splits, referred to as detector A and detector B (or simply A and B). Note that in the NPIPE A and B data splits the systematic effects between the sets are expected to be uncorrelated \cite{Planck:2020olo}. However, we remark that the PR4 polarisation data at the largest angular scale are affected by a transfer function that reduces the power of the polarised signal (see section 4.3 of \cite{Planck:2020olo}). Therefore, to ensure the fidelity of the scientific results, we correct for the transfer function the values of the theoretical power spectra TE and EE which enter in the equations of NAPS, Eqs. (\ref{eq:defNAPS2}),(\ref{eq:defNAPSEE}) and (\ref{eq:defNAPSTE}),  see appendix~\ref{sec:appendixB} and Figure~\ref{fig:transf_func} for more details.

\subsection{Large angular scale maps and power spectra}
In our analysis for the PR3 dataset we use the odd-even (OE) splits for both data and FFP10 simulations for all the four \textit{Planck} component separation methods. Combining the first 300 CMB realisations with the two different OE splits of 300 noise maps we build two MC sets of 300 signal-plus-noise simulations which also include residuals of known systematic effects. For the PR4 dataset we use {\tt SEVEM} maps, data and 600 MC simulations, divided into detector A-B splits. In order to use the NPIPE {\tt Commander} products, since the simulated maps for detector-splits are not available for this component separation method in temperature, we build a hybrid set, and we refer to it as {\tt hybrid}. In particular, we consider the {\tt SEVEM} temperature maps and {\tt Commander} polarisation maps, split into detector A and detector B, to build the TQU data map with the corresponding 400 MC simulated maps.

The PR3 and PR4 maps, available from the \textit{Planck} Legacy Archive\footnote{\url{https://www.esa.int/Planck}} (PLA), are provided at \texttt{HEALPix}\footnote{\url{https://healpix.sourceforge.net}} \cite{Gorski:2004by} resolution $\mathrm{N_{side}}=2048$, with a Gaussian beam with a full width at half maximum (FWHM) of $\mathrm{FWHM}=5^\prime$. Given our interest in large angular scales, the full resolution maps are downgraded in the harmonic space to lower resolution $\mathrm{N_{side}}=16$, according to the following equation:
\begin{equation}
a_{\ell m}^{\mathrm{out}} = \frac{b_{\ell}^{\mathrm{out}} p_{\ell}^{\mathrm{out}}}{b_{\ell}^{\mathrm{in}} \,p_{\ell}^{\mathrm{in}}} \, a_{\ell m}^{\mathrm{in}} \, ,
\label{eq:alm_LR}
\end{equation}
where superscripts 'in' and 'out' stand for input (full resolution) and output (low resolution) maps, respectively; the coefficient $b_{\ell}$ is the Gaussian beam and $p_{\ell}$ the HEALPix pixel window functions. 
Once the aforementioned procedure has been completed, the harmonic coefficients $a_{\ell m}^{\mathrm{out}}$ are used to generate maps at resolution $\mathrm{N_{side}}=16$ with a Gaussian beam with $\mathrm{FWHM}=527.69^\prime$, that corresponds to 2.4 times the pixel size.

\begin{figure}[htbp]
\centering
\includegraphics[width=75mm]{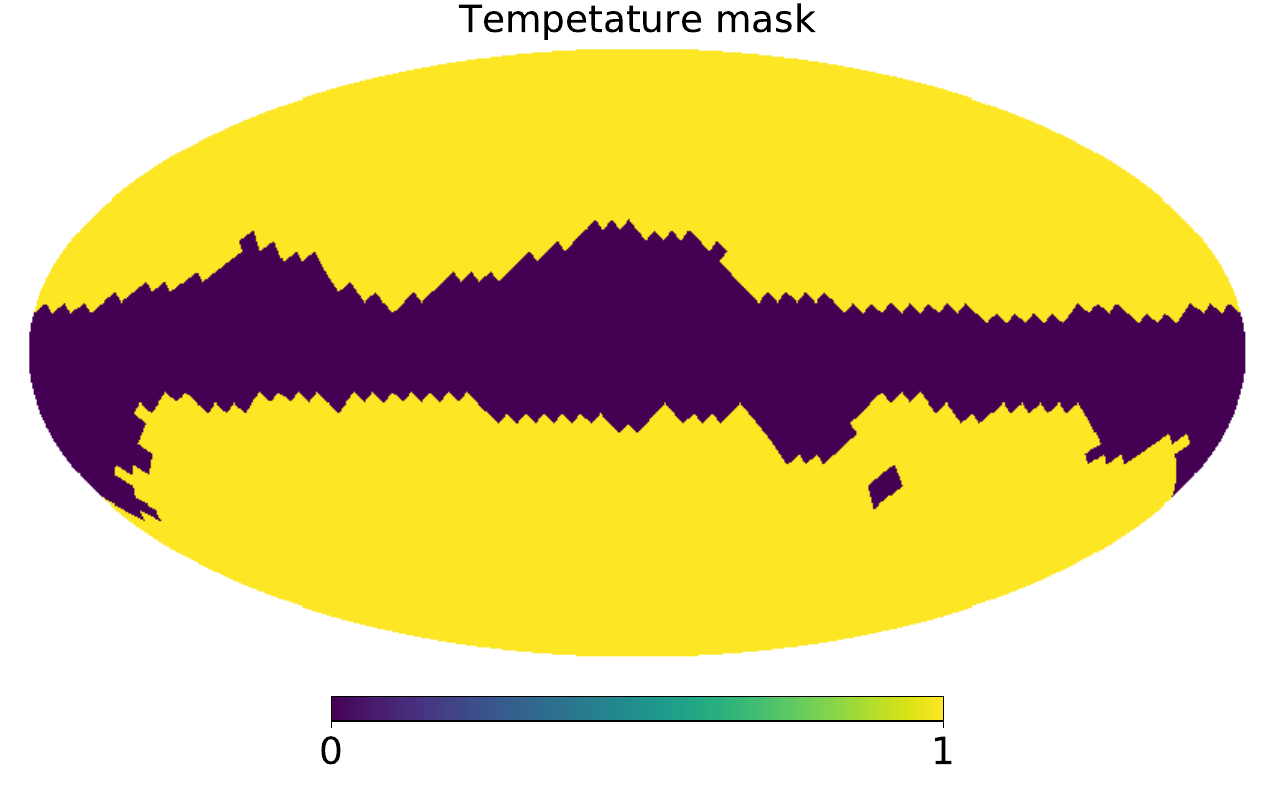} 
\includegraphics[width=75mm]{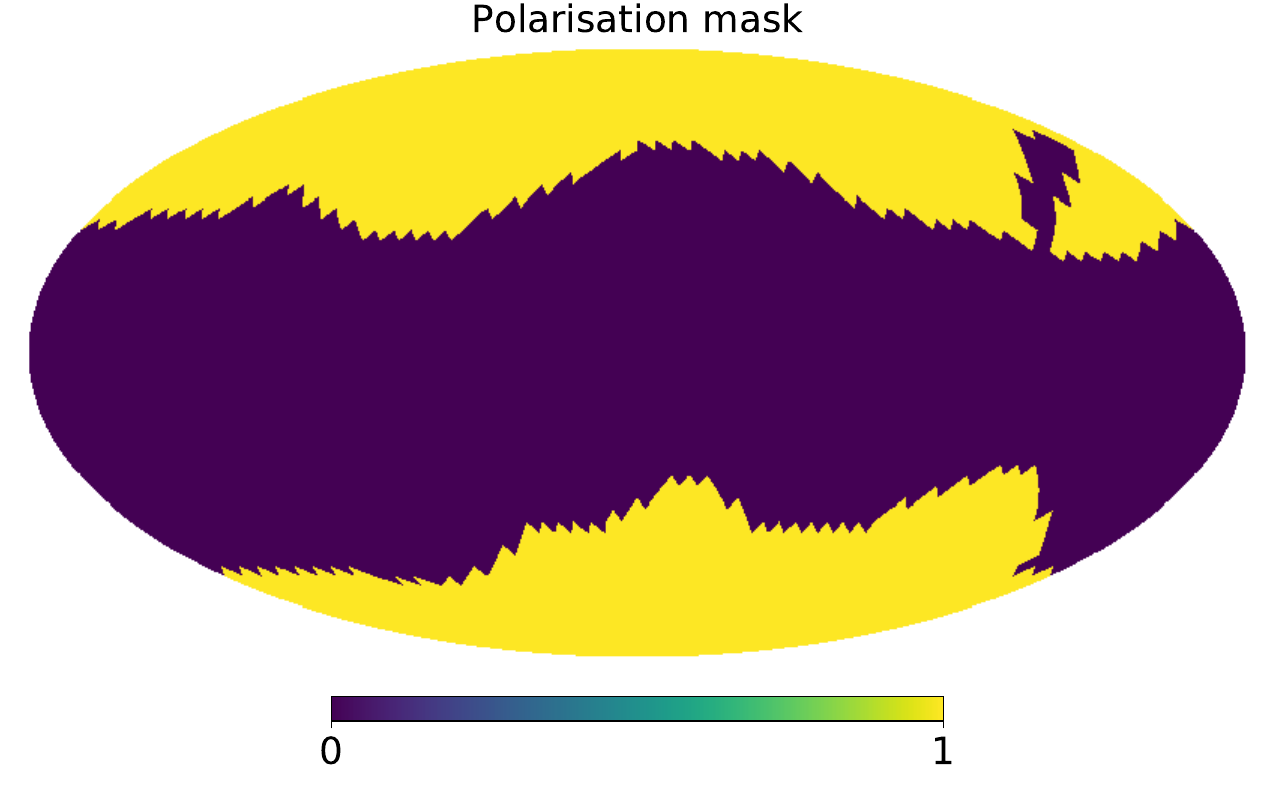}
\caption{T (left) and P (right) masks at resolution $\mathrm{N_{side}}=16$ which leave uncovered the $71.4\%$ and $34.7\%$ of the sky, respectively.}
\label{fig:TQU_masks} 
\end{figure}

\begin{figure}[htbp]
\centering
\includegraphics[width=75mm]{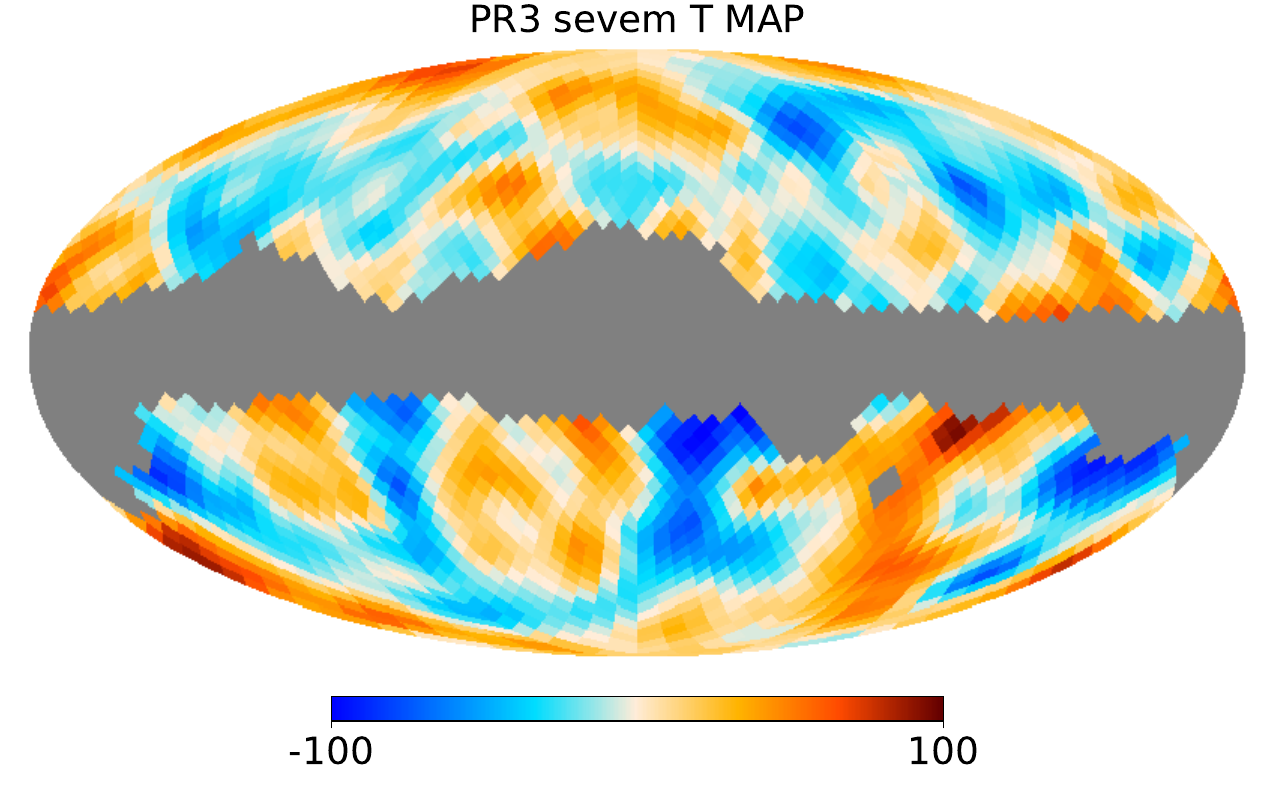} 
\includegraphics[width=75mm]{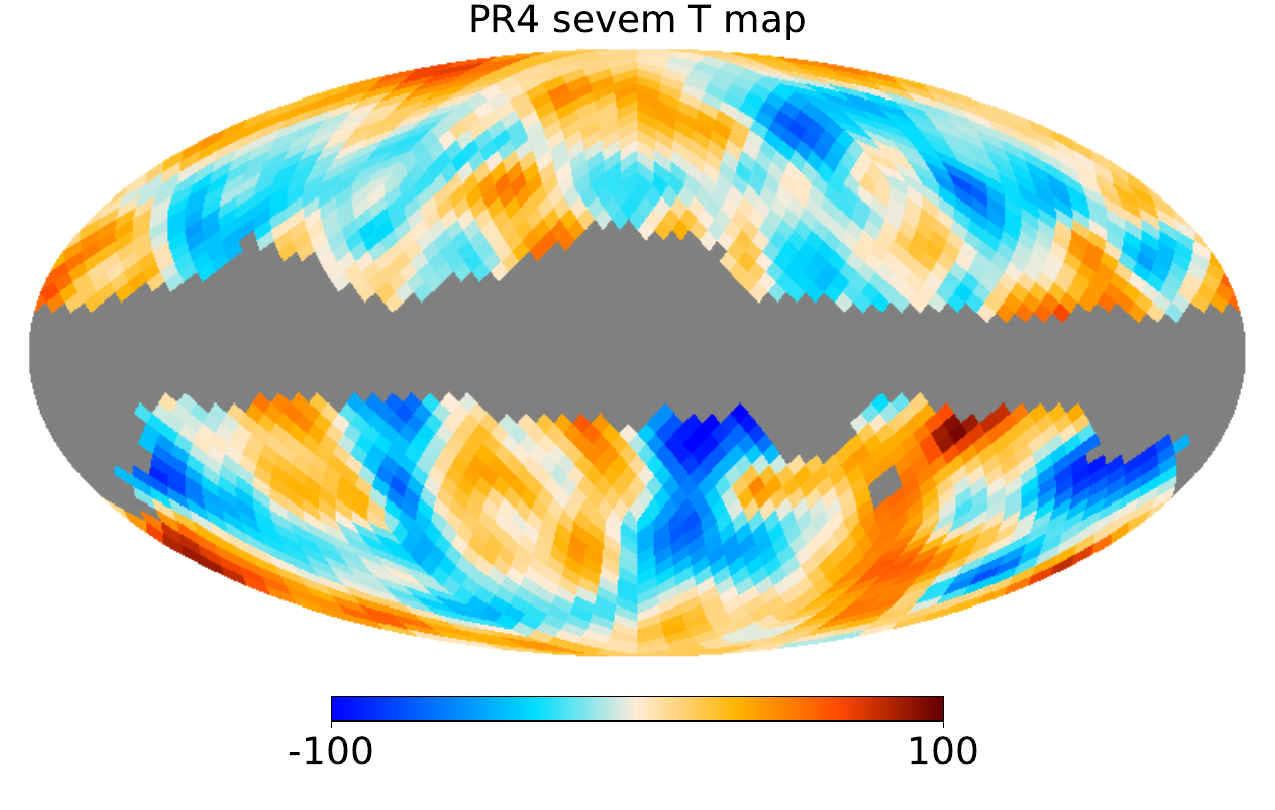}\\
\includegraphics[width=75mm]{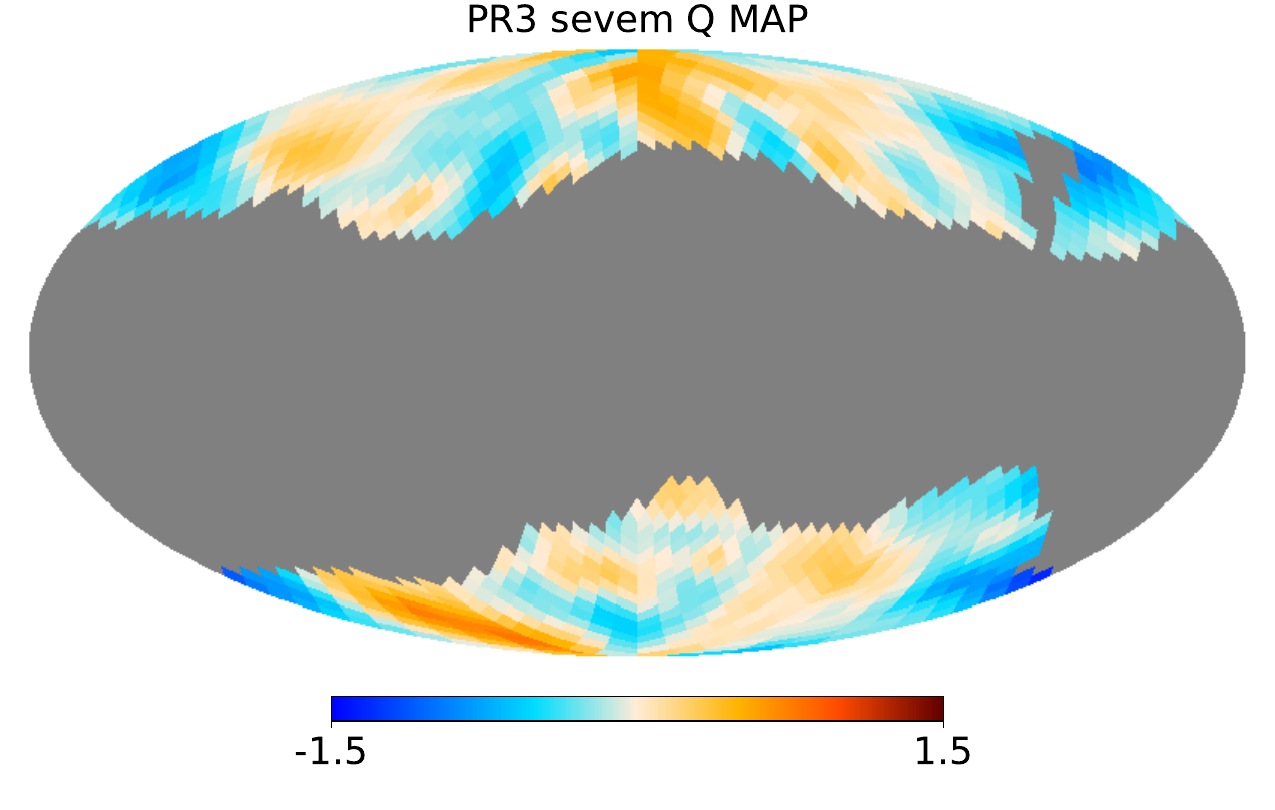} 
\includegraphics[width=75mm]{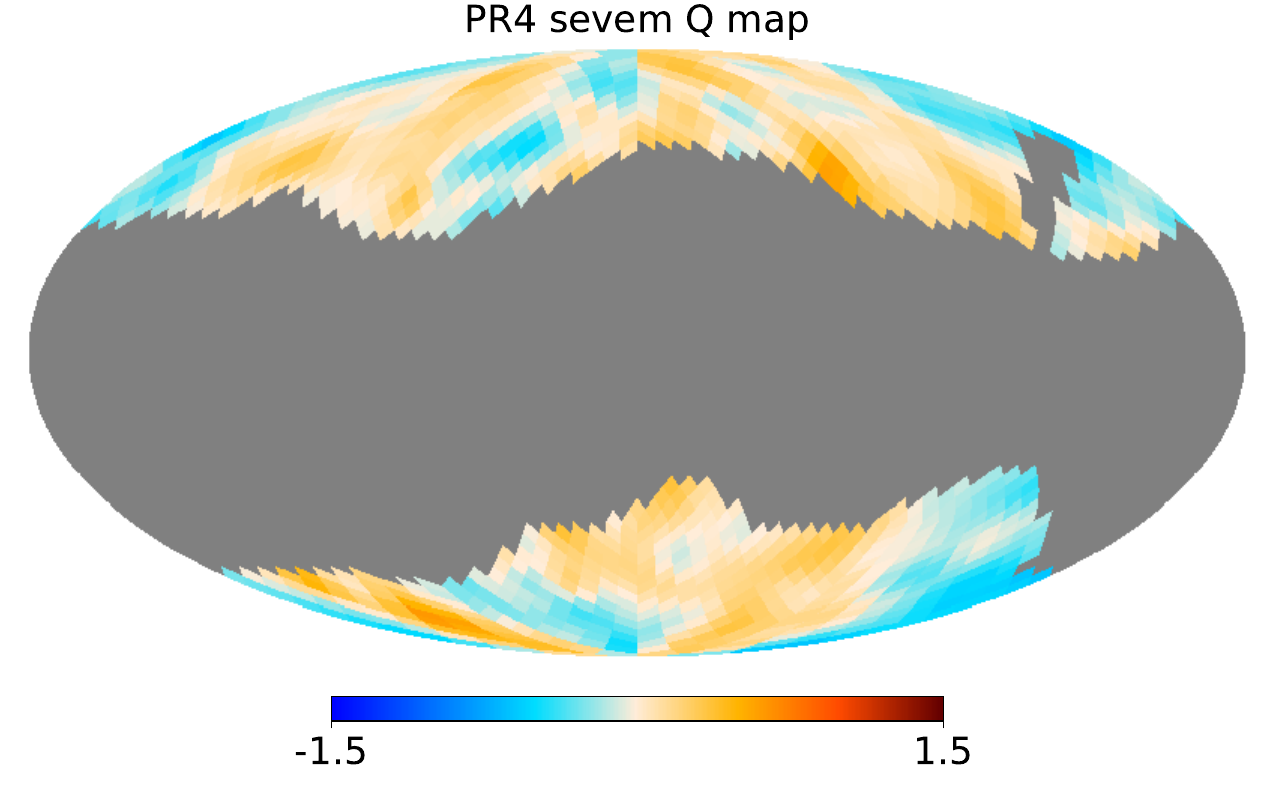}\\
\includegraphics[width=75mm]{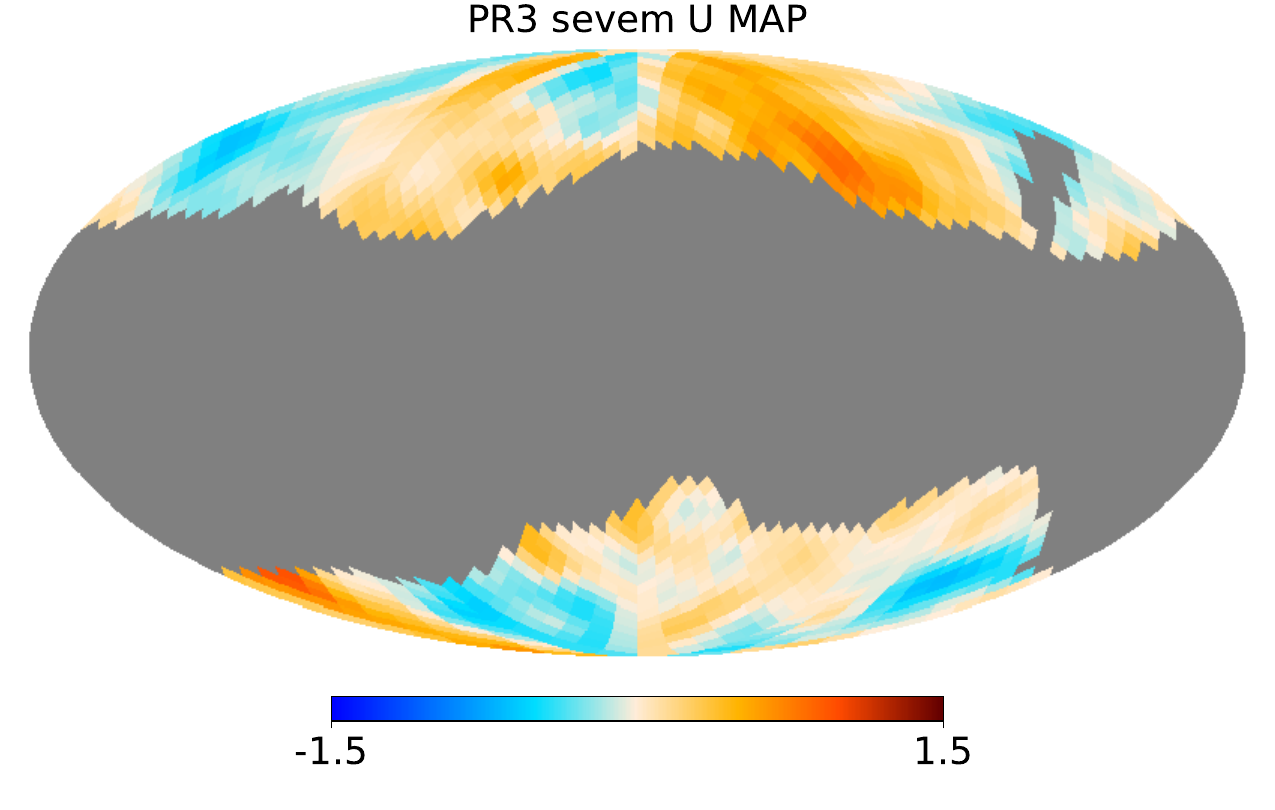} 
\includegraphics[width=75mm]{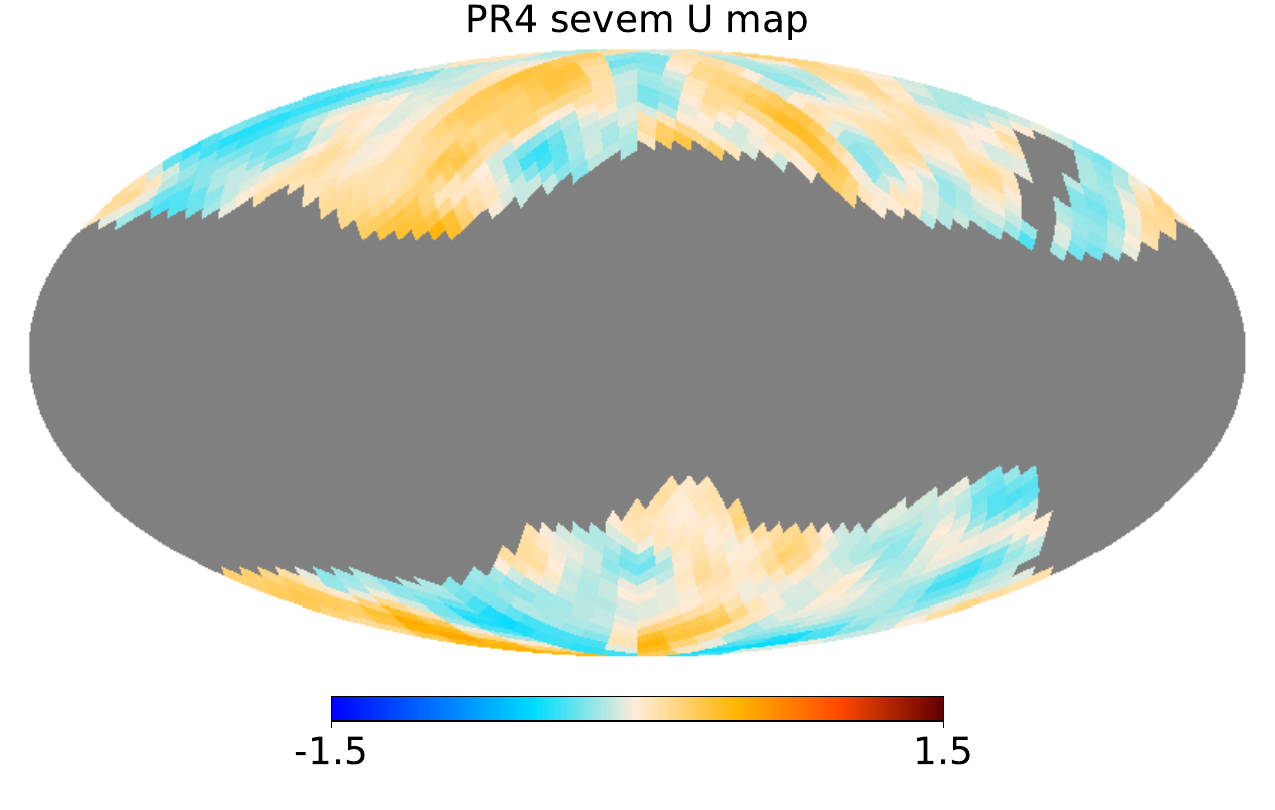}
\caption{CMB {\tt SEVEM} data maps ($\mu \mathrm{K}$) at resolution $\mathrm{N_{side}}=16$. The first column shows PR3 T (upper panel), Q (middle panel), U (lower panel) maps for the $odd$ split, while the second column displays the PR4 maps for detector A. The masks in Figure \ref{fig:TQU_masks} have been applied to these maps.}
\label{fig:sevem_maps} 
\end{figure}

\begin{figure}
\centering
\includegraphics[width=130mm]{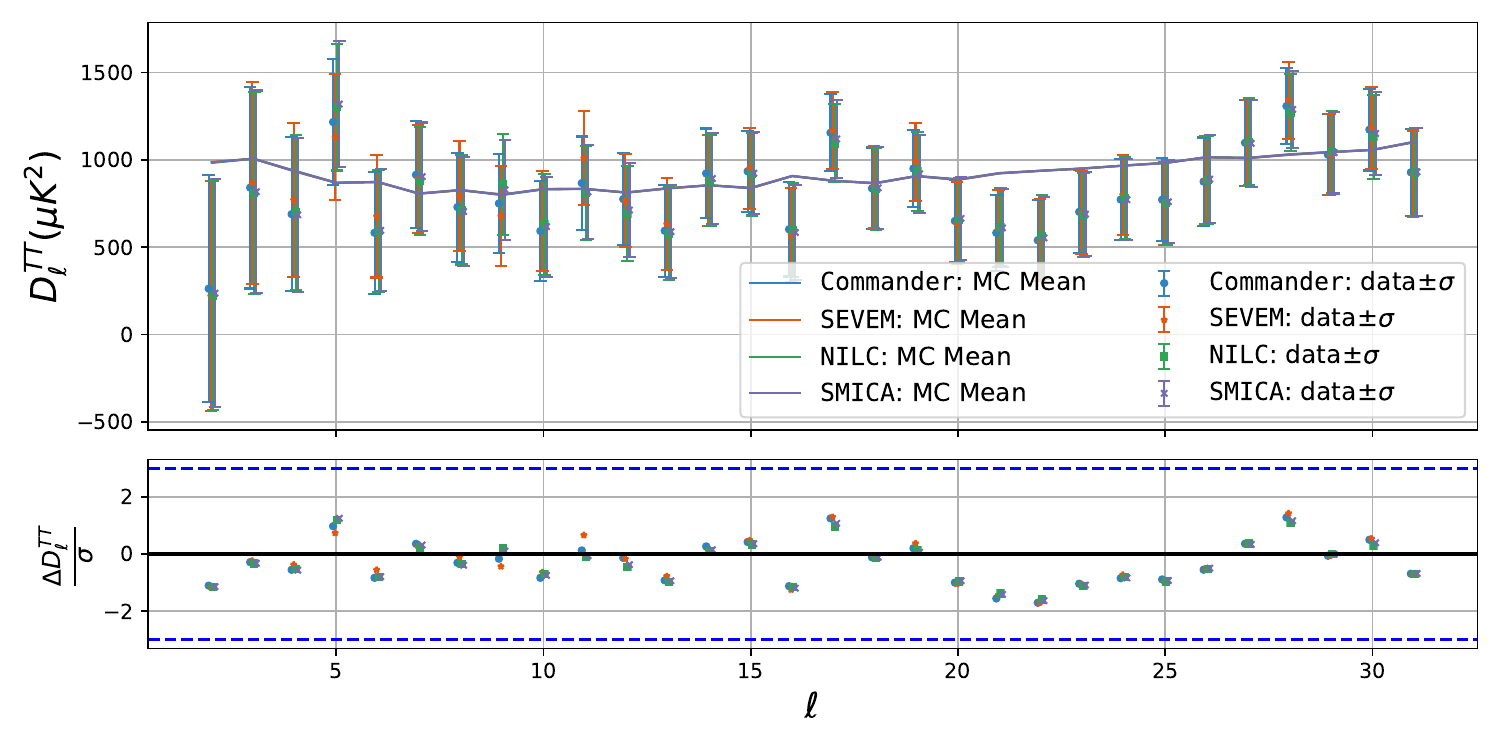} \\
\includegraphics[width=130mm]{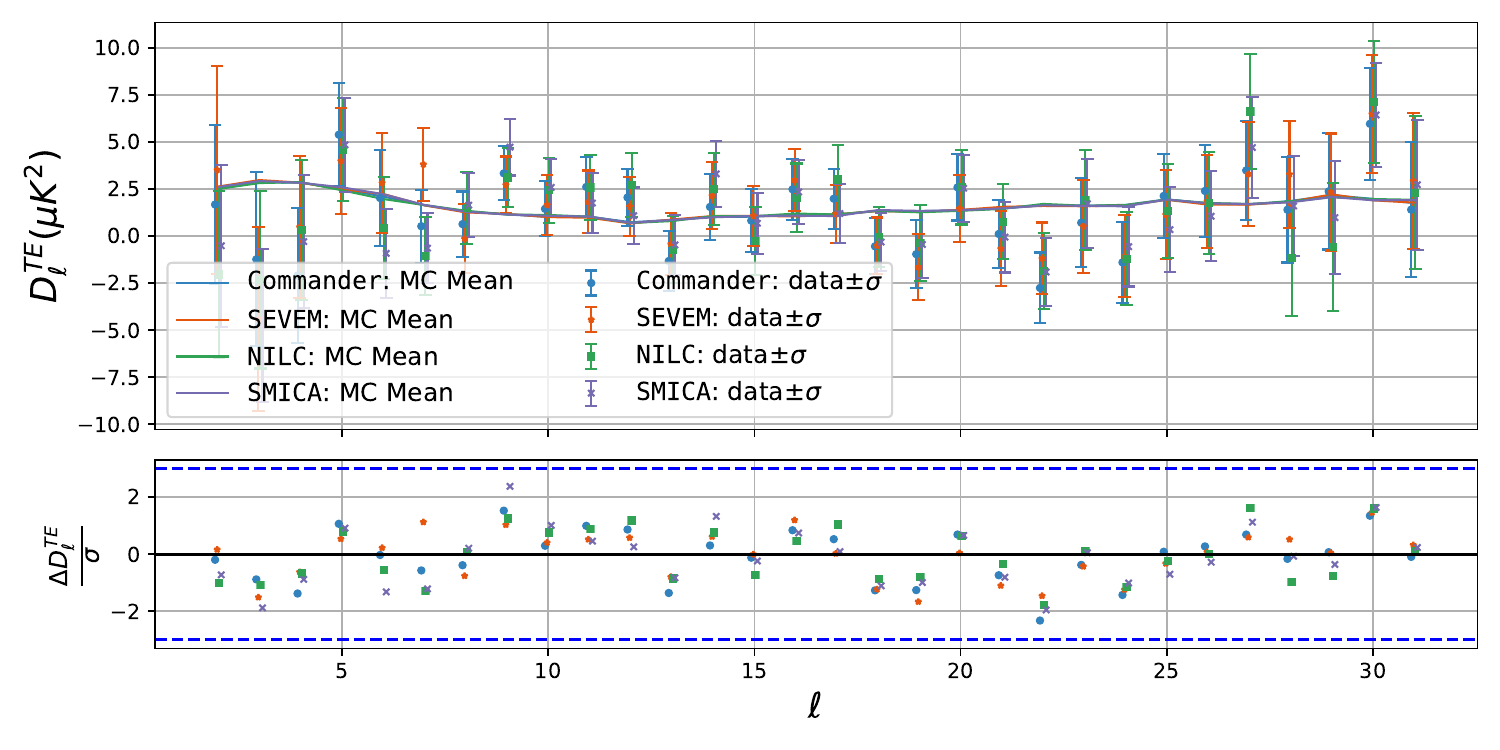}\\
\includegraphics[width=130mm]{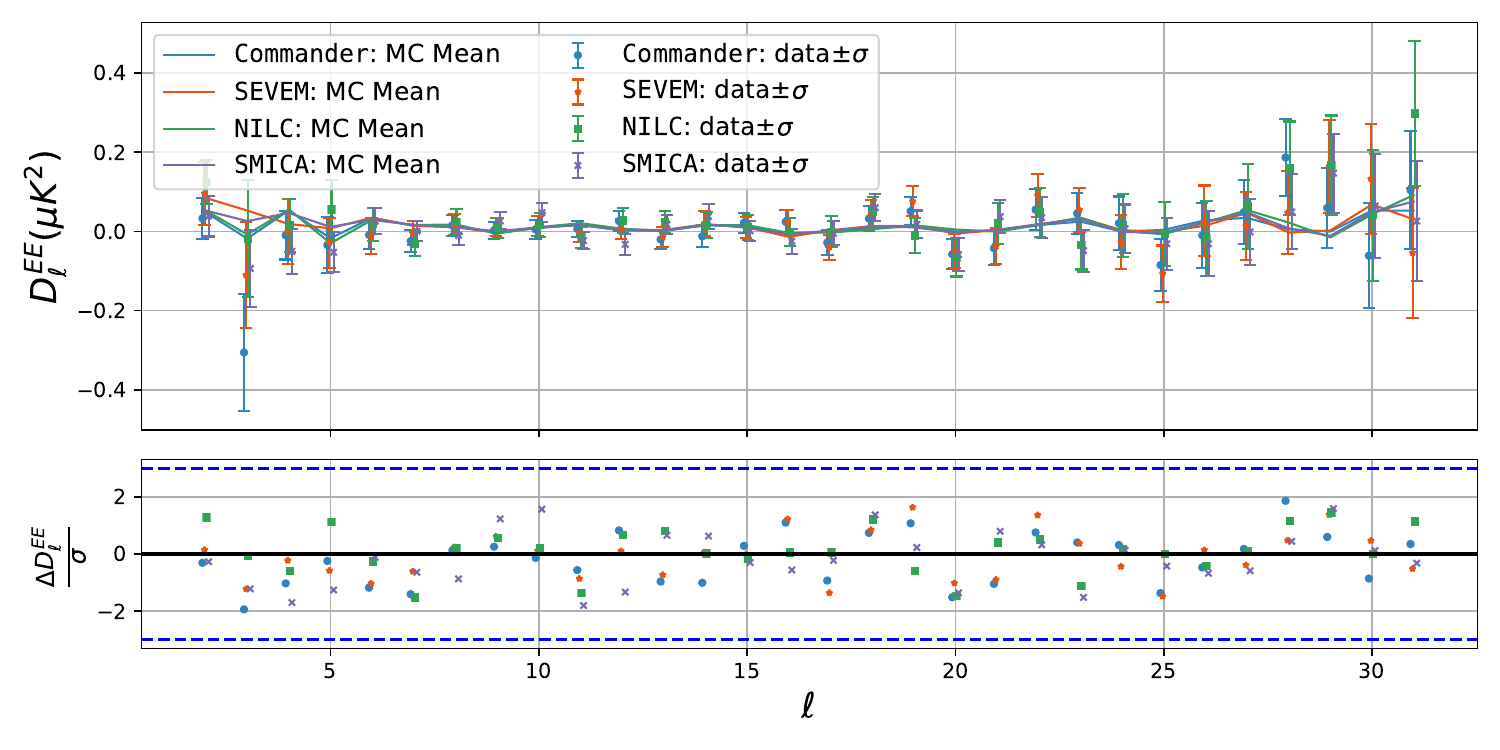}
\caption{The upper boxes of each panel show the $D_{\ell}^\mathrm{{TT}}$ (upper panel), $D_{\ell}^\mathrm{{TE}}$ (middle panel) and $D_{\ell}^\mathrm{{EE}}$ (lower panel), where $D_{\ell} = \frac{\ell(\ell+1)}{2\pi} C_{\ell}$, as a function of $\ell$ (for $2 \le \ell \le 31$), obtained from the PR3 dataset for {\tt Commander} (blue),  {\tt NILC} (green), {\tt SEVEM} (orange), {\tt SMICA} (violet). The error bars represent the standard deviation of the estimates, i.e. $\sigma$. The solid lines represent the MC average of the spectra for each pipelines. Each panel also displays a lower box, which shows the distance of the estimates from the MC mean in units of $\sigma$, for each value of $\ell$.}
\label{fig:PR3_spectra} 
\end{figure}

\begin{figure}[htbp]
\centering
\includegraphics[width=130mm]{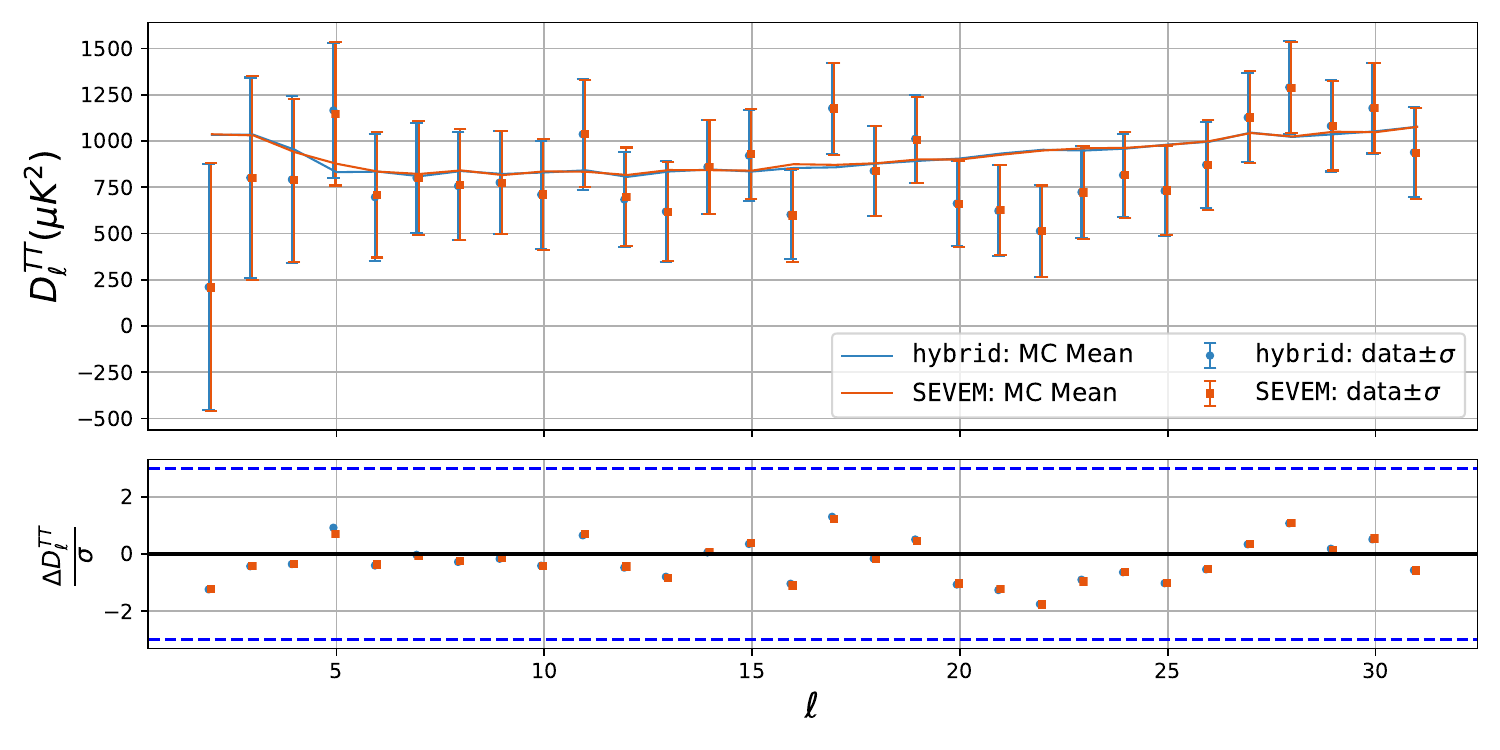} \\
\includegraphics[width=130mm]{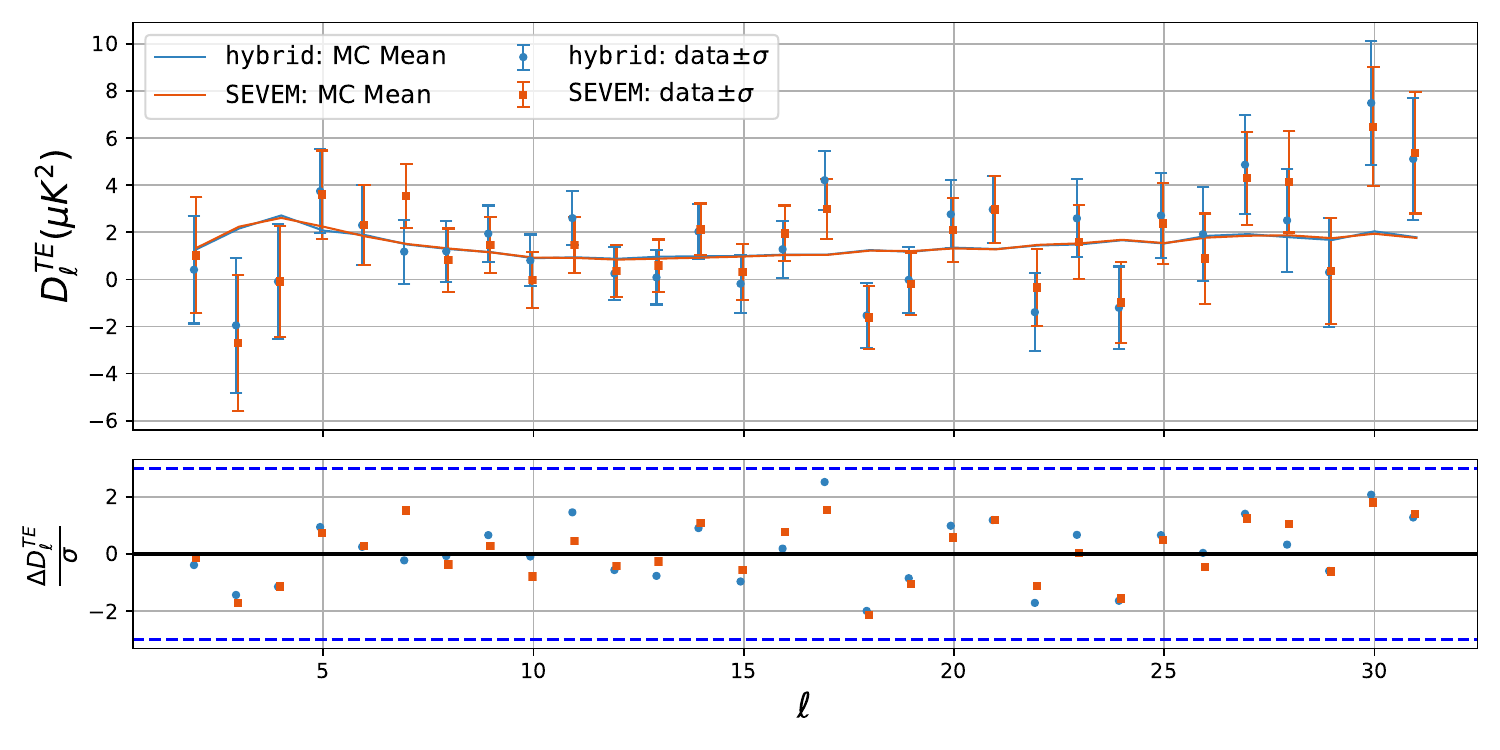}\\
\includegraphics[width=130mm]{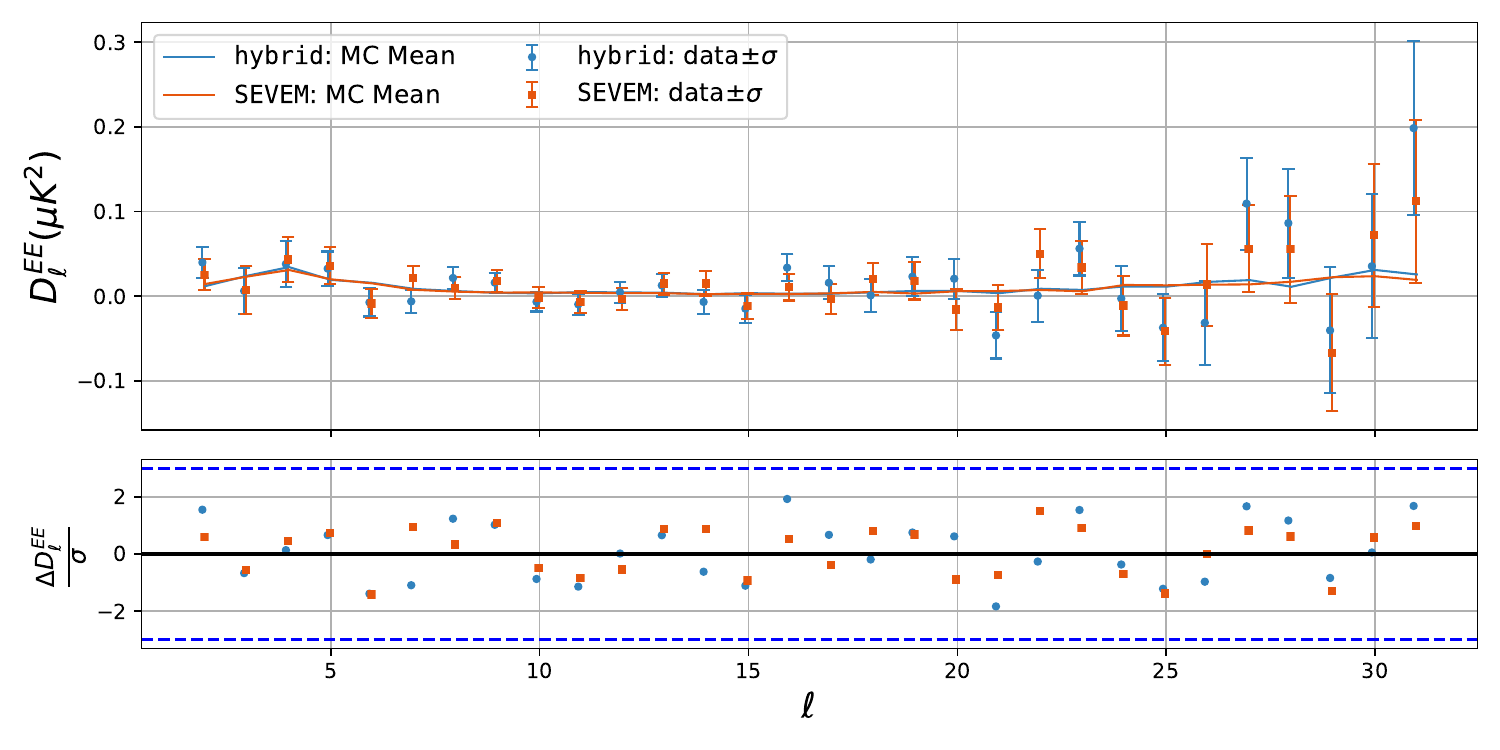}
\caption{Same as Figure~\ref{fig:PR3_spectra} but for the PR4 data.}
\label{fig:PR4_spectra} 
\end{figure}

The \textit{Planck} confidence mask is applied to the temperature maps, while a combination of the polarisation confidence mask and the $40\%$ galactic sky mask, provided by Planck, is used for the polarisation maps.
Figure \ref{fig:TQU_masks} shows the T (left) and P (right) masks at resolution $\mathrm{N_{side}}=16$ which leave uncovered the $71.4\%$ and $34.7\%$ of the sky, respectively. These masks are derived from the full resolution masks \footnote{The masks at resolution $\mathrm{N_{side}}=2048$ are available from the \textit{Planck} Legacy Archive (PLA).} by applying the same downgrading procedure used with the CMB maps and setting a threshold of 0.8 \footnote{All the pixels with a value below the threshold are set to 0, while the remaining ones are set to 1}. It should be noted that the larger mask employed for the polarisation maps is motivated by the observation that using the confidence mask we obtained significant differences in the EE data spectra between different pipelines, which is likely due to the impact of the foreground residuals. However, as can be observed in Figures \ref{fig:PR3_spectra} and \ref{fig:PR4_spectra}, this is not the case when the mask in Figure \ref{fig:TQU_masks} is used. Figure \ref{fig:sevem_maps} displays the {\tt SEVEM} maps at resolution $\mathrm{N_{side}}=16$ for both \textit{Planck} releases. 

\begin{figure}[htbp]
\centering
\includegraphics[width=130mm]{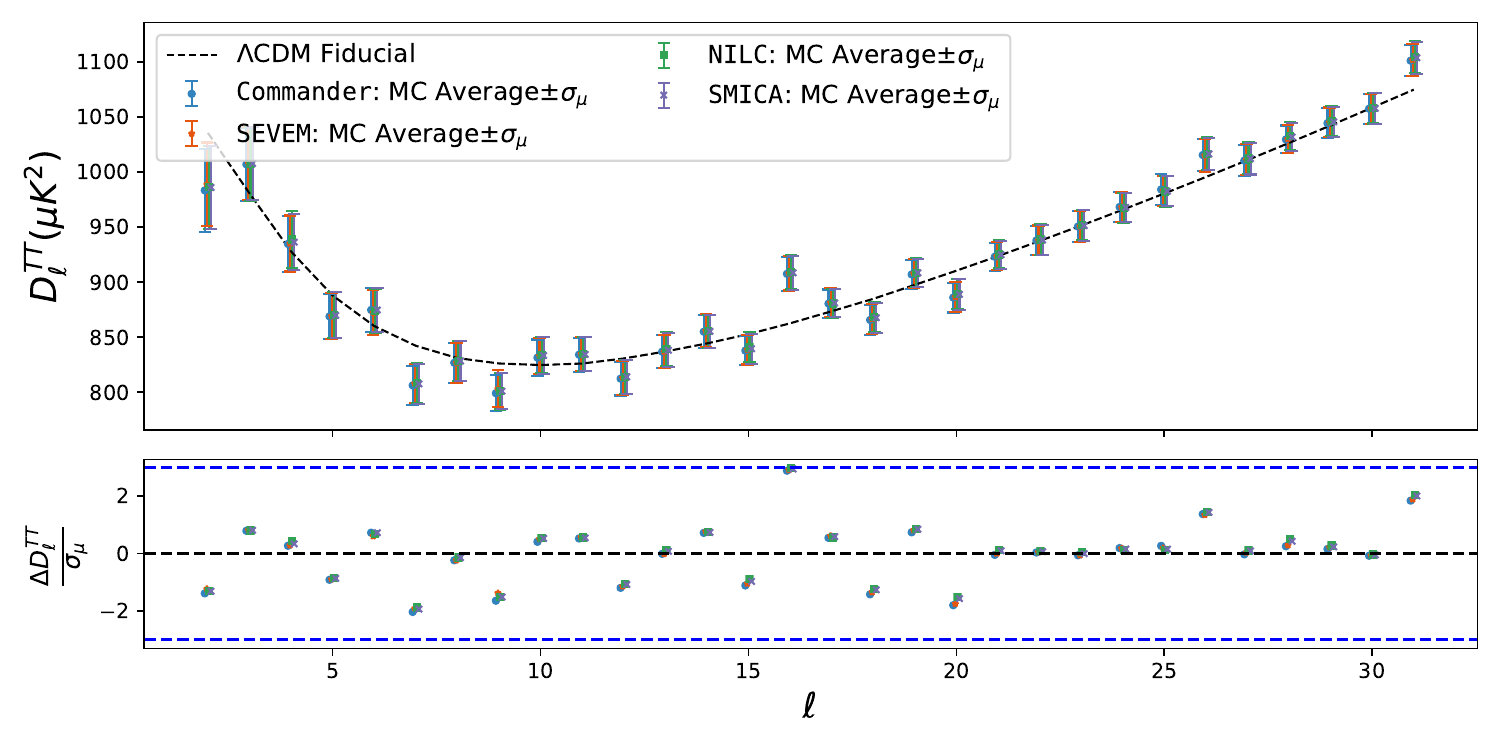} \\
\includegraphics[width=130mm]{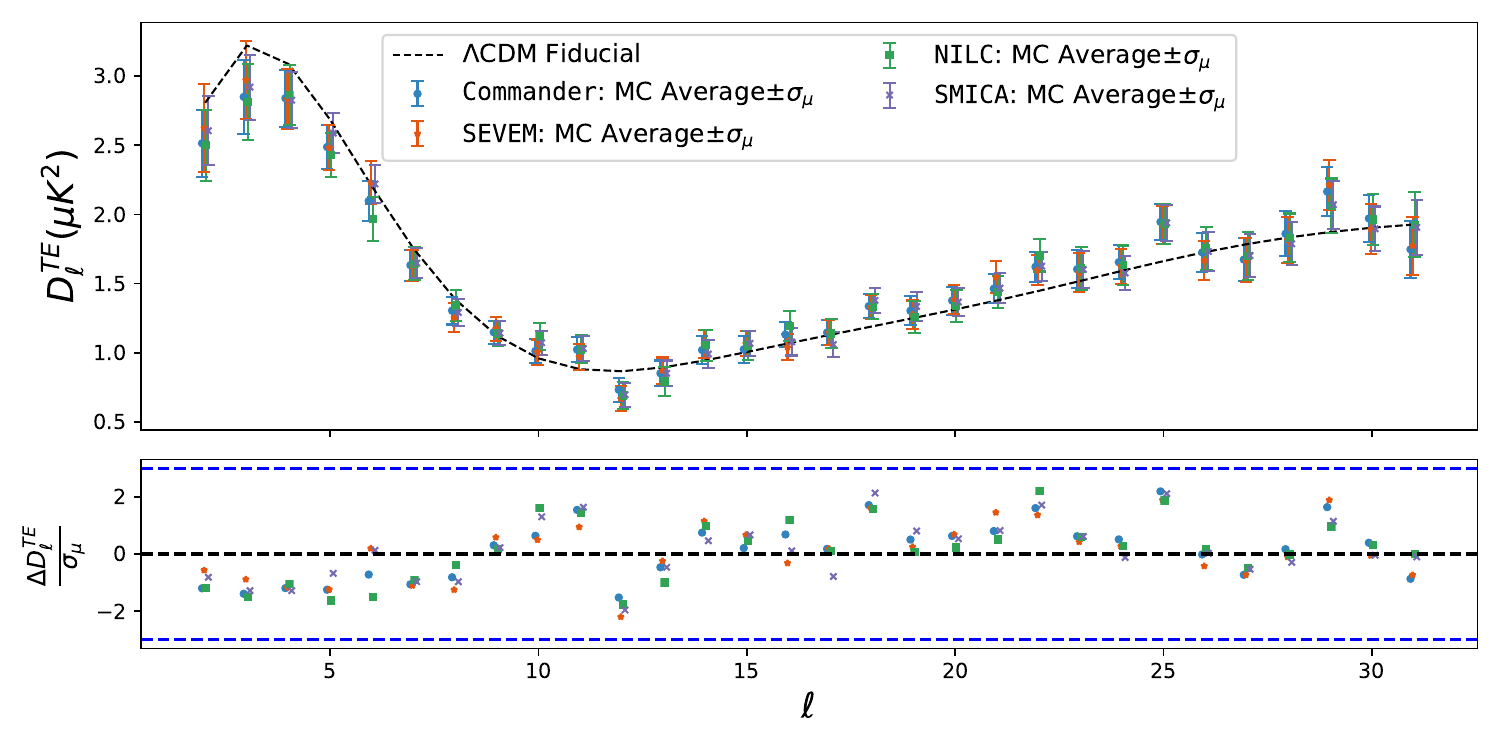}\\
\includegraphics[width=130mm]{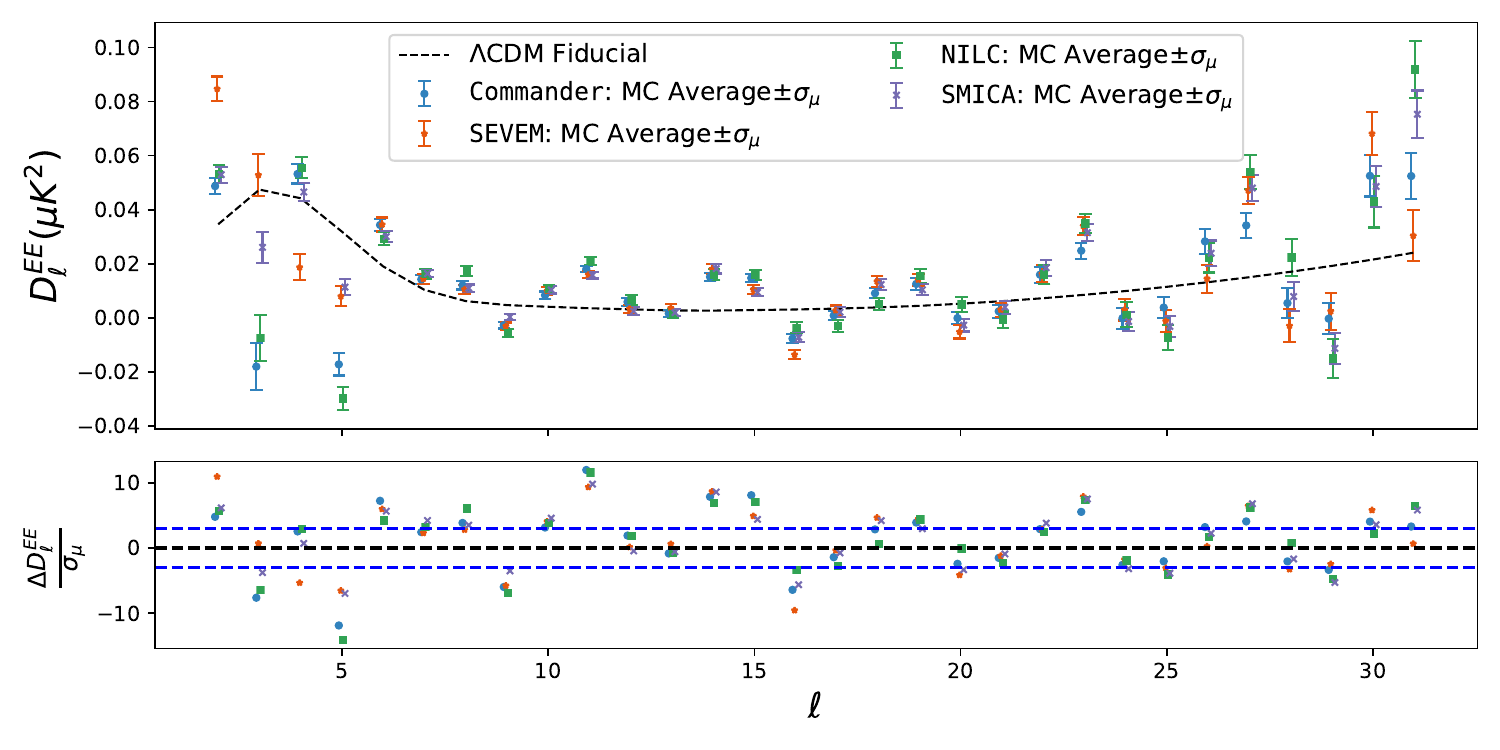}
\caption{The upper boxes of each panel show MC average of the $D_{\ell}^\mathrm{{TT}}$ (upper panel), $D_{\ell}^\mathrm{{TE}}$ (middle panel) and $D_{\ell}^\mathrm{{EE}}$ (lower panel), where $D_{\ell} = \frac{\ell(\ell+1)}{2\pi} C_{\ell}$, as a function of $\ell \in [2,31]$, obtained from the FFP10 MC set for {\tt Commander} (blue),  {\tt NILC} (green), {\tt SEVEM} (orange), {\tt SMICA} (violet). The error bars represent the uncertainties associated with the mean. The dashed black lines represent the \textit{Planck} $\Lambda$CDM best-fit model. Additionally, each panel displays a lower box in which the distance between the MC mean and the fiducial power spectrum is shown for each $\ell$ in units of the standard deviation of the mean, $\sigma_{\mu}$.}
\label{fig:PR3_MCspectra} 
\end{figure}

\begin{figure}[htbp]
\centering
\includegraphics[width=130mm]{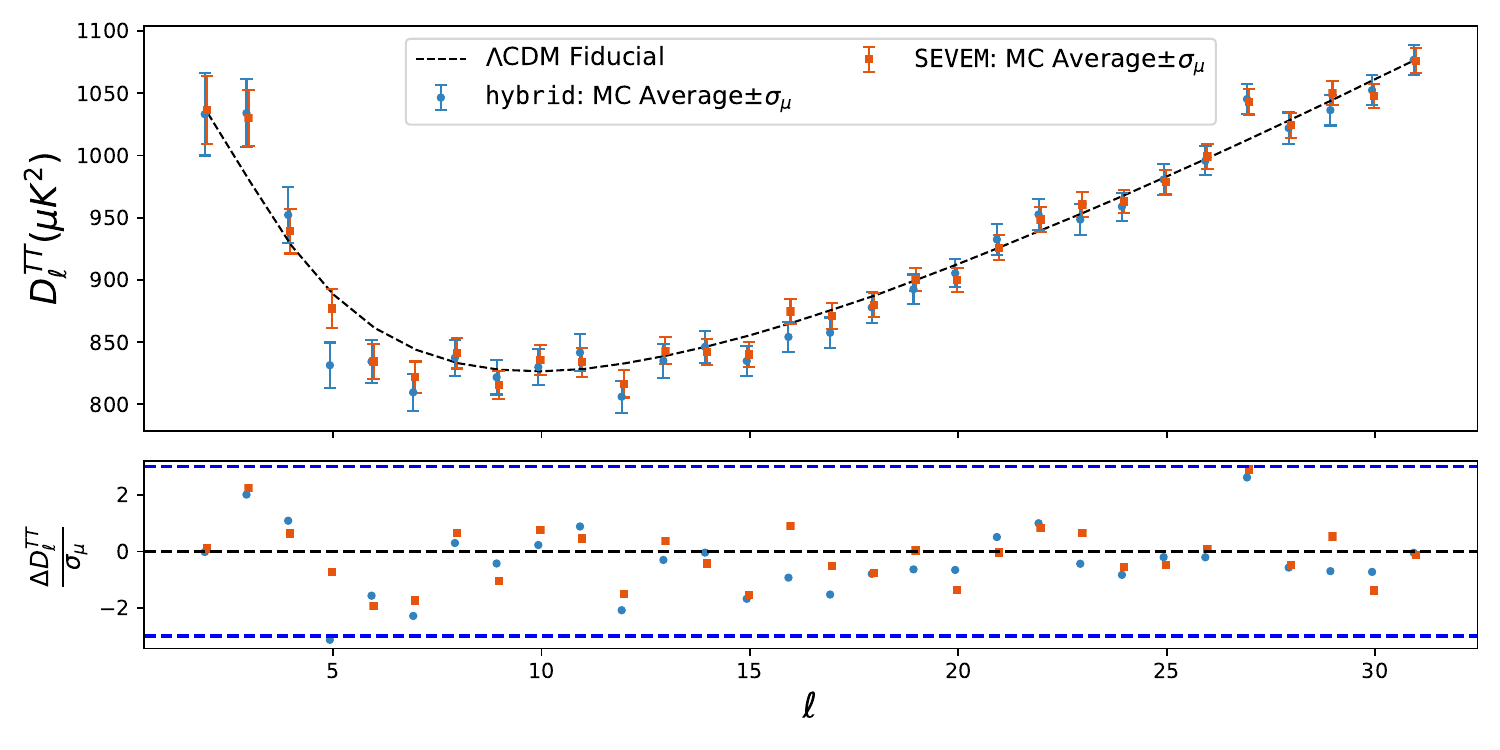} \\
\includegraphics[width=130mm]{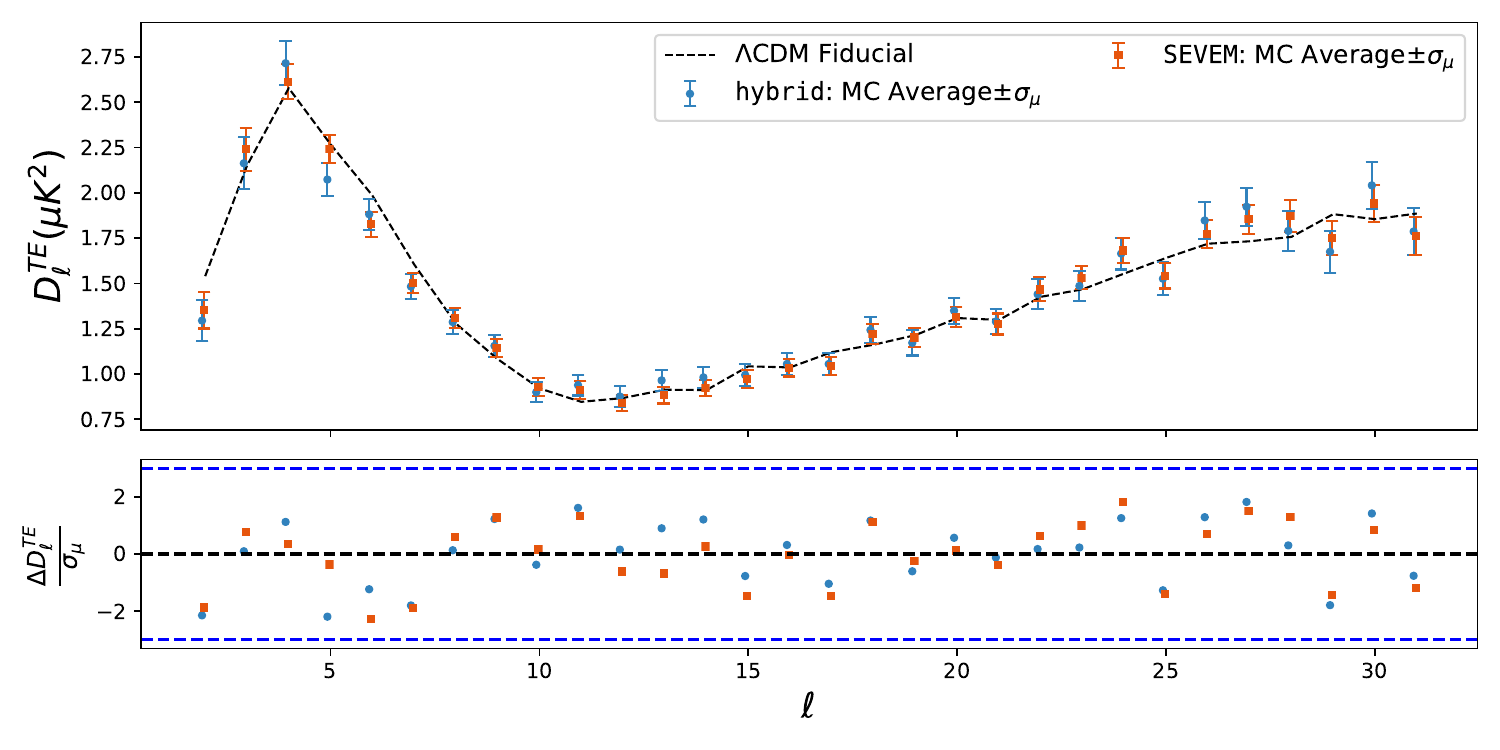}\\
\includegraphics[width=130mm]{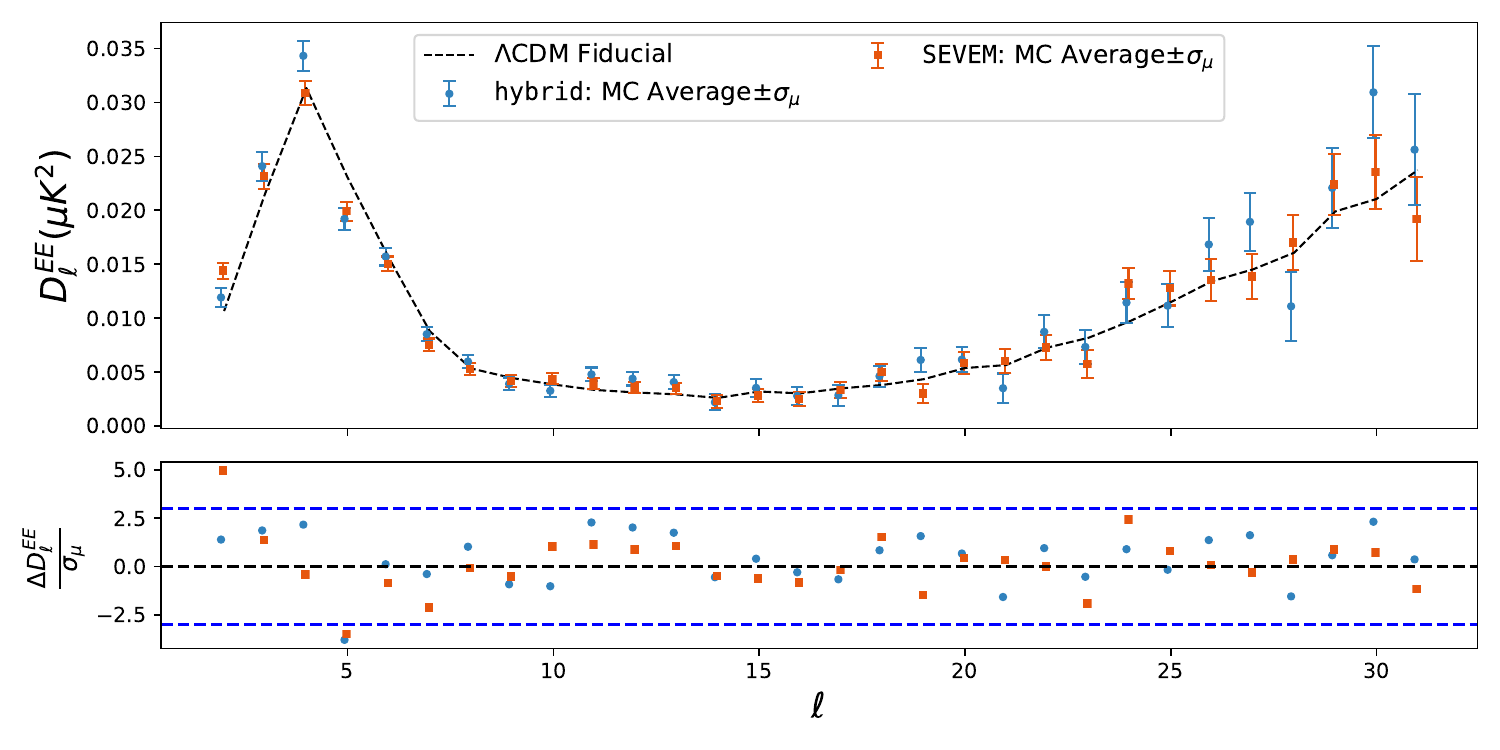}
\caption{Same as Figure~\ref{fig:PR3_MCspectra} but for the PR4 dataset, {\tt hybrid} (blue) and {\tt SEVEM} (orange). We remark that the {\tt hybrid} set is composed by 400 simulated TQU maps built with the {\tt SEVEM} temperature maps and the {\tt Commander} polarisation maps, split in detector A and detector B, while 600 simulations are used for the {\tt SEVEM} case.}
\label{fig:PR4_MCspectra} 
\end{figure}

From these maps we estimate the six CMB spectra in cross-mode over the considered sky fraction, in the harmonic range $\ell \in [2,31]$, with the Quadratic-Maximum-Likelihood (QML) estimator {\sc ECLIPSE} \footnote{\url{https://github.com/CosmoTool/ECLIPSE}} \cite{Bilbao-Ahedo:2021jhn}. This approach allows us to reduce the residuals of systematic effects and noise bias in the auto-spectra. Figures \ref{fig:PR3_MCspectra} and \ref{fig:PR4_MCspectra} characterise the distributions of the TT, TE, EE power spectra obtained from the PR3 and PR4 MC simulations, respectively. In each panel of these figures, the upper box shows the comparison of the MC mean with $ \sigma_{\mu} = \sigma_{\mathrm{MC}}/\sqrt{\mathrm{N_{sims}}}$ as error bar of the mean, with respect to the theoretical power spectra. The lower box displays the fluctuations from the theoretical model of the mean in units of $ \sigma_{\mu}$. As theoretical model we take into account the \textit{Planck} 2018 $\Lambda$CDM best-fit model\cite{Planck:2019evm}, henceforth referred to as fiducial power spectrum, defined by the following parameters: $\Omega_b h^2 = 0.022166$, $\Omega_c h^2 = 0.12029 $, $\Omega_{\nu} h^2 = 0.000645 $, $\Omega_{\Lambda} = 0.68139$, $h = 0.67019$, $\tau = 0.06018$,  $A_s = 2.1196 \,\, 10^{-9}$, $n_s = 0.96369$, where $h = H_{0}/100 \,\, \mathrm{km\,s^{-1}\,Mpc^{-1}}$. This spectrum is the one used to generate the CMB component in the simulated maps. It should be noted that for the PR4 dataset, the TE and EE fiducial spectra have been properly corrected for the NPIPE transfer function (see appendix~\ref{sec:appendixB}). These figures provide insight into the systematic effects present in the end-to-end simulations. In particular, for TT and TE spectra the MC mean follows the fiducial for both releases. 
However, for the EE power spectrum, significant discrepancies are observed for the PR3 dataset, reflecting the presence of systematic effects. Conversely, the PR4 EE spectrum follows the fiducial model for almost the full range of considered multipoles, which indicates the improvement in the level of systematics of the PR4 dataset.

Finally, Figures \ref{fig:PR3_spectra} and \ref{fig:PR4_spectra} present the TT (upper panel), TE (middle panel), and EE (lower panel) CMB power spectra obtained from the PR3 and PR4 data as a function of $\ell$, for $2 \le \ell \le 31$. For each plot, the points in the upper box represent the values of the $D_{\ell}$, where $D_{\ell} \equiv \frac{\ell(\ell+1)}{2\pi} C_{\ell}$, obtained from the data with the standard deviation of the MC distribution as error bar, while the dashed horizontal lines indicate the mean of the MC simulations. Each panel also displays a lower box in which the distance of the estimates from the MC mean in units of $\sigma$ is shown for each $\ell$. 
From these figures it can be observed that the data generally follow the mean of the simulations, indicating that the systematics are well characterised in the simulations for both the datasets.

%% file: 4results.tex
\section{Analysis Results}
\label{sec:Results}

\subsection{NAPS with \textit{Planck} dataset}

\begin{figure}[htbp]
\centering
\includegraphics[width=130mm]{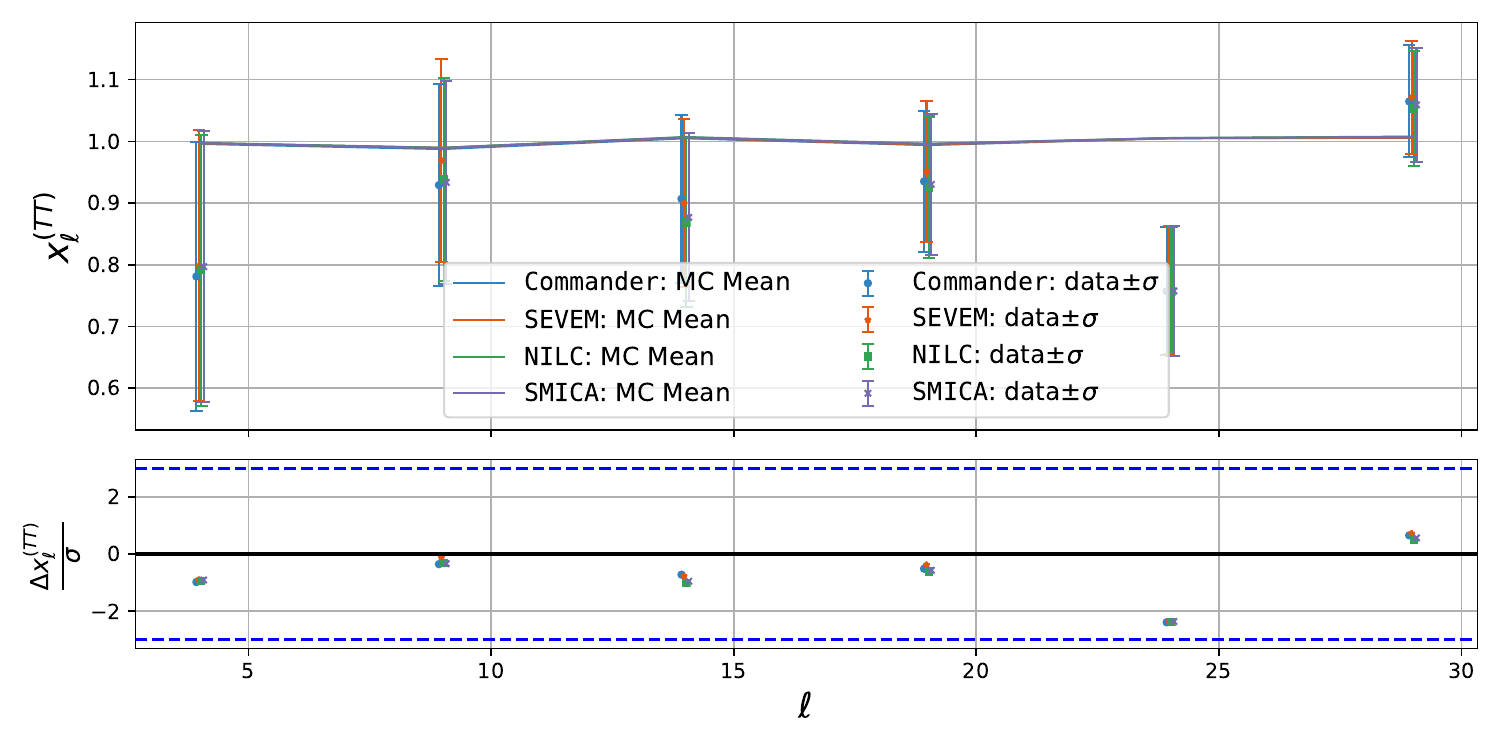} \\
\includegraphics[width=130mm]{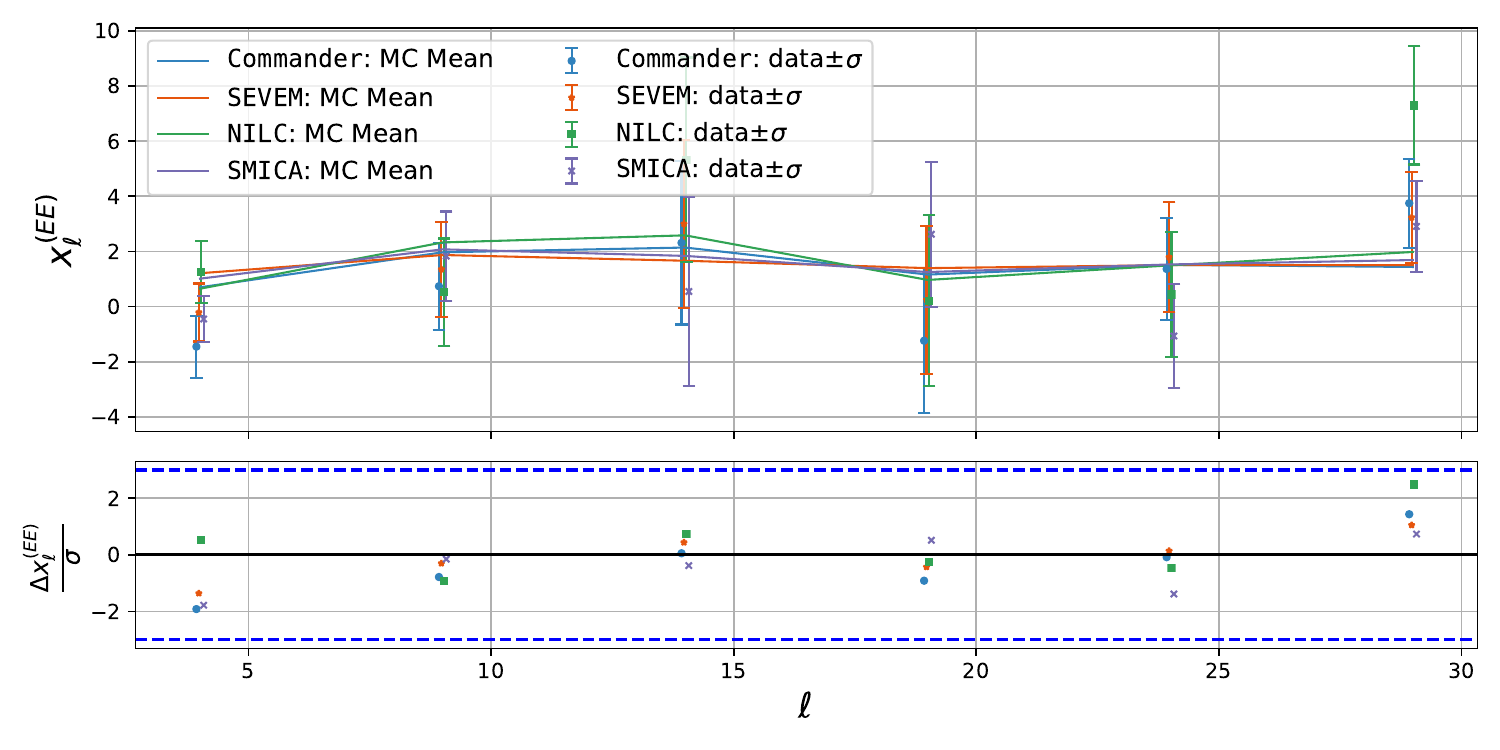} \\
\includegraphics[width=130mm]{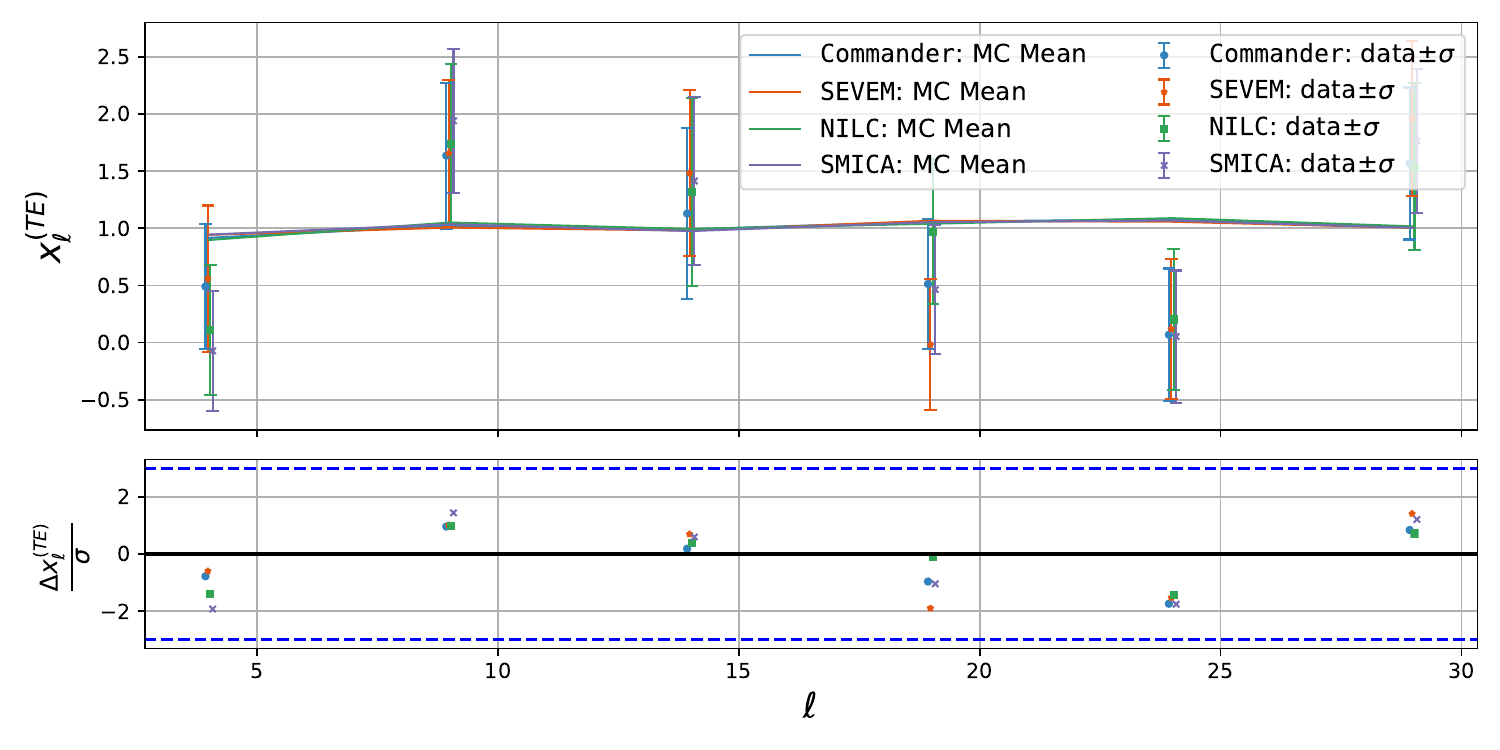}

\caption{The upper boxes of each panel show the $x^{\mathrm{(TT)}}$ (upper panel), $x^{\mathrm{(EE)}}$ (middle panel) and $x^{\mathrm{(TE)}}$ (lower panel), as defined in Eqs.~(\ref{eq:defNAPS1bin}), (\ref{eq:defNAPSEEbin}) and (\ref{eq:defNAPSTEbin}), for the PR3 data as a function of $\ell$, for $2 \le \ell \le 31$ and a bin of $\Delta \ell = 5$. The error bars represent the statistical uncertainties associated with the estimates, computed from the MC simulations. The solid lines represent the MC average of the spectra for each pipeline. Each panel also displays a lower box, in which the distance of the estimates from the MC mean in units of the standard deviation of the estimate is shown for each $\ell$.}
\label{fig:NAPS_PR3} 
\end{figure}

\begin{figure}[htbp]
\centering
\includegraphics[width=130mm]{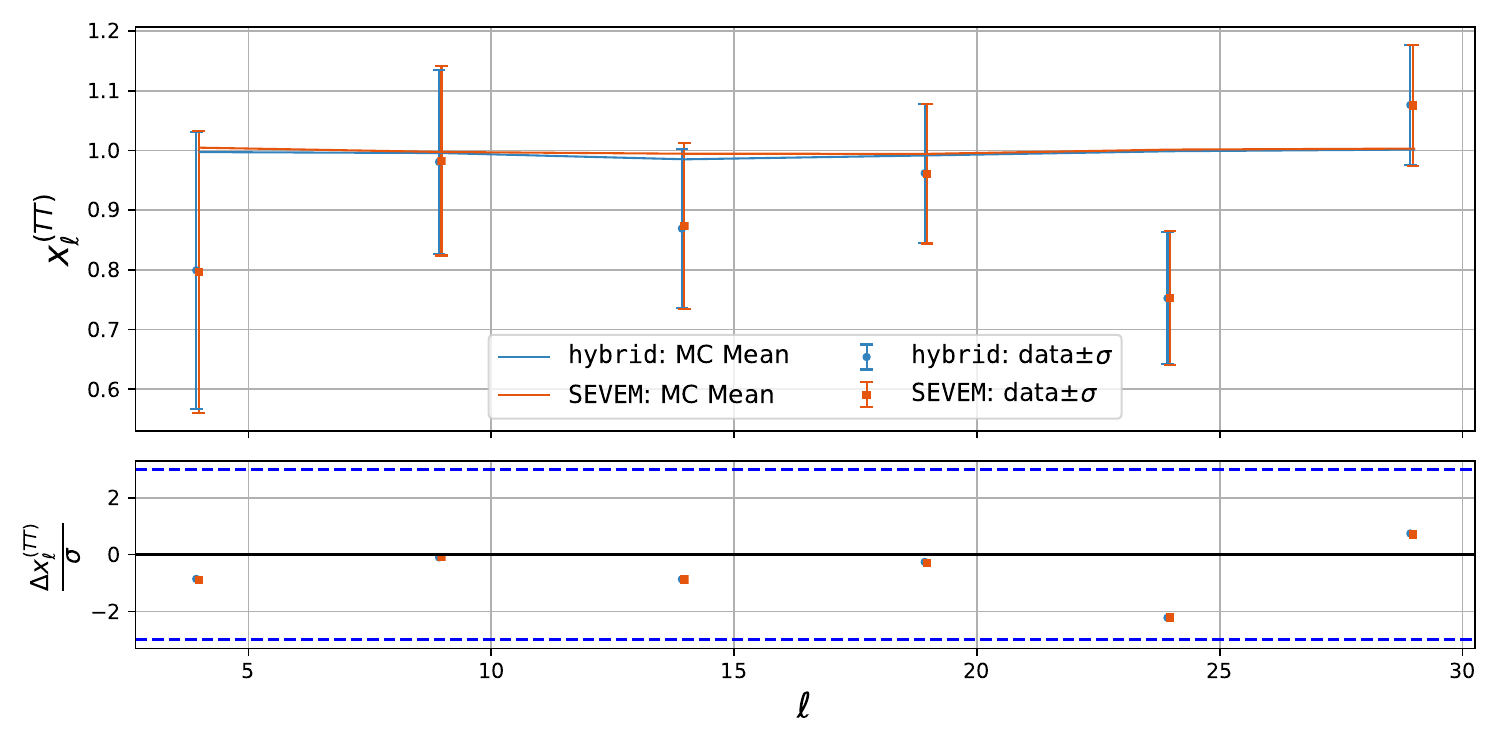} \\
\includegraphics[width=130mm]{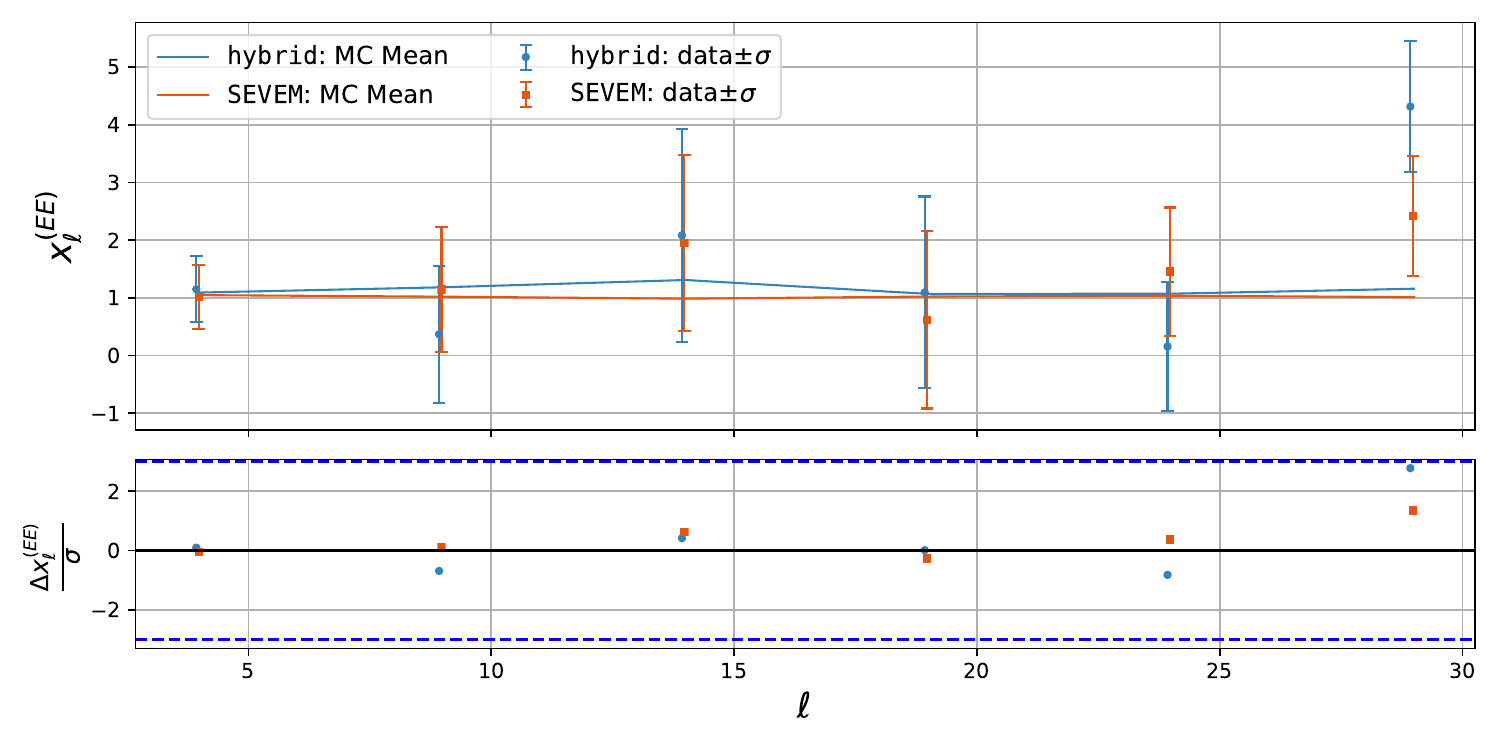} \\
\includegraphics[width=130mm]{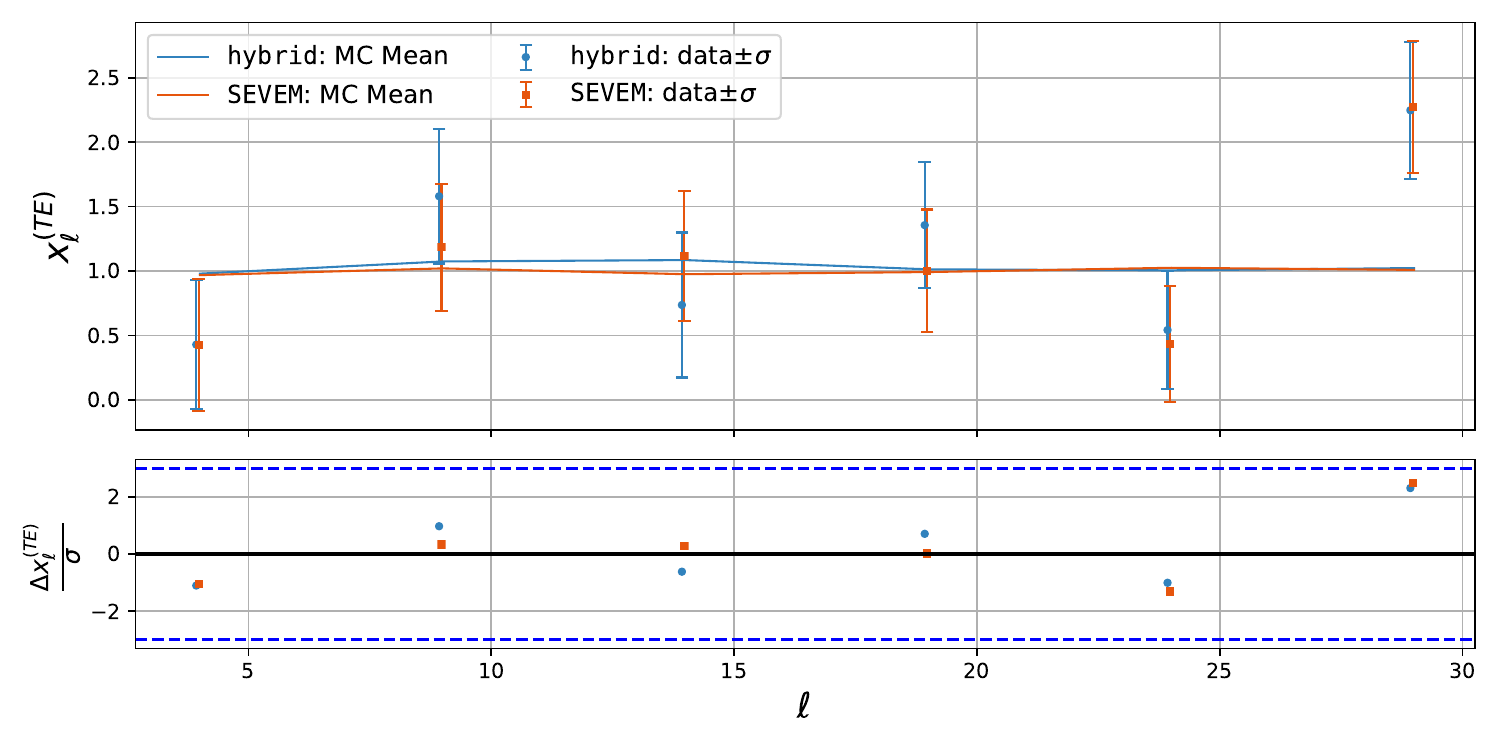}
\caption{Same as Figure~\ref{fig:NAPS_PR3} for the PR4 dataset.}
\label{fig:NAPS_PR4} 
\end{figure}

Starting from the angular spectra, computed in the previous section, we build the NAPS, as defined in Eqs.~(\ref{eq:defNAPS1bin}),(\ref{eq:defNAPSEEbin}) and (\ref{eq:defNAPSTEbin}), for each MC simulation and for the data, in the harmonic range $\ell \in [2,31]$. In order to reduce any possible correlation effect between different multipoles arising from the mask, we have studied different combinations of binning schemes on the CMB multipoles, choosing a constant bin of $\Delta \ell = 5$. Figures~\ref{fig:NAPS_PR3} and ~\ref{fig:NAPS_PR4} show the values of the NAPS $x^{(\mathrm{TT})}_{\ell}$ (upper panel), $x^{(\mathrm{EE})}_{\ell}$ (middle panel) and $x^{(\mathrm{TE})}_{\ell}$ (lower panel) for the PR3 and PR4 datasets, respectively. In each panel, the upper box displays the values of the data along with the standard deviation of the MC distribution. Each panel also presents a lower box, in which the distance between the data and the MC mean is shown in units of the uncertainties of the estimates themselves for each $\ell$. For both PR3 and PR4, the data follow the mean of the simulations, which is around the expected value of 1.  However, it is interesting to note that the estimated error of the binned NAPS for $x^{(\mathrm{EE})}_{\ell}$ -- obtained as the dispersion of the simulations -- in PR4 has decreased. This reflects the improvement in noise and systematics in PR4 with respect to the previous data release.

\subsection {Variance-based estimators}
\begin{figure}[htbp]
\centering
\includegraphics[width=90mm]{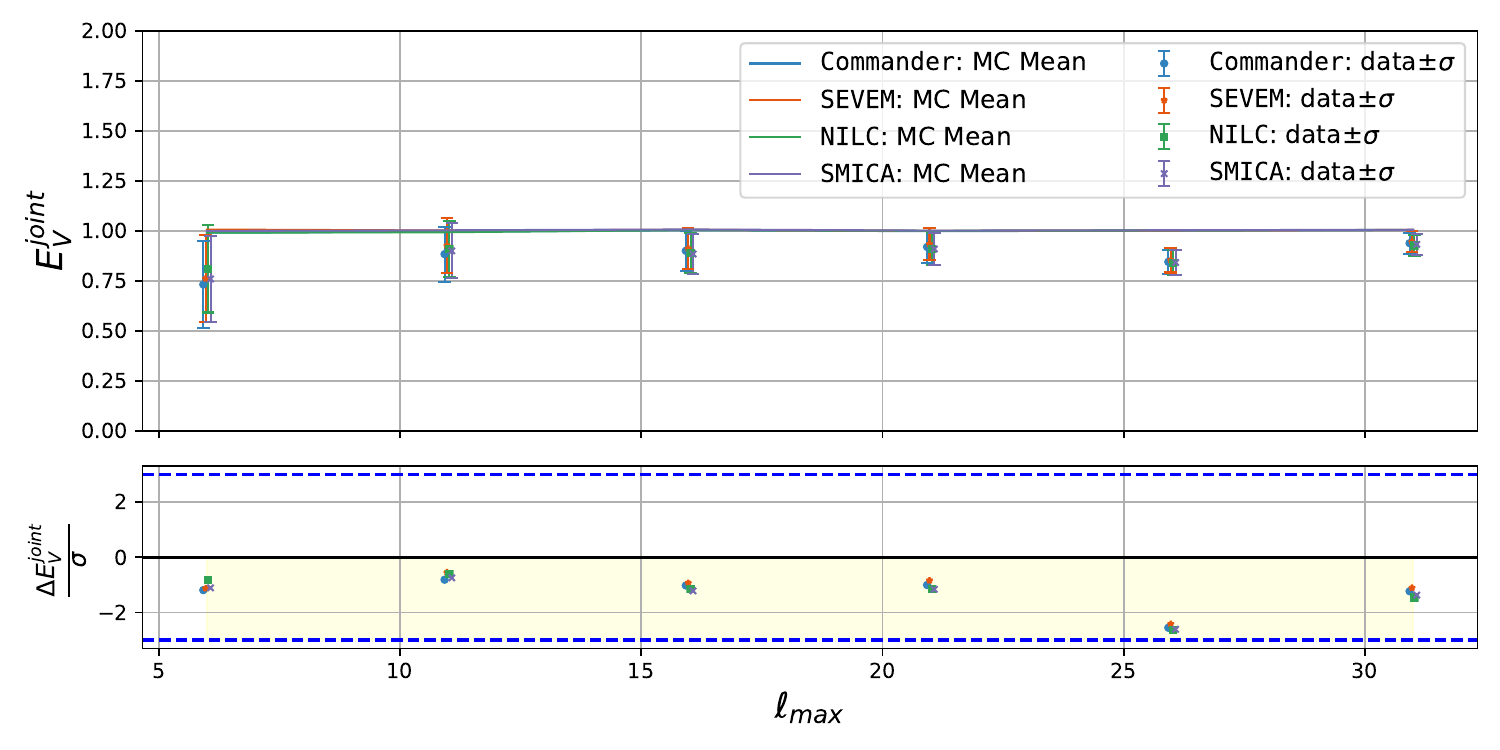}\\
\includegraphics[width=90mm]{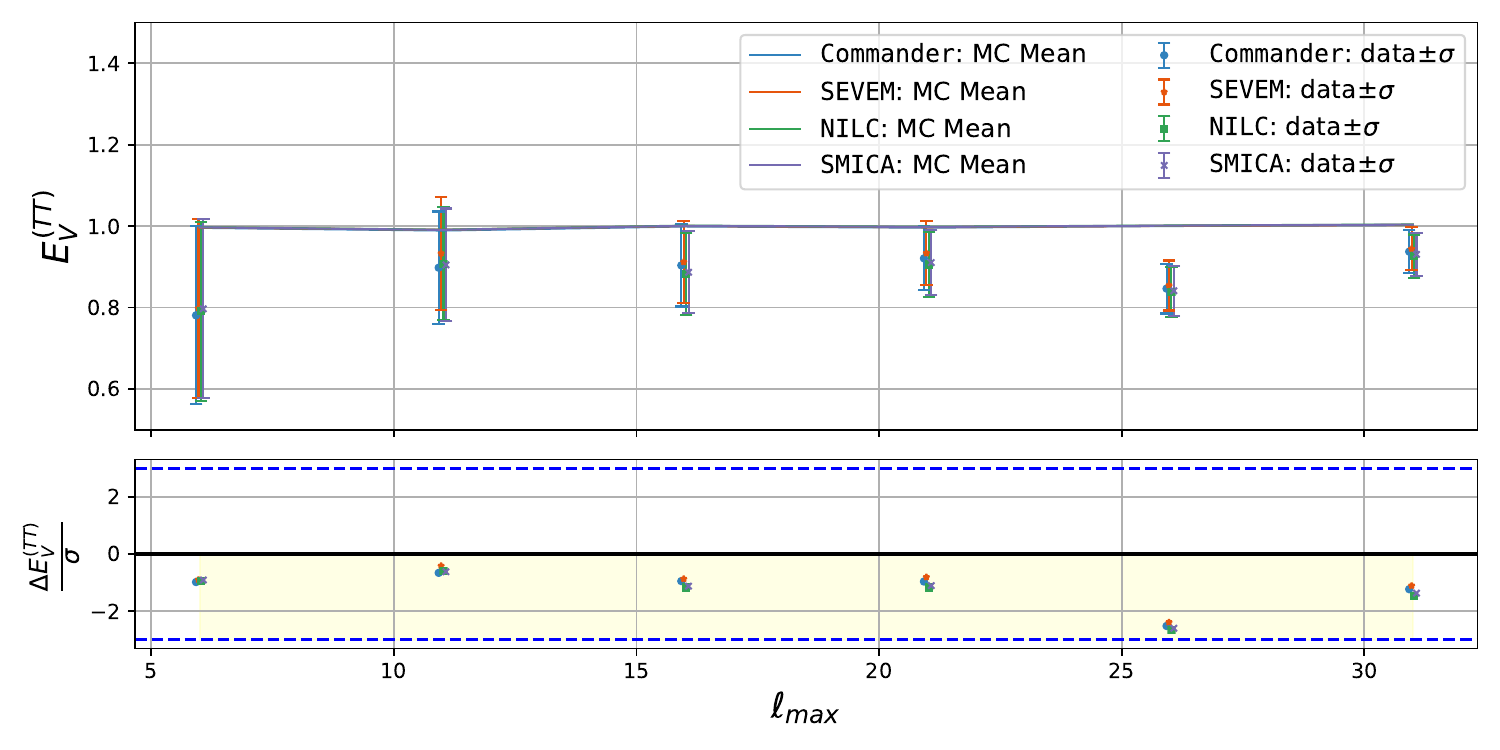}\\
\includegraphics[width=90mm]{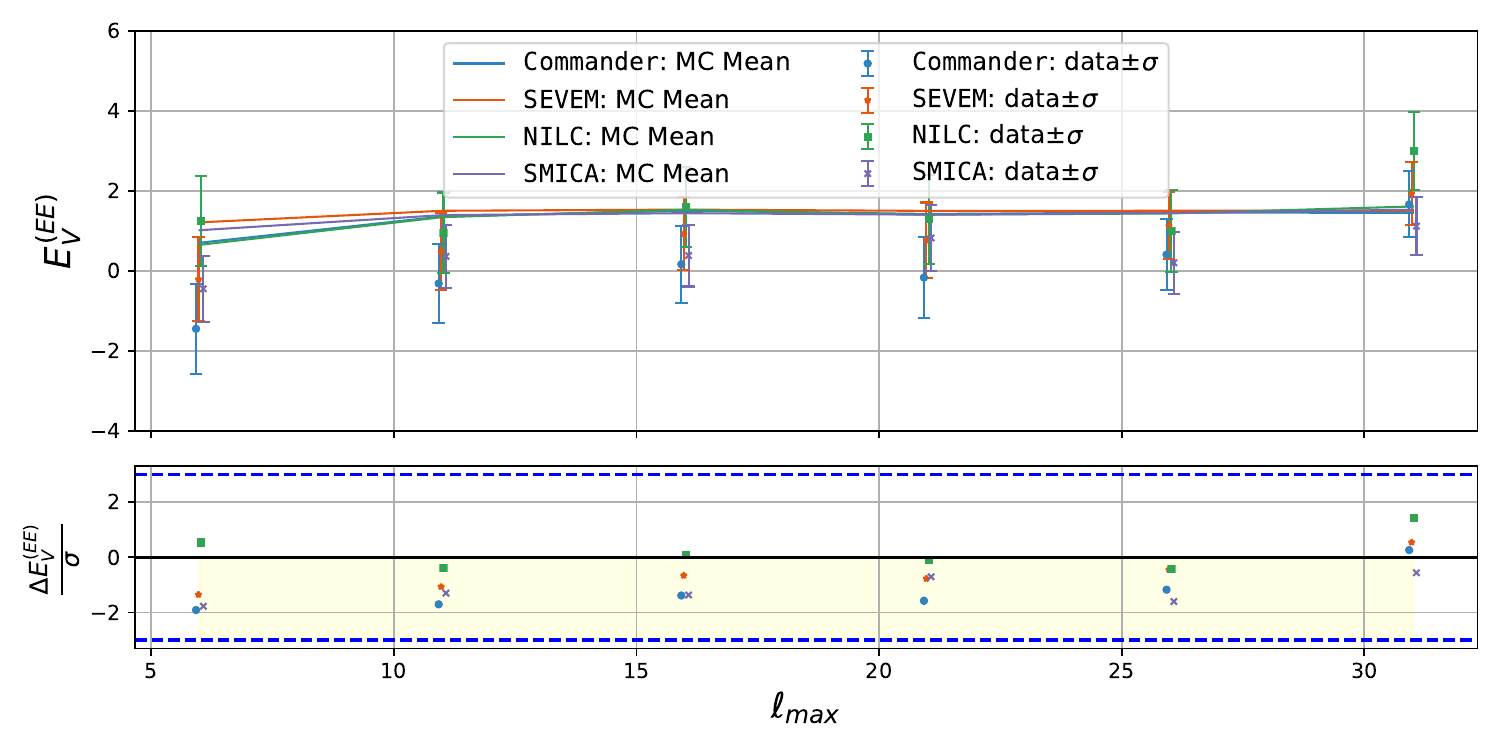}  \\
\includegraphics[width=90mm]{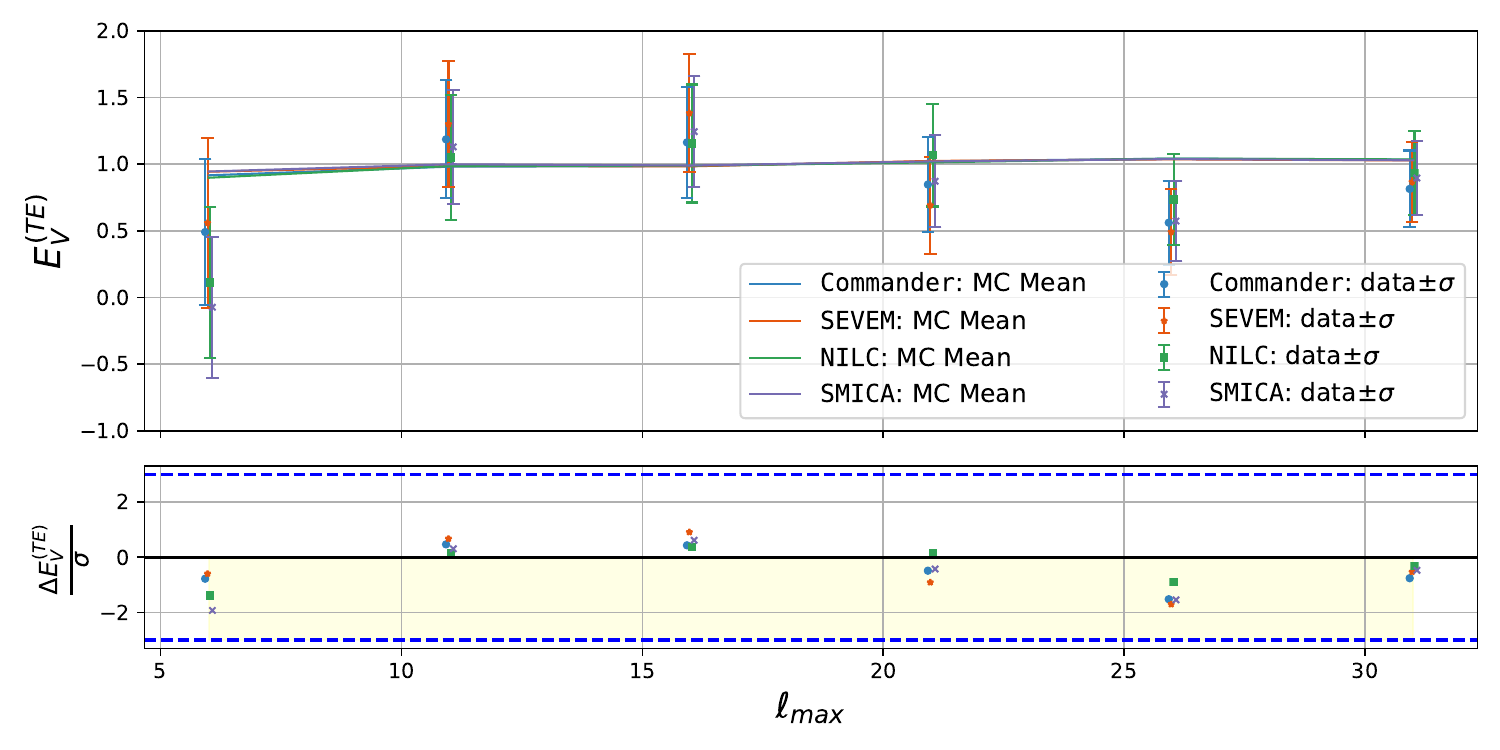} 
\caption{$E_\mathrm{V}^{\mathrm{joint}}$ (first row), $E_\mathrm{V}^{\mathrm{(TT)}}$ (second row), $E_\mathrm{V}^{\mathrm{(EE)}}$ (third row), $E_\mathrm{V}^{\mathrm{(TE)}}$ (fourth row), defined in Eqs. (\ref{eq:EVjoint_bin}), (\ref{eq:EV1_bin}), (\ref{eq:EVEE_bin}) and (\ref{eq:EVTE_bin}), when applied on the PR3 dataset as a function of $\ell_{\mathrm{max}} \in [2,31]$.  In the upper boxes of each panel the values of the data with the standard deviation of the MC distribution as error bars is shown for all the pipelines ({\tt Commander} in blue,  {\tt NILC} in green, {\tt SEVEM} in orange, {\tt SMICA} in violet). The solid lines represent the mean of the MC distribution for each pipeline. In the lower panel the difference between the data and the mean in units of standard deviation of the MC distribution is shown for each $\ell_{\mathrm{max}}$. The yellow regions represent a lack of power with respect to that predicted by the $\Lambda$CDM model.}
\label{fig:EV_estimators_PR3} 
\end{figure}

\begin{figure}[htbp]
\centering
\includegraphics[width=90mm]{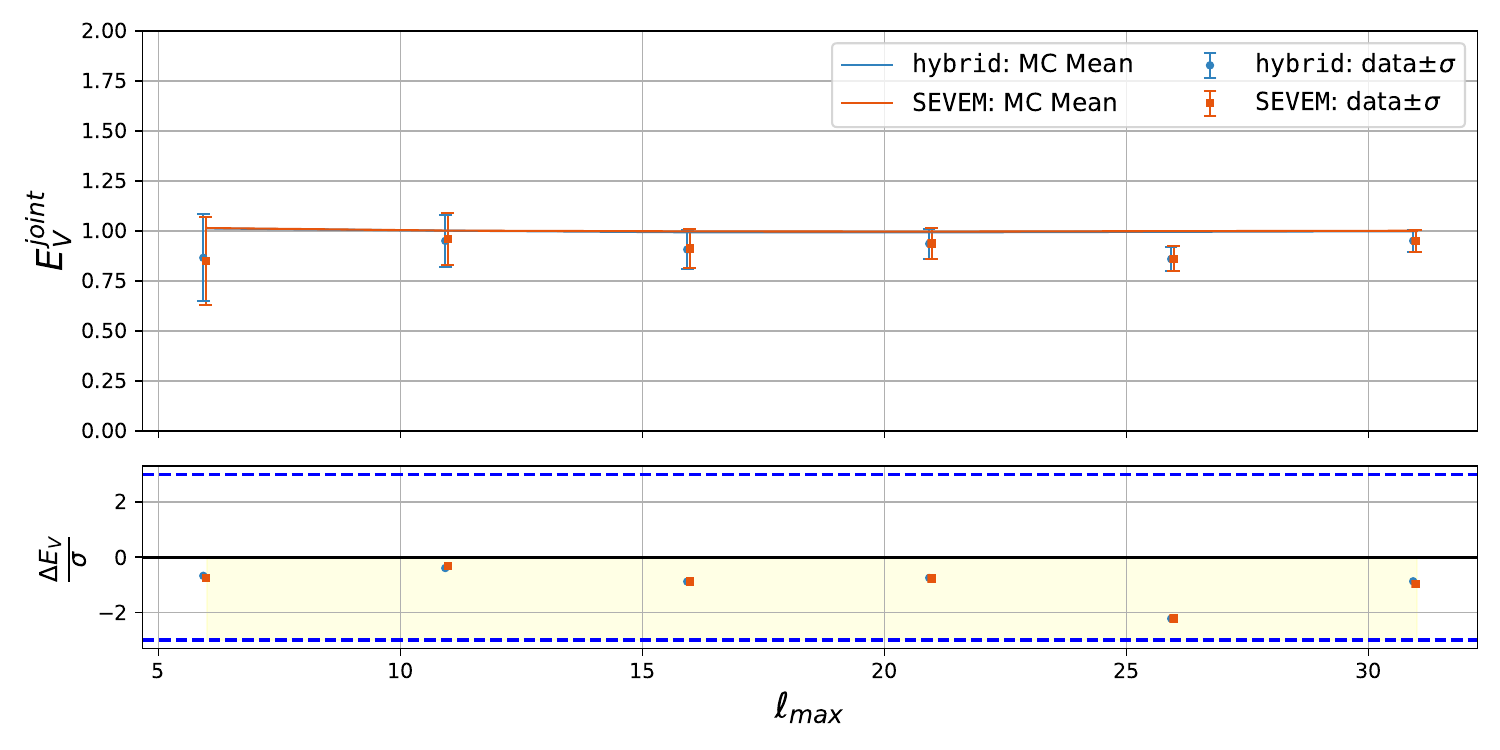}\\
\includegraphics[width=90mm]{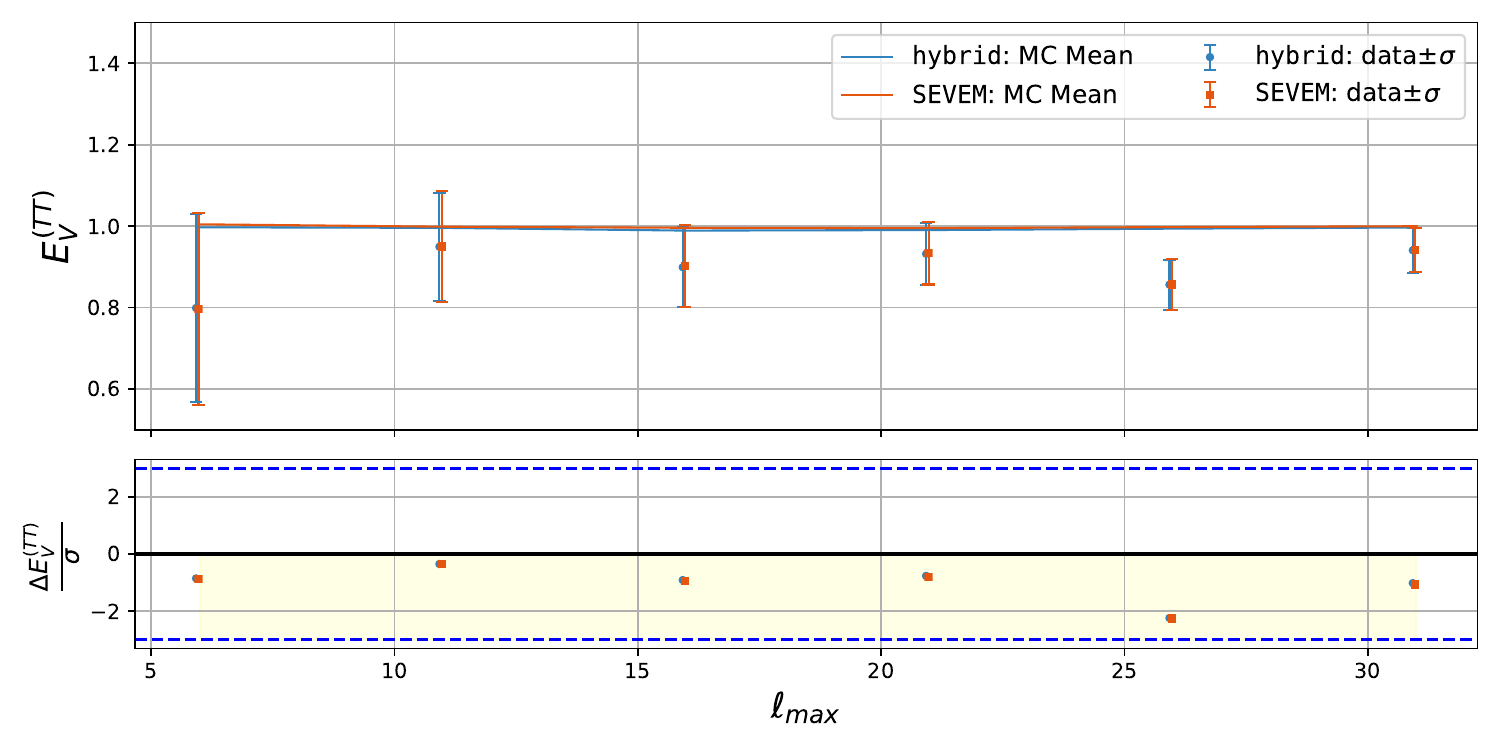}\\
\includegraphics[width=90mm]{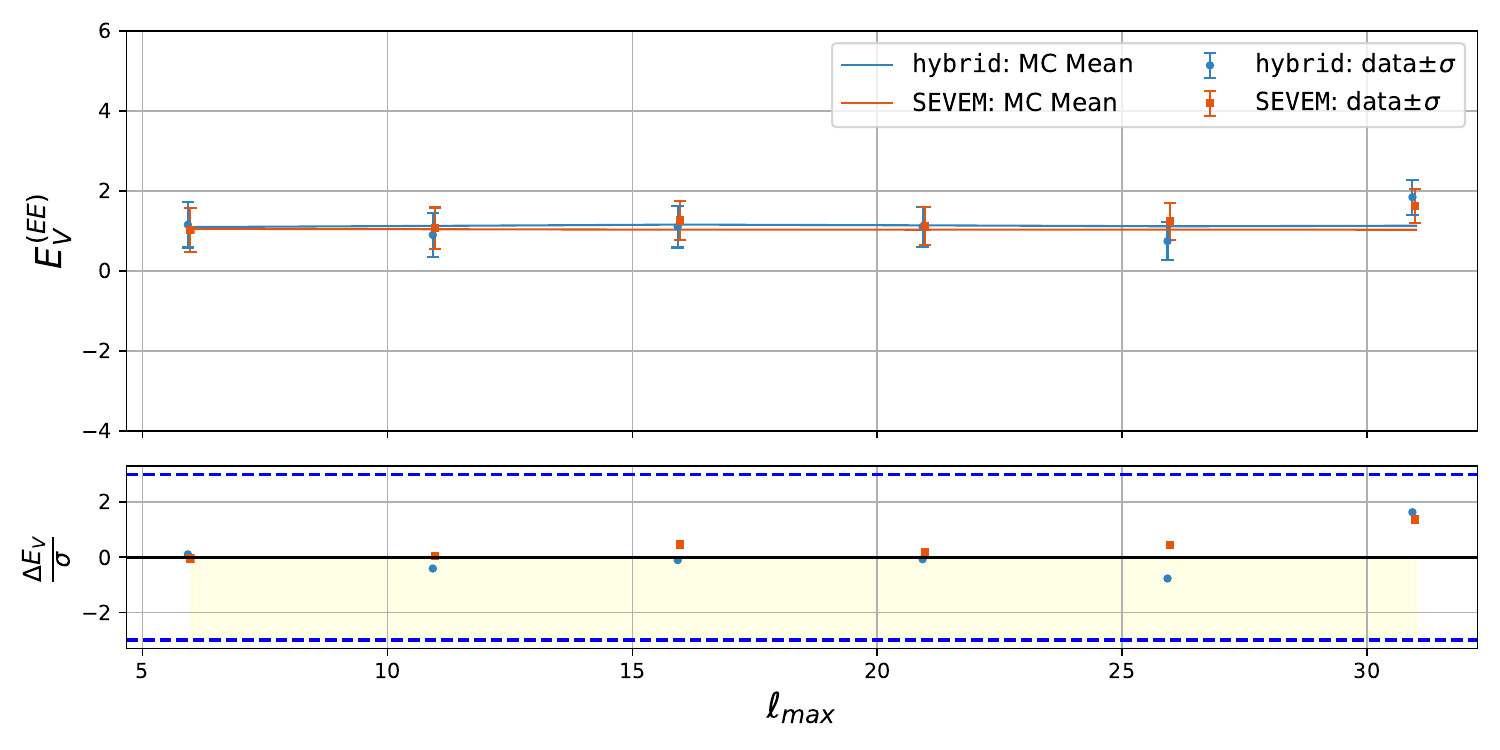} \\
\includegraphics[width=90mm]{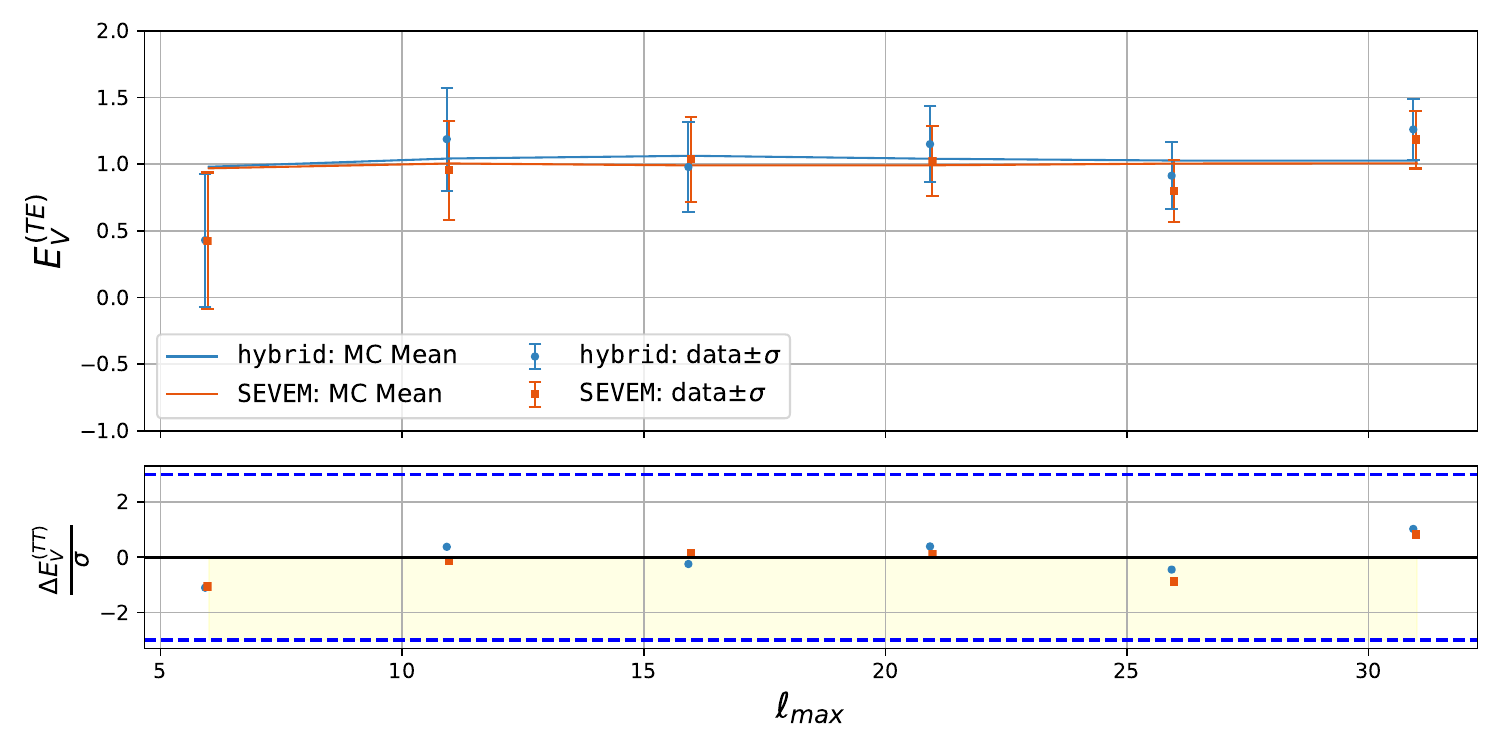}
\caption{Same as Figure~\ref{fig:EV_estimators_PR3} but for the PR4 data.}
\label{fig:EV_estimators_PR4} 
\end{figure}

Figures \ref{fig:EV_estimators_PR3} and ~\ref{fig:EV_estimators_PR4} show $E_\mathrm{V}^{\mathrm{joint}}$ (first row), $E_\mathrm{V}^{\mathrm{(TT)}}$ (second row), $E_\mathrm{V}^{\mathrm{(EE)}}$ (third row), $E_\mathrm{V}^{\mathrm{(TE)}}$ (fourth row), defined in Eqs. (\ref{eq:EVjoint_bin}), (\ref{eq:EV1_bin}), (\ref{eq:EVEE_bin}) and (\ref{eq:EVTE_bin}), as a function of the maximum multipole considered in the analysis, when applied to the PR3 and PR4 dataset respectively. In each panel the upper box displays the values of the data for all the pipelines, with the standard deviation of the distribution of the simulations as error bars, together with the mean of the MC distribution (solid lines). In the lower panel for each $\ell_{\mathrm{max}}$, the difference between the data and the MC mean is shown in units of standard deviation of the MC distribution. If the points are inside the yellow region it means that the observed power is less than that predicted by the $\mathrm{\Lambda CDM}$ model. Note that by applying $E_\mathrm{V}^{\mathrm{joint}}$, $E_\mathrm{V}^{\mathrm{(TT)}}$, $E_\mathrm{V}^{\mathrm{(EE)}}$ and $E_\mathrm{V}^{\mathrm{(TE)}}$ to PR3 and PR4 datasets, all the values for the first two estimators and the majority for the last two ones fall within the yellow regions. Consequently, the \textit{Planck} data seem to exhibit less power than that expected in the $\mathrm{\Lambda CDM}$ scenario.

In order to quantify the statistical significance of this feature we consider the lower-tail-probability (LTP), defined as: 
\begin{equation}
\mathrm{LTP}(\%) = 100* \frac{N_{sim}^{(data)}}{N_{sims}}
\label{eq:defLTP}
\end{equation}
where $\mathrm{N_{sim}^{(data)}}$ is the number of simulations with a value equal or smaller than the data value and $\mathrm{N_{sims}}$ is the total number of simulations. Consequently, Eq.(\ref{eq:defLTP}) provides the probability to obtain a value as low as the data. However, this analysis is limited by the number of MC simulations we have. In particular, it is not possible to distinguish between $\mathrm{LTP}<0.33\%$ for the PR3 dataset and between $\mathrm{LTP}<0.25\%$ and $\mathrm{LTP}<0.17\%$ for the {\tt hybrid} and {\tt SEVEM} PR4 pipelines \footnote{We remark that we have 300 simulations for the PR3 pipelines and 400 and 600 simulations for the {\tt hybrid} and {\tt SEVEM} PR4 pipelines, respectively.}. If the LTP value is below the aforementioned thresholds, it indicates that the data is lower than all the simulations. 

\begin{figure}[htbp]
\centering

\includegraphics[width=130mm]{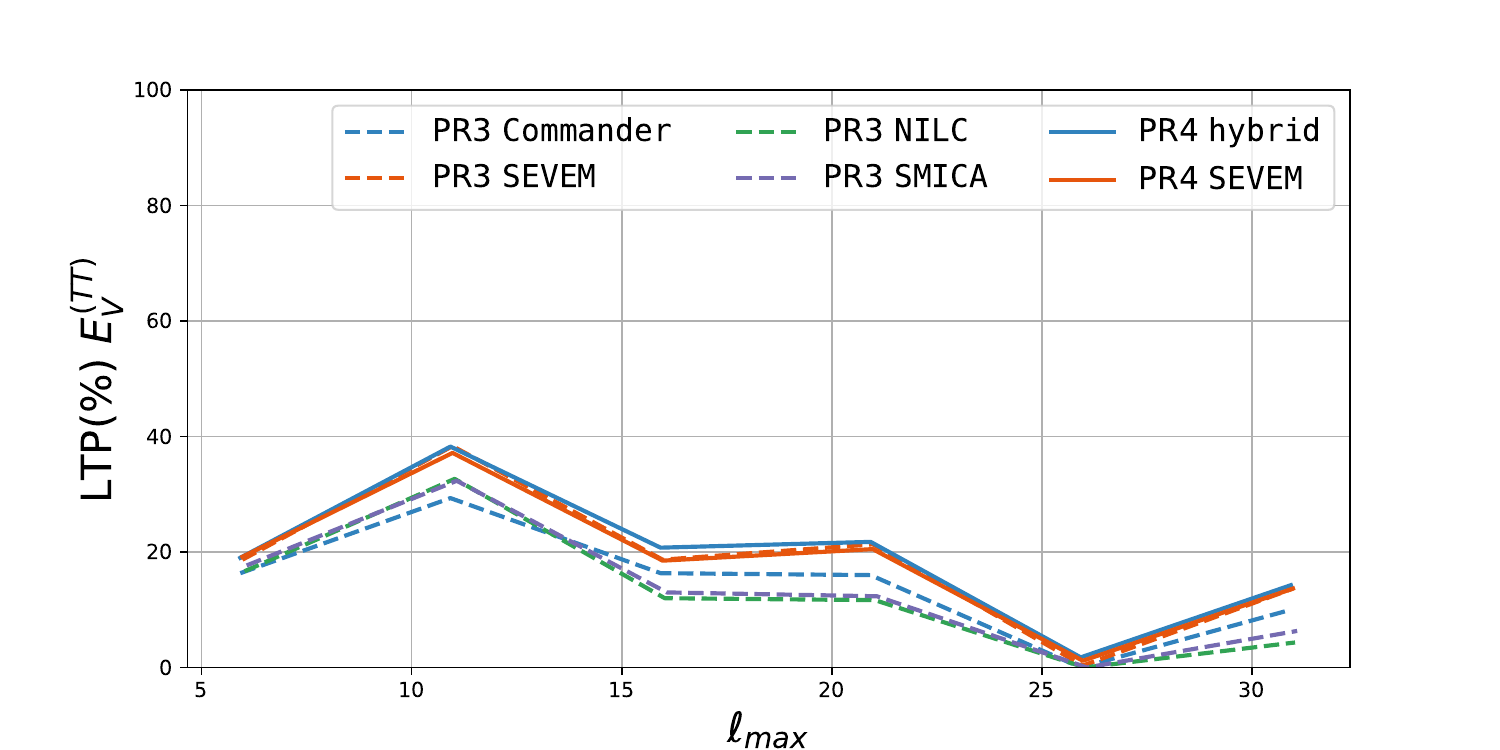} 
\includegraphics[width=130mm]{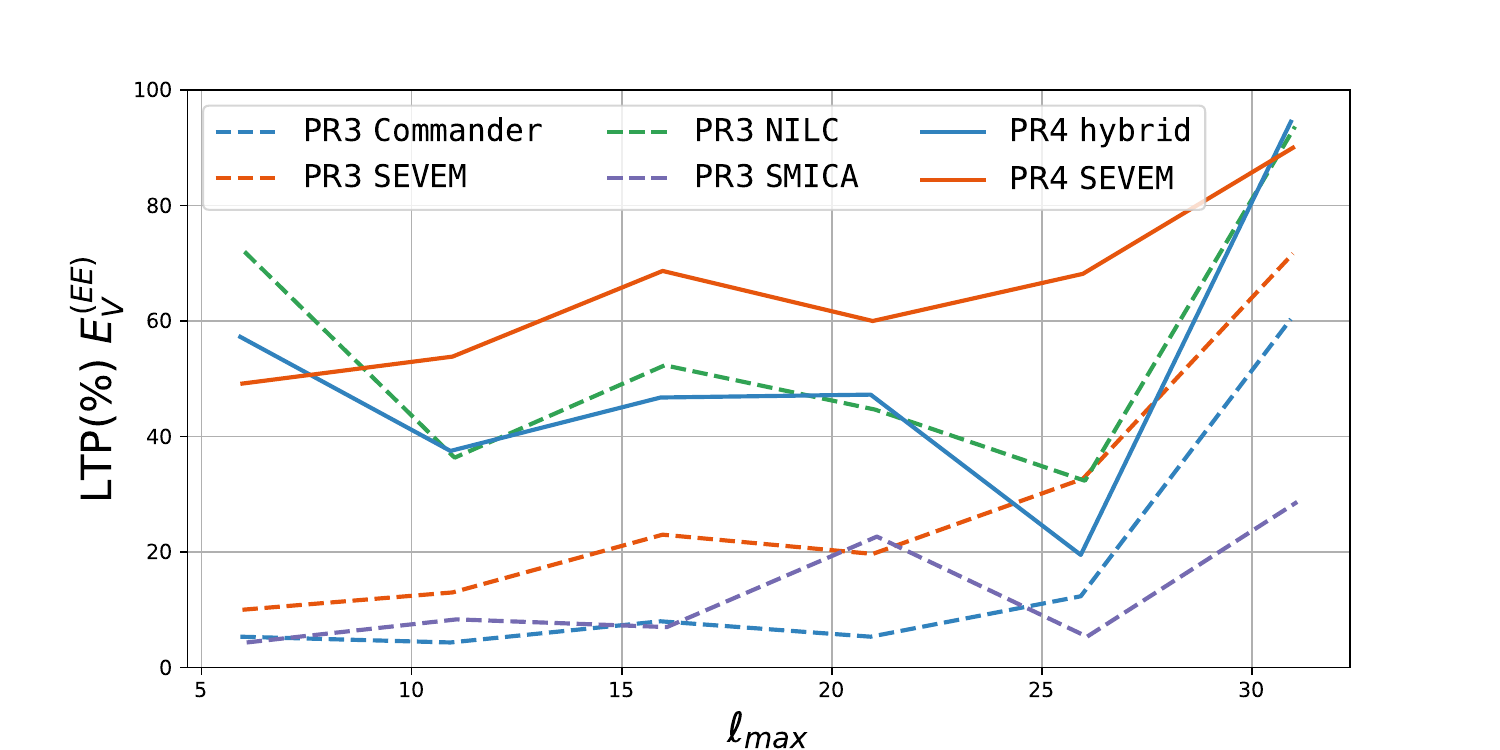} 
\includegraphics[width=130mm]{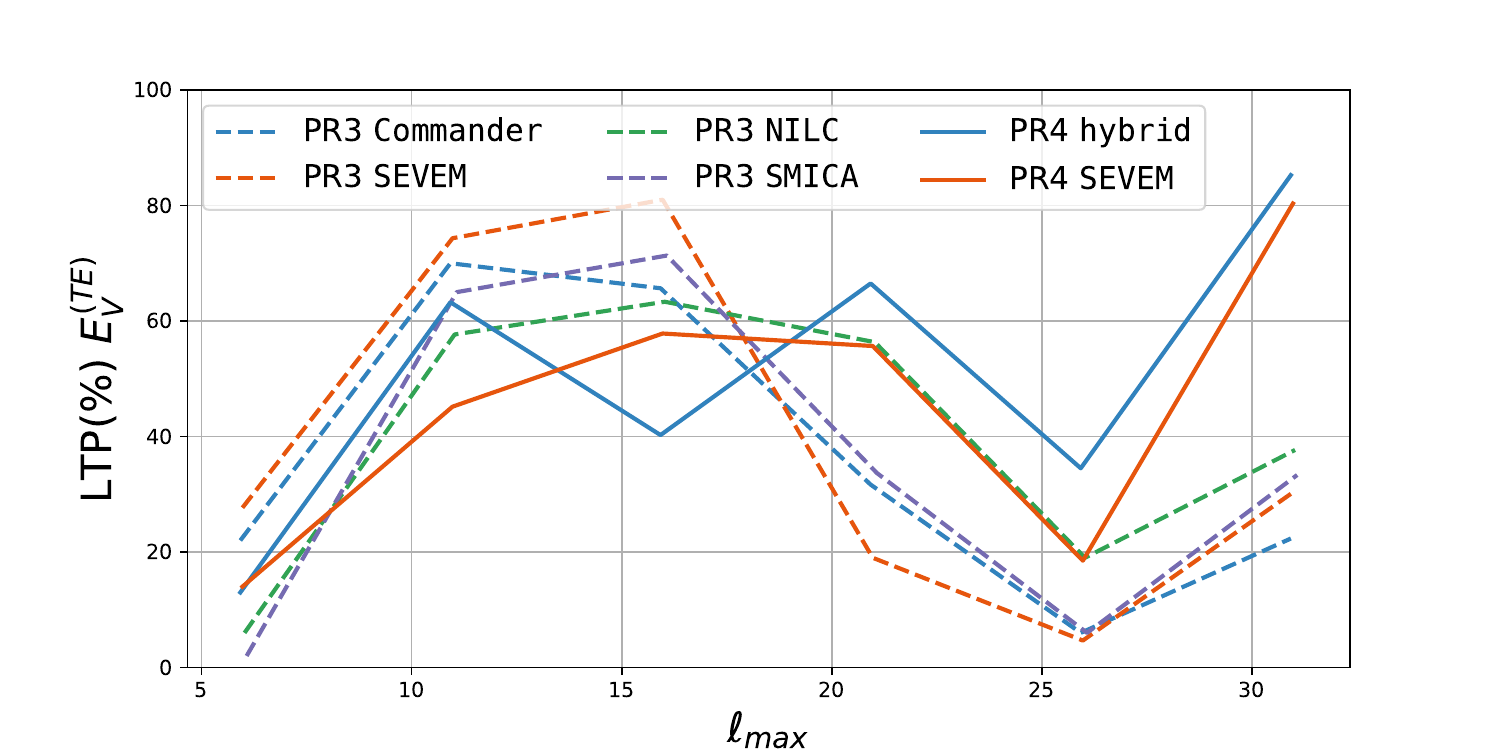}

\caption{LTP$(\%)$ of the $E_\mathrm{V}^{\mathrm{(TT)}}$ (upper panel), $E_\mathrm{V}^{\mathrm{(EE)}}$ (middle panel) and $E_\mathrm{V}^{\mathrm{(TE)}}$ (lower panel) applied on the PR3 (dashed lines) and PR4 (solid lines) dataset as a function of $\ell_{\mathrm{max}} \in [2,31]$. The LTP is obtained as the percentage of simulations with a value smaller than that of the data. However, note that, in practice, due to the limited number of simulations, the maximum sensitivity that we can reach for the LTP is 0.33\% (PR3 dataset), 0.25\% (PR4, {\tt hybrid} pipeline) and 0.17\% (PR4, {\tt SEVEM} pipeline). Values below those limits in the plot indicate that none of the simulations has a lower value than the data.}
\label{fig:ltpEV12} 
\end{figure}

Figure \ref{fig:ltpEV12} displays the LTP$(\%)$ computed for the $E_\mathrm{V}^{\mathrm{(TT)}}$ (upper panel), $E_\mathrm{V}^{\mathrm{(EE)}}$ (middle panel) and $E_\mathrm{V}^{\mathrm{(TE)}}$ (lower panel), applied on the PR3 (dashed lines) and PR4 (solid lines) datasets as a function of $\ell_{\mathrm{max}}$. 
From these plots it can be observed that:
\begin{itemize}
    \item with the estimator based only on temperature data, $E_\mathrm{V}^{\mathrm{(TT)}}$, we find well compatible results between different pipelines and between PR3 and PR4 datasets. The minimum value of the LTP is reached at $\ell_{\mathrm{max}} = 26$: we confirm the presence of a lack of power with a LTP $\leq 0.33\%$ (PR3) and $\leq 1.76\%$ (PR4), see table \ref{tab:LTP}. To the best of our knowledge, the probability for the PR3 dataset is the lowest one present in the literature obtained from this dataset applying the \textit{Planck} confidence mask (see Figure \ref{fig:TQU_masks}). Note that, as shown in \cite{Cruz:2010ud,Gruppuso:2013xba}, this anomaly tends to become less significant with increasing sky coverage: in the previous works \cite{Planck:2015igc, Planck:2019evm} a LTP below the $1\%$ was measured only when high-latitude galactic regions were taken into account; 
    \item we find significant differences between PR3 and PR4 dataset when polarisation is taken into account through the estimators $E_\mathrm{V}^{\mathrm{(EE)}}$ and $E_\mathrm{V}^{\mathrm{(TE)}}$. These differences are most likely due to a combination of two factors which are out of our control: a different level of systematics between the PR3 and PR4 dataset and the transfer function which impacts on the PR4 dataset. Indeed, PR4 data, with respect to PR3 release, better describe the component in polarisation of CMB, providing reduced levels of noise and systematics at all angular scales. However,  on the other hand the PR4 polarisation at the largest angular scales are affected by the transfer function which cut the power of the polarised signal, reducing their contribution at those scales;
    \item note that the estimator based only on EE power spectrum, $E_\mathrm{V}^{\mathrm{(EE)}}$, at very large scales may exhibit a hint of lack of power when applied to the PR3 data. In particular for the {\tt Commander} and {\tt SMICA} pipelines the LTP is $\leq 8\%$ for $\ell_{\mathrm{max}} \le 16$, smaller than the values obtained with the estimator based only on temperature data. 
\end{itemize}

\begin{figure}[htbp]
\centering
\includegraphics[width=150mm]{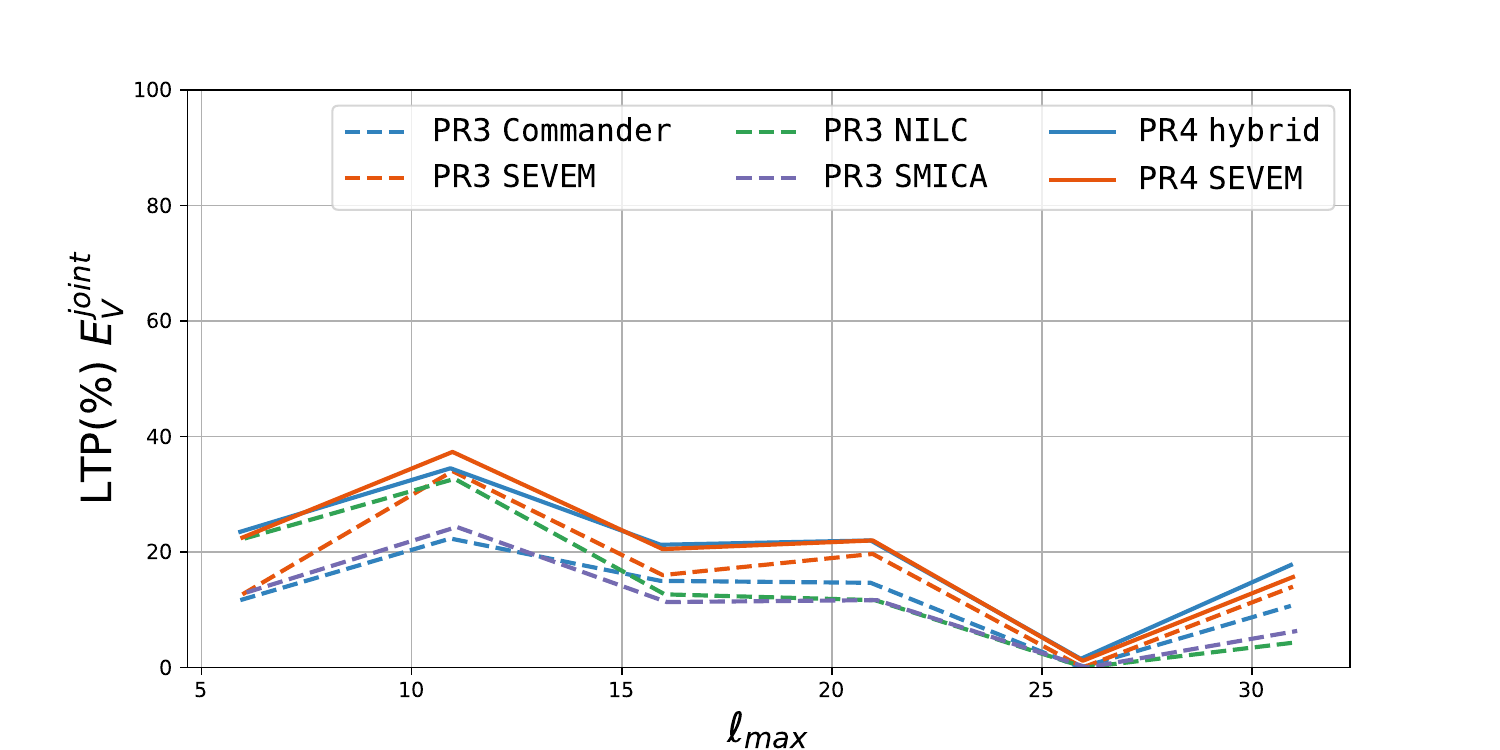} \\
\caption{$LTP(\%)$ of the $E_\mathrm{V}^{\mathrm{joint}}$ applied to the PR3 (dashed lines) and PR4 (solid lines) dataset as a function of $\ell_{\mathrm{max}}\in [2,31]$. The LTP is obtained as the percentage of simulations with a value smaller than that of the data. However, note that, in practice, due to the limited number of simulations, the maximum sensitivity that we can reach for the LTP is 0.33\% (PR3 dataset), 0.25\% (PR4, {\tt hybrid} pipeline) and 0.17\% (PR4, {\tt SEVEM} pipeline). Values below those limits in the plot indicate that none of the simulations has a lower value than the data.}
\label{fig:ltpEVs} 
\end{figure}

Figure \ref{fig:ltpEVs} displays the LTP$(\%)$ computed for the $E_\mathrm{V}^{\mathrm{joint}}$ when applied to the PR3 (dashed lines) and PR4 (solid lines) datasets as a function of $\ell_{\mathrm{max}}$. 
We find well compatible results between different pipelines and between PR3 and PR4 dataset. The trend of the LTP is very similar to the one obtained from $E_\mathrm{V}^{\mathrm{(TT)}}$, reaching the minimun at $\ell_{\mathrm{max}} = 26$ (see table \ref{tab:LTP}). Furthermore, for the PR3 dataset, the LTP is smaller than the one measured with $E_\mathrm{V}^{\mathrm{(TT)}}$ at each scale, except for the {\tt NILC} pipeline. In particular at $\ell_{\mathrm{max}} = 26$, applying $E_\mathrm{V}^{\mathrm{joint}}$ to the PR3 dataset, we find that no simulation has a value as low as the data for all the pipelines.

\begin{table}
\centering
\begin{tabular}{|l||c|c|c|c||c|c|}
\hline
\multicolumn{7}{|c|}{LTP ($\%$) at $\ell_{\mathrm{max}} = 26$ } \\
\hline \hline
& \multicolumn{4}{|c||}{\textbf{PR3}} & \multicolumn{2}{|c|}{\textbf{PR4}} \\
\hline  \hline
&{\tt Commander} &{\tt NILC} & {\tt SEVEM} & {\tt SMICA} & {\tt hybrid} & {\tt SEVEM} \\
\hline \hline 
$E^{\mathrm{joint}}_{\mathrm{V}}$ & $<0.33$& $<0.33$& $<0.33$ & $<0.33$ & $1.50$ & $1.16$ \\ 
\hline
$E^{\mathrm{(TT)}}_{\mathrm{V}}$  & $<0.33$& $<0.33$& $0.33$  & $<0.33$ & $1.76$ & $1.16$ \\
\hline
$E^{\mathrm{(EE)}}_{\mathrm{V}}$  & $12.33$& $32.33$& $32.66$  & $5.33$ & $19.50$ & $68.16$ \\
\hline
$E^{\mathrm{(TE)}}_{\mathrm{V}}$  & $6.00$& $19.00$& $4.66$  & $6.00$ & $34.50$ & $18.50$ \\
\hline
\end{tabular}
\caption{LTP ($\%$) computed at $\ell_{\mathrm{max}} = 26$ for the estimators $E_\mathrm{V}^{\mathrm{joint}}$, $E_\mathrm{V}^{\mathrm{(TT)}}$, $E_\mathrm{V}^{\mathrm{(EE)}}$ and $E_\mathrm{V}^{\mathrm{(TE)}}$ when applied to the PR3 and PR4 dataset. Note that where the LTP is $<0.33 \%$, it indicates that the data value is smaller than all the simulations.}
\label{tab:LTP}
\end{table}

\begin{figure}
\centering
\begin{tabular}{c}
\includegraphics[width=150mm]{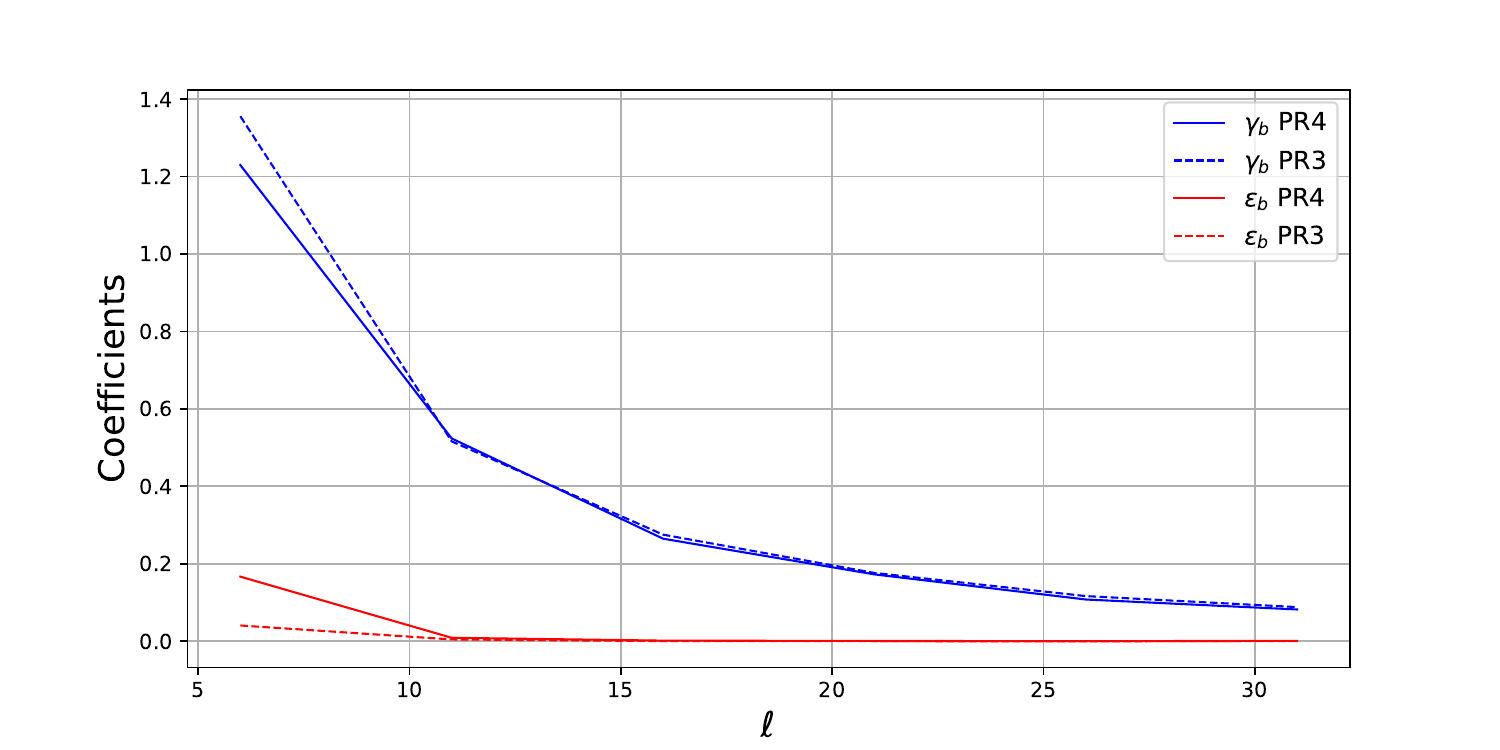}
\end{tabular}
\caption{The behaviour of the coefficients $\gamma_{b}$ (blue lines) and $\epsilon_{b}$ (red lines), as defined in Eqs.~(\ref{eq:EVjoint_bin}),(\ref{eq:gammab}) and (\ref{eq:epsilonb}), as a function of $\ell \in [2,31]$ for the {\tt SEVEM} PR3 (dashed lines) and PR4 dataset (solid lines). }
\label{fig:EVcoeff_sevem}
\end{figure}

Figure \ref{fig:EVcoeff_sevem} displays $\gamma_{b}$ and $\epsilon_b$ (see equations (\ref{eq:gammab}) and (\ref{eq:epsilonb})) as a function of $\ell$ for the PR3 (dashed lines) and PR4 datasets (solid lines) for the {\tt SEVEM} pipeline\footnote{Similar results are found for all the other pipelines.}. Note how $\epsilon_b$ for $\ell > 10$ goes to zero because of the noise level in polarisation.
Moreover this plot indicates that in this estimator the largest scales weight much more than small scales, for both the NAPS. 

\begin{figure}[htbp]
\centering
\includegraphics[width=150mm]{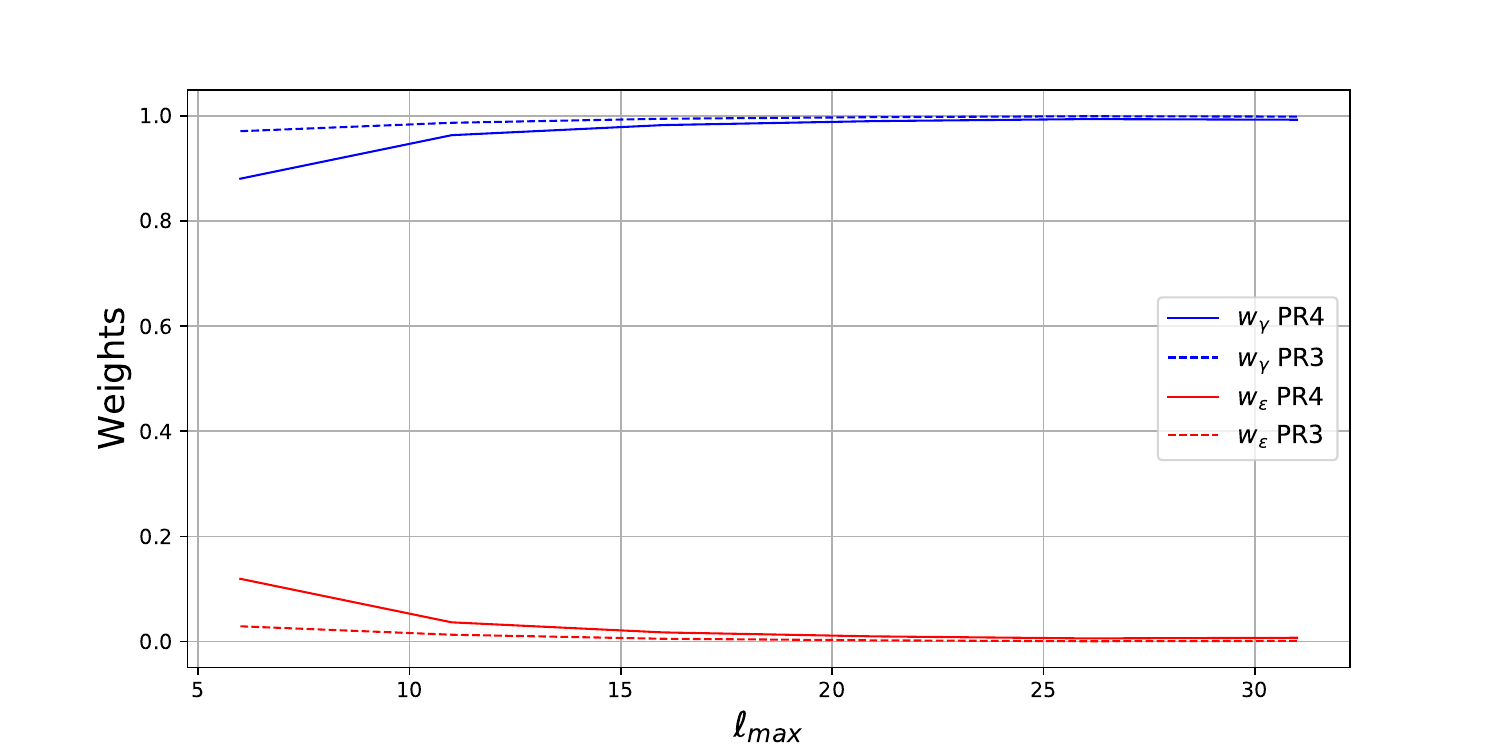}
\caption{The behaviour of $w_{\gamma}$ (in blue) and $w_{\epsilon}$ (in red), defined in Eqs. (\ref{eq:weightg}) and (\ref{eq:weighte}), as a function of $\ell_{\mathrm{max}} \in [2,31]$, for the SEVEM PR3 (dashed lines) and PR4 dataset (solid lines).  }
\label{fig:EVweights_sevem} 
\end{figure}

In order to better evaluate the impact of polarisation and temperature data on $E_\mathrm{V}^{\mathrm{joint}}$ we define the following weights

\begin{eqnarray}
w_{\gamma} (b_{\mathrm{max}})&=& 0.5 \sum_{b} \frac{2\ell_{\mathrm{eff}}^{(b)}+1}{4\pi } \gamma_{b} \, , \label{eq:weightg}\\
w_{\epsilon} (b_{\mathrm{max}}) &=& 0.5 \sum_{b} \frac{2\ell_{\mathrm{eff}}^{(b)}+1}{4\pi }  \epsilon_{b} \label{eq:weighte}\, ,
\end{eqnarray}
such that $w_{\gamma} (b_{\mathrm{max}}) + w_{\epsilon} (b_{\mathrm{max}}) = 1$ for every $b_{\mathrm{max}}$. 
For $b_{\mathrm{max}}=5$ \footnote{where $b_{\mathrm{max}}$ stands for the maximum bin considered. In this analysis we consider $\ell \in [2,31]$, with a constant bin of $\Delta \ell = 5$, therefore we have six bins and $b_{\mathrm{max}} \in [1,6]$.}, which contains information up to $\ell_{\mathrm{max}}=26$ where the LTP is minimum, we find that polarised PR3 data contribute at the level of $0.1\%$ to the building of $E_\mathrm{V}^{\mathrm{joint}}$. This value increases to $0.6\%$ for PR4 polarised data at the same maximum multipole. The behaviour of $w_{\gamma}$ and $w_{\epsilon}$ for each $\ell_{\mathrm{max}}$ for the {\tt SEVEM} PR3 (dashed lines) and PR4 dataset (solid lines) is given in Figure \ref{fig:EVweights_sevem}.
Even if the contribution of polarisation increases using PR4 data by a factor $\sim 6$, the joint analysis is still limited by the noise in \textit{Planck} polarisation. However, for the PR3 dataset it is interesting to note how the inclusion of the subdominant polarisation part in the analysis makes the LTP of $E_\mathrm{V}^{\mathrm{joint}}$ smaller than the one of $E_\mathrm{V}^{\mathrm{(TT)}}$ for the whole harmonic range considered.

%% file: 5conclusions.tex
\section{Conclusions}
\label{sec:Conclusion}
In this paper we evaluated the lack of power anomaly present at large angular scales of the CMB map considering the latest \textit{Planck} data, both PR3 (2018) and PR4 (2020) releases. Up to now lack of power has never been studied with PR4 dataset. In particular, our aim was to deeply investigate how CMB polarisation improves the understanding of this anomaly, expanding on the previous work \cite{Billi:2019vvg}.

We have defined and applied on the latest \textit{Planck} data releases in the harmonic range $2 \le \ell \le 31$, a new class of optimised estimators, see Eqs. (\ref{eq:EVjoint}), (\ref{eq:EV1}), (\ref{eq:EV2}),(\ref{eq:EVEE}) and  (\ref{eq:EVTE}), able to test this signature considering temperature and polarisation data both separately and in a jointly way. In order to critically evaluate this feature, taking into account the residuals of known systematic effects present in the \textit{Planck} datasets, we have considered the cleaned CMB maps (data and end-to-end simulations). Moreover, with the aim of reducing any possible correlation effect between different multipoles arising from the sky mask, we have binned the CMB multipoles with $\Delta \ell = 5$.  The main outcomes of this analysis are listed below.

\begin{enumerate}
\item The estimator based only on temperature data has confirmed the presence of a lack of power with a LTP $\leq 0.33\%$ (PR3) and $\leq 1.76\%$ (PR4). Note that the LTP$ \leq 0.33\%$ for the PR3 dataset is the lowest LTP present in the literature obtained from \textit{Planck} 2018 data considering the confidence sky mask.
\item We found significant differences between PR3 and PR4 datasets when polarisation is taken into account, most likely due to a combination of two factors: a different level of systematics between the PR3 and PR4 datasets and the transfer function which impacts the PR4 data. However, we have also shown that when the estimator based only on EE power spectrum, $E_\mathrm{V}^{\mathrm{(EE)}} $, is applied to the PR3 data, and in particular to the {\tt Commander} and {\tt SMICA} pipelines, at very large scale we found a LTP $\leq 8\%$ for $\ell_{\mathrm{max}}  \le 16$, smaller than the probability we got with the estimator based only on temperature data. Although this behaviour is not statistically significant, it is interesting to note that could point in the direction of a lack of power in large scales also for polarisation. Future observations, such as those from the LiteBIRD satellite, will help to shed light on this result.

\item Considering the joint estimator, $E_\mathrm{V}^{\mathrm{joint}} $, the statement of the LTP is very similar to the one of the probability we got from $E_\mathrm{V}^{\mathrm{(TT)}} $. At $\ell_{\mathrm{max}}  = 26$, where the LTP reaches its minimum, we found that the polarised PR3 data contribute at the level of $0.1\%$ to the total information budget of the estimator. This value increases to $0.6\%$ for the PR4 polarised data at the same maximum multipole, see Figure \ref{fig:EVweights_sevem}. Even though the contribution of polarisation increases using PR4 data by factor $\sim 6$, the joint analysis is still limited by the noise in \textit{Planck} polarisation. However, for the PR3 dataset it is interesting to note that the inclusion of polarisation information through the joint estimator has provided estimates which are less likely accepted in a $\Lambda$CDM model than the corresponding only-temperature version of the same estimator. In particular, at $\ell_{\mathrm{max}}  = 26$, we found that no simulation has a value as low as the data for all the pipelines.
\end{enumerate}

%% file: appendixA.tex
\section{Computation of the coefficients for the joint estimator}
\label{sec:appendixA}
The method of the Lagrange multipliers is employed to minimise the variance of the joint estimator $E_\mathrm{V}^\mathrm{{joint}}$, keeping fixed its expected value. This can be achieved by requiring that:
\begin{equation}
\sum_{\ell=2}^{\ell_\mathrm{{max}}} \frac{2\ell+1}{4\pi }( \gamma_{\ell} + \epsilon_{\ell} ) = \mathrm{const} = 1  \, .\label{eq:EVconstraint}
\end{equation}
for each maximum multipole $\ell_\mathrm{{max}}$. It should be noted that the value selected for the constant is totally arbitrary. Replacing the definition of  $E_\mathrm{V}^\mathrm{{joint}}$, as defined in Eq. (\ref{eq:EVjoint}), within the expression of $\mathrm{var}(E_\mathrm{V}^\mathrm{{joint}})$, we obtain:
\begin{eqnarray}
\mathrm{var}(E_\mathrm{V}^\mathrm{{joint}}) & \equiv &\langle \left( E_\mathrm{V}^\mathrm{{joint}} - \langle E_\mathrm{V}^\mathrm{{joint}} \rangle \right)^{2} \rangle = \langle \left( E_\mathrm{V}^\mathrm{{joint}} \right)^{2}\rangle - \langle E_\mathrm{V}^\mathrm{{joint}} \rangle^{2} = \nonumber \\ 
&=& \sum_{\ell} \mathrm{var} \left(E_\mathrm{V}^\mathrm{{joint}}(\ell) \right) \, ,
\label{eq:varEVjoint}
\end {eqnarray}
where the cross-terms among different multipoles vanish in the full sky case. Here $E_\mathrm{V}^\mathrm{{joint}}(\ell) $  is defined as:
\begin{equation}
E_\mathrm{V}^\mathrm{{joint}}(\ell)  =  \tilde \gamma_{\ell}  x_{\ell}^{\mathrm{(TT)}} + \tilde \epsilon_{\ell} x_{\ell}^{\mathrm{(uEE)}},
\label{eq:EVjoint_ell}
\end{equation}
with a variance of:
\begin{equation}
\mathrm{var} \left(E_\mathrm{V}^\mathrm{{joint}}(\ell) \right) =  \tilde \gamma_{\ell} ^{2} \, \mathrm{var}(x_{\ell}^{\mathrm{(TT)}}) + \tilde \epsilon_{\ell}^{2} \, \mathrm{var}(x_{\ell}^{\mathrm{(uEE)}}) + 2 \tilde \gamma_{\ell} \tilde \epsilon_{\ell} \, \mathrm{cov}(x_{\ell}^{\mathrm{(TT)}},x_{\ell}^{\mathrm{(uEE)}}) \, .
\label{eq:varEVjoint_ell}
\end{equation}
In Eqs. (\ref{eq:EVjoint_ell}) and (\ref{eq:varEVjoint_ell}), the tilde quantities are defined as $\tilde a = \frac{2\ell+1}{4\pi } a$. 

Given Eq.~(\ref{eq:varEVjoint}), the minimisation of $\mathrm{var}( E_\mathrm{V}^\mathrm{{joint}})$ is equivalent to minimise each $\mathrm{var}( E_\mathrm{V}^\mathrm{{joint}}(\ell))$. As is customary in the Lagrange multiplier method, for each multipole $\ell$ we introduce a new variable 
\begin{equation}
\tilde \kappa_{\ell} = \frac{2\ell+1}{4\pi } \kappa \,, 
\end{equation}
known as the Lagrange multiplier, and we minimise the function $F(\tilde \gamma_{\ell},\tilde \epsilon_{\ell},\tilde \kappa_{\ell})$, defined as:
\begin{equation}
F(\tilde \gamma_{\ell},\tilde \epsilon_{\ell},\tilde \kappa_{\ell}) = \mathrm{var} \left(E_\mathrm{V}^\mathrm{{joint}}(\ell) \right) + \tilde \kappa_{\ell} \left[ \left(\sum_{\ell} \tilde \gamma_{\ell} + \tilde \epsilon_{\ell} \right) -1\right] \, \label{eq:Ffunx}. 
\end{equation}
This is equivalent to minimise the variance of $E_\mathrm{V}^\mathrm{{joint}}(\ell)$ on the constraint given by equation (\ref{eq:EVconstraint}).
Therefore, we compute the partial derivatives with respect to the coefficients $\tilde \gamma_{\ell}$, $\tilde \epsilon_{\ell}$ and $\tilde \kappa_{\ell}$ and set them to be zero:
\begin{eqnarray}
\frac{\partial F(\tilde \gamma_{\ell},\tilde \epsilon_{\ell},\tilde \kappa_{\ell})}{\partial \tilde \gamma_{\ell}} & = & 2\tilde \gamma_{\ell} \, \mathrm{var}(x_{\ell}^{\mathrm{(TT)}}) + 2\tilde \epsilon_{\ell} \, \mathrm{cov}(x_{\ell}^{\mathrm{(TT)}},x_{\ell}^{\mathrm{(uEE)}}) + \tilde \kappa_{\ell} = 0, \\
\frac{\partial F(\tilde \gamma_{\ell},\tilde \epsilon_{\ell},\tilde \kappa_{\ell})}{\partial \tilde \epsilon_{\ell}} & = & 2\tilde \epsilon_{\ell} \, \mathrm{var}(x_{\ell}^{\mathrm{(uEE)}}) + 2 \tilde \gamma_{\ell} \, \mathrm{cov}(x_{\ell}^{\mathrm{(TT)}},x_{\ell}^{\mathrm{(uEE)}}) + \tilde \kappa_{\ell} = 0, \\
\frac{\partial F(\tilde \gamma_{\ell},\tilde \epsilon_{\ell},\tilde \kappa_{\ell})}{\partial \tilde \kappa_{\ell}} & = & \left(\sum_{\ell} \tilde \gamma_{\ell} + \tilde \epsilon_{\ell} \right) -1 = 0 \, .
\end{eqnarray}
Solving the equations we get:
\begin{eqnarray}
&& \tilde \gamma_{\ell} = - \frac{\tilde \kappa_{\ell}}{2} \frac{\mathrm{var}(x_{\ell}^{\mathrm{(uEE)}})-\mathrm{cov}(x_{\ell}^{\mathrm{(TT)}} ,x_{\ell}^{\mathrm{(uEE)}})} {\mathrm{var}( x_{\ell}^{\mathrm{(TT)}} )\mathrm{var}( x_{\ell}^{\mathrm{(uEE)}} )-[\mathrm{cov}(x_{\ell}^{\mathrm{(TT)}} ,x_{\ell}^{\mathrm{(uEE)}})]^2 } \, , \label{eq:fd_gammaell} \\
&& \tilde \epsilon_{\ell} = - \frac{\tilde \kappa_{\ell}}{2}   \frac{\mathrm{var}(x_{\ell}^{\mathrm{(TT)}})-\mathrm{cov}(x_{\ell}^{\mathrm{(TT)}} ,x_{\ell}^{\mathrm{(uEE)}})} {\mathrm{var}( x_{\ell}^{\mathrm{(TT)}} )\mathrm{var}( x_{\ell}^{\mathrm{(uEE)}} )-[\mathrm{cov}(x_{\ell}^{\mathrm{(TT)}} ,x_{\ell}^{\mathrm{(uEE)}})]^2 } \, , \label{eq:fd_epsilonell}  \\
&& - \sum_{\ell=2}^{\ell_{max}} \frac{\tilde \kappa_{\ell}}{2} \, \frac{\mathrm{var}(x_{\ell}^{\mathrm{(TT)}}) + \mathrm{var}( x_{\ell}^{\mathrm{(uEE)}})-2\mathrm{cov}(x_{\ell}^{\mathrm{(TT)}} ,x_{\ell}^{\mathrm{(uEE)}})} {\mathrm{var}( x_{\ell}^{\mathrm{(TT)}} )\mathrm{var}( x_{\ell}^{\mathrm{(uEE)}} )-[\mathrm{cov}(x_{\ell}^{\mathrm{(TT)}} ,x_{\ell}^{\mathrm{(uEE)}})]^2 } = 1\label{eq:kappaell} \,.
\end{eqnarray}
Substituting the definition of variable $\tilde \kappa_{\ell} = \frac{2\ell+1}{4\pi } \kappa$ in Eq.(\ref{eq:kappaell}), we get:
\begin{equation}
\kappa = - \frac{2}{\sum_{\ell} \frac{2\ell+1}{4\pi } \frac{\mathrm{var}(x_{\ell}^{\mathrm{(TT)}}) + \mathrm{var}( x_{\ell}^{\mathrm{(uEE)}})-2\mathrm{cov}(x_{\ell}^{\mathrm{(TT)}} ,x_{\ell}^{\mathrm{(uEE)}})} {\mathrm{var}( x_{\ell}^{\mathrm{(TT)}} )\mathrm{var}( x_{\ell}^{\mathrm{(uEE)}} )-[\mathrm{cov}(x_{\ell}^{\mathrm{(TT)}} ,x_{\ell}^{\mathrm{(uEE)}})]^2 }} \label{eq:kappa}\,. 
\end{equation}

Finally, multiplying the tilde quantities in Eqs. (\ref{eq:fd_gammaell}) and (\ref{eq:fd_epsilonell}) by the factor of $\frac{4\pi }{2\ell+1}$ and then substituting Eq.(\ref{eq:kappa}), we get the coefficients  $\gamma_{\ell}$ and $\epsilon_{\ell}$ as defined in Eqs.(\ref{eq:gammaell}) and (\ref{eq:epsilonell}).

%% file: appendixB.tex
\section{Transfer function for PR4 dataset}
\label{sec:appendixB}
The PR4 {\tt SEVEM} dataset provides the simulations divided into signal and noise maps and therefore we are able to obtain the E-mode transfer function (TF) from the MC simulations as:
\begin{equation}
\mathrm{TF} = \frac{C_{\ell, \,\mathrm{S} \otimes \mathrm{(S+N)}}^{\mathrm{EE}}}{C_{\ell, \, \mathrm{S}}^{\mathrm{EE}}}   \,, 
\label{eq:def_TFEE}
\end{equation}
where at the numerator we have the cross-mode EE spectra obtained from a signal map and a signal-plus-noise map, while at the denominator is present the EE auto-spectra of the signal component. The transfer function (TF) is computed for all the 600 Monte Carlo (MC) simulations for both the detector splits (A and B). The average over the MC simulations is then taken into account, resulting in the mean transfer functions for detector A $\mathrm{(TF^{(A)})}$ and for detector B $\mathrm{(TF^{(B)})}$.

The PR4 {\tt Commander} dataset does not provide the simulations divided in signal and noise maps, therefore we cannot compute the transfer function by ourselves. However in \cite{Planck:2020olo}, the E-mode transfer function over 60\% of the sky is provided for this pipeline.

\begin{figure}[htbp]
\centering
\includegraphics[width=150mm]{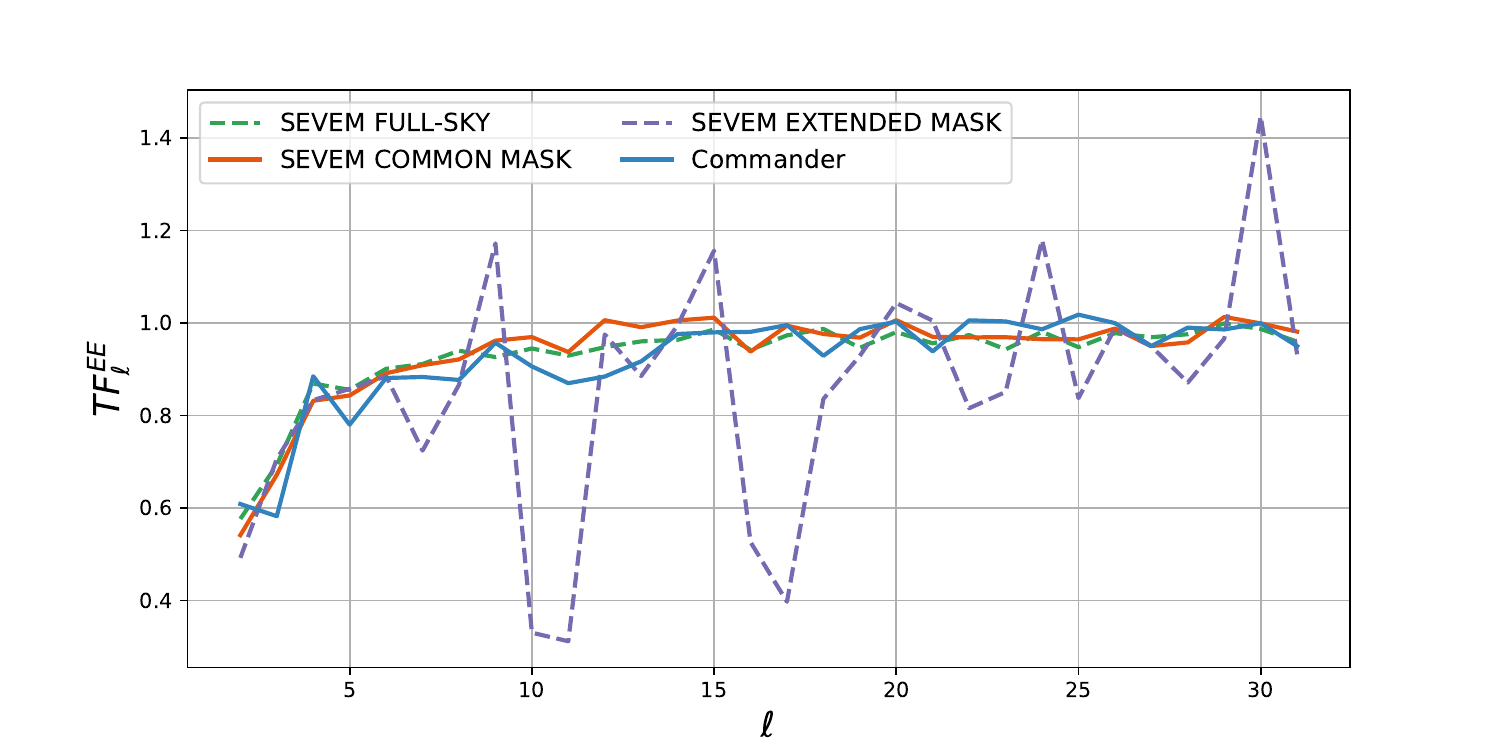} 
\caption{E-mode transfer function provided for {\tt Commander} (blue solid line), and  the $\mathrm{(TF^{(A)})}$ we compute from the {\tt SEVEM} detector A maps for three different cases of sky-coverage: full-sky (green dashed line), \textit{Planck} common polarisation mask (orange solid line) and extended mask used in this work(violet dashed line), as shown in Fig.~\ref{fig:TQU_masks}. The E-mode transfer functions shown with solid lines are those used in this analysis.}
\label{fig:transf_func} 
\end{figure}

Figure~\ref{fig:transf_func} shows the comparison between the E-mode transfer function provided by the {\tt Commander} team (blue solid line) and the $\mathrm{TF^{(A)}}$ we compute from the {\tt SEVEM} detector A maps for three different cases of sky coverage: full-sky (green dashed line), \textit{Planck} polarisation common mask (orange solid line) and extended mask used in this work, right panel of Figure~\ref{fig:TQU_masks}, (violet dashed line). 

 It can be observed that the {\tt Commander} TF and both the {\tt SEVEM} $\mathrm{TF^{(A)}}$ computed considering full-sky and the \textit{Planck} common mask are highly similar. The {\tt SEVEM} $\mathrm{TF^{(A)}}$ obtained with the extended mask is very noisy but follows the behaviour of the one computed with the standard mask.  Therefore, we apply the latter to the theoretical power spectra.   

 The values of the theoretical power spectra TE and EE, which enter in the equation of NAPS (\ref{eq:defNAPS2}), (\ref{eq:defNAPSEE}) and (\ref{eq:defNAPSTE}), are corrected for these transfer functions. In the NAPS computed for the {\tt SEVEM} dataset the following correction is applied: 
\begin{eqnarray} 
C_{\ell}^{\mathrm{EE,rescaled}} &=& C_{\ell}^{\mathrm{EE,th}} *\left(\mathrm{TF^{(A)}}*\mathrm{TF^{(B)}}\right) \,, \\
C_{\ell}^{\mathrm{TE,rescaled}} &=& C_{\ell}^{\mathrm{TE,th}} * \left(\frac{\mathrm{TF^{(A)}}+\mathrm{TF^{(B)}}}{2} \right) \,.
\label{eq:corr_TF_sevem}
\end{eqnarray}
In the case of the {\tt Commander} dataset the correction is:
\begin{eqnarray} 
C_{\ell}^{\mathrm{EE,rescaled}} &=& C_{\ell}^{\mathrm{EE,th}} *(\mathrm{TF}^2) \,, \\
C_{\ell}^{\mathrm{TE,rescaled}} &=& C_{\ell}^{\mathrm{TE,th}} *(\mathrm{TF}) \, ,
\label{eq:corr_TF_comm}
\end{eqnarray}
where TF is the E-mode transfer function provided for {\tt Commander} by the \textit{Planck} team. 

%% file: appendixC.tex
\section{Comparison with previous estimator}
\label{sec:appendixC}
This appendix presents the application of the joint estimator for the lack of power developed in \cite{Billi:2019vvg} to latest \textit{Planck} {\tt SEVEM} datasets. The NAPS $x^{(\mathrm{TT})}_{\ell}$ and $x^{(\mathrm{uEE})}_{\ell}$ were combined to define the optimal, i.e. with minimum variance, estimator $\mathrm{\tilde P}$ as follow:
\begin{equation}
\mathrm{\tilde P} \equiv \frac{1}{(\ell_{max}-1) }\sum_{\ell=2}^{\ell_{max}} ( \alpha_{\ell} x^{(\mathrm{TT})}_{\ell} + \beta_{\ell} x^{(\mathrm{uEE})}_{\ell} ) \,. 
\label{eq:def_Ptilde}
\end{equation}
with:
\begin{eqnarray}
\alpha_{\ell} &=& 2 \frac{\mathrm{var(}x^{(\mathrm{uEE})}_{\ell})-\mathrm{cov(}x^{(\mathrm{TT})}_{\ell} ,x^{(\mathrm{uEE})}_{\ell})} {\mathrm{var(} x^{(\mathrm{TT})}_{\ell} ) + \mathrm{var(} x^{(\mathrm{uEE})}_{\ell} )-2\mathrm{cov(}x^{(\mathrm{TT})}_{\ell} ,x^{(\mathrm{uEE})}_{\ell}) } \, , \label{alphasol} \\
\beta_{\ell} &=& 2 \frac{\mathrm{var(} x^{(\mathrm{TT})}_{\ell} )-\mathrm{cov(}x^{(\mathrm{TT})}_{\ell} ,x^{(\mathrm{uEE})}_{\ell})} {\mathrm{var(} x^{(\mathrm{TT})}_{\ell} ) + \mathrm{var(} x^{(\mathrm{uEE})}_{\ell} )-2\mathrm{cov(}x^{(\mathrm{TT})}_{\ell} ,x^{(\mathrm{uEE})}_{\ell}) } \, , \label{betasol} \,, 
\end{eqnarray}
 where $\mathrm{var(}x_{\ell}^{\mathrm{(TT)}})$ and $\mathrm{var(}x_{\ell}^{\mathrm{(uEE)}})$ are the variance of $x_{\ell}^{\mathrm{(TT)}}$ and $x_{\ell}^{\mathrm{(uEE)}}$ respectively, and $\mathrm{cov(}x_{\ell}^{\mathrm{(TT)}},x_{\ell}^{\mathrm{(uEE)}})$ is their covariance
\begin{equation}
\mathrm{cov(}x_{\ell}^{\mathrm{(TT)}},x_{\ell}^{\mathrm{(uEE)}}) = \langle(x_{\ell}^{\mathrm{(TT)}} - \langle x_{\ell}^{\mathrm{(TT)}} \rangle)(x_{\ell}^{\mathrm{(uEE)}} -\langle x_{\ell}^{\mathrm{(uEE)}}\rangle) \rangle.
\end{equation}
The coefficients $\alpha_{\ell}$ and $\beta_{\ell}$ are obtained with the method of the Lagrange multipliers by minimising the variance of $\mathrm{\tilde P}$, at each $\ell$, on the constraint $\alpha_{\ell} + \beta_{\ell} = \mathrm{const} = 2 $. The estimator $\mathrm{\tilde{P}}$ could be interpreted as a dimensionless normalised mean power, which jointly combines the temperature and polarisation data, with an expectation value of:
\begin{equation}
\langle\mathrm{\mathrm{ \mathrm{\tilde {P}}}} \rangle = 2 \, , 
\label{expectedvalueP}
\end{equation}
regardless of the value of $\ell_{max}$. 

\begin{figure}[htbp]
\centering
\includegraphics[width=150mm]{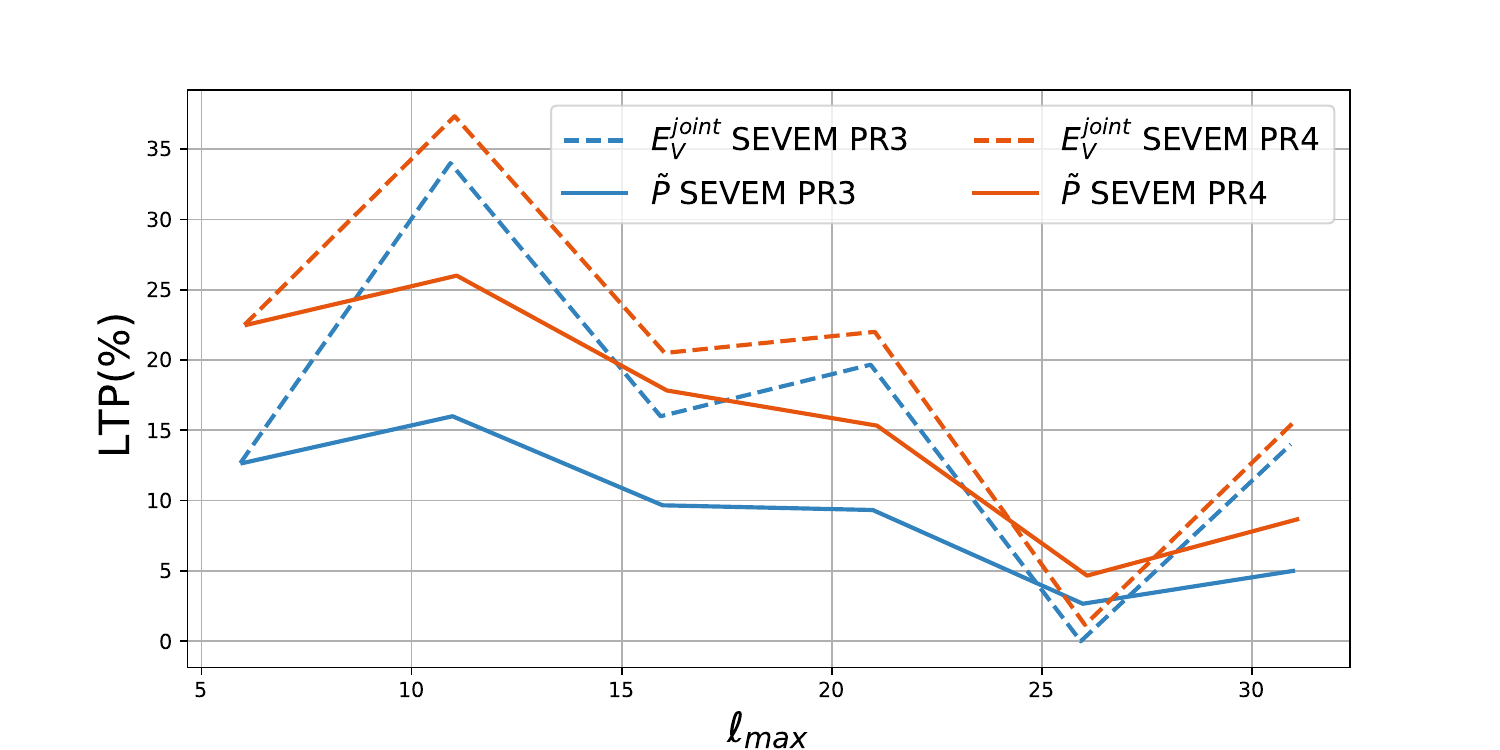} 
\caption{$\mathrm{ LTP(\%)}$ of the $\mathrm{\tilde {P}}$ (solid lines) and the $E_V^{(joint)}$ (dashed lines) estimators when applied to the PR3 (blue lines) and PR4 (orange lines) {\tt SEVEM} datasets as a function of $\ell_{max} \in [2,31]$. The LTP is obtained as the percentage of simulations with a value smaller than that of the data. However, note that, in practice, due to the limited number of simulations, the maximum sensitivity that we can reach for the LTP is 0.33\% (PR3 dataset) and 0.17\% (PR4, {\tt SEVEM} pipeline). Values below those limits in the plot indicate that none of the simulations has a lower value than the data.}
\label{fig:ltpPtilde} 
\end{figure}

The estimator is now applied to the PR3 and PR4 {\tt SEVEM} dataset, MC simulations and data, in the harmonic range $\ell \in [2,31]$, with a constant bin of $\Delta \ell = 5$. Figure \ref{fig:ltpPtilde} displays the $\mathrm{ LTP(\%)}$ computed for the $\mathrm{\tilde {P}}$ applied to the PR3 (blue solid line) and PR4 (orange solid line) {\tt SEVEM} datasets as a function of $\ell_{max}$. Furthermore, the comparison with the $E_\mathrm{V}^{\mathrm{(joint)}}$ computed with both PR3 (blue dashed line) and PR4 (orange dashed line) {\tt SEVEM} datasets is presented.
Both the estimators when applied to the PR3 dataset provide estimates that are less likely accepted in a $\Lambda$CDM model than when applied to the PR4 dataset. Moreover, it can be noted that the trends of these two estimators are comparable, exhibiting peaks and valleys at the same values of $\ell_{max}$.

